\documentclass[12pt]{article}
\pdfoutput=1 
\usepackage{tikz}
\usepackage{amssymb}
\usepackage{epsfig}
\usepackage{bbm}
\usepackage{bbold}
\usepackage{graphicx}
\usepackage{hyperref}
\usepackage{romannum}
\usepackage{graphics,graphpap}
\usepackage{amsmath,exscale,amsthm}
\usepackage{yfonts}
\usepackage{mathrsfs}
\usepackage{float}
\usepackage{hyperref}
\usepackage{subcaption}
\usepackage{cite}
\usepackage[T1]{fontenc}
\usepackage[utf8]{inputenc}

\usepackage[none]{hyphenat}
\usetikzlibrary{matrix,positioning,decorations.pathreplacing}

\pdfoutput=1
\usepackage{color}
\usepackage{amssymb}
\usepackage{epsfig}
\usepackage{romannum}

\def\e{\varepsilon}

\newcommand{\wt}{\widetilde}

\usepackage{amsmath}
\usepackage{amsfonts}

\begin{document}

\def\a{\alpha}
\def\b{\beta}
\def\c{\chi}
\def\d{\delta}
\def\e{\epsilon}
\def\f{\phi}
\def\g{\gamma}
\def\h{\eta}
\def\i{\iota}
\def\j{\psi}
\def\k{\kappa}
\def\la{\lambda}
\def\m{\mu}
\def\n{\nu}
\def\o{\omega}
\def\p{\pi}
\def\q{\theta}
\def\r{\rho}
\def\s{\sigma}
\def\t{\tau}
\def\u{\upsilon}
\def\x{\xi}
\def\z{\zeta}
\def\D{\Delta}
\def\F{\Phi}
\def\G{\Gamma}
\def\J{\Psi}
\def\L{\Lambda}
\def\O{\Omega}
\def\P{\Pi}
\def\Q{\Theta}
\def\S{\Sigma}
\def\U{\Upsilon}
\def\X{\Xi}

\def\ve{\varepsilon}
\def\vf{\varphi}
\def\vr{\varrho}
\def\vs{\varsigma}
\def\vq{\vartheta}

\def\dg{\dagger}                                     
\def\ddg{\ddagger}                                   
\def\wt#1{\widetilde{#1}}                    
\def\mt{\widetilde{m}_1}
\def\mti{\widetilde{m}_i}
\def\rt{\widetilde{r}_1}
\def\mtt{\widetilde{m}_2}
\def\mttt{\widetilde{m}_3}
\def\rtt{\widetilde{r}_2}
\def\mb{\overline{m}}
\def\VEV#1{\left\langle #1\right\rangle}        
\def\be{\begin{equation}}
\def\ee{\end{equation}}
\def\ds{\displaystyle}
\def\ra{\rightarrow}

\def\bea{\begin{eqnarray}}
\def\eea{\end{eqnarray}}
\def\NO{\nonumber}
\def\Bar#1{\overline{#1}}


\def\pl#1#2#3{Phys.~Lett.~{\bf B {#1}} ({#2}) #3}
\def\np#1#2#3{Nucl.~Phys.~{\bf B {#1}} ({#2}) #3}
\def\prl#1#2#3{Phys.~Rev.~Lett.~{\bf #1} ({#2}) #3}
\def\pr#1#2#3{Phys.~Rev.~{\bf D {#1}} ({#2}) #3}
\def\zp#1#2#3{Z.~Phys.~{\bf C {#1}} ({#2}) #3}
\def\cqg#1#2#3{Class.~and Quantum Grav.~{\bf {#1}} ({#2}) #3}
\def\cmp#1#2#3{Commun.~Math.~Phys.~{\bf {#1}} ({#2}) #3}
\def\jmp#1#2#3{J.~Math.~Phys.~{\bf {#1}} ({#2}) #3}
\def\ap#1#2#3{Ann.~of Phys.~{\bf {#1}} ({#2}) #3}
\def\prep#1#2#3{Phys.~Rep.~{\bf {#1}C} ({#2}) #3}
\def\ptp#1#2#3{Progr.~Theor.~Phys.~{\bf {#1}} ({#2}) #3}
\def\ijmp#1#2#3{Int.~J.~Mod.~Phys.~{\bf A {#1}} ({#2}) #3}
\def\mpl#1#2#3{Mod.~Phys.~Lett.~{\bf A {#1}} ({#2}) #3}
\def\nc#1#2#3{Nuovo Cim.~{\bf {#1}} ({#2}) #3}
\def\ibid#1#2#3{{\it ibid.}~{\bf {#1}} ({#2}) #3}

%
\title{On the origin of matter in the Universe}
\author{Pasquale Di Bari \\
\\
{\it\small School of Physics and Astronomy},
{\it\small University of Southampton,} \\
{\it\small  Southampton, SO17 1BJ, U.K.}
}
\maketitle \thispagestyle{empty}
\pagenumbering{arabic}

\begin{abstract}
The understanding of the physical processes that lead to the origin of matter  in the early Universe,  
creating both an excess of matter over anti-matter and a dark matter abundance that survived until the present, 
is one of the most  fascinating  challenges in modern science. 
The problem cannot be addressed within our current description of fundamental physics
and, therefore, it currently provides a very strong evidence of new physics.
Solutions  can either reside in a modification of the standard model of elementary particle physics
or in a modification of the way we describe gravity, based on general relativity, or at the interface of both. We will mainly discuss the first class of solutions.  Traditionally, models that separately explain either the matter-antimatter asymmetry of the Universe or dark matter 
have been proposed.  However, in the last years there has also been an accreted interest and 
intense activity on scenarios able to provide a unified picture of the origin of matter in the early universe. 
In this review we discuss some of the main ideas emphasising primarily those 
models that have more chances to be experimentally tested during  next years.  Moreover, after a general discussion, we will focus on extensions of the standard model that can also address neutrino masses and mixing. 
Since this is currently the only evidence of physics beyond the  standard model coming directly from particle physics experiments, it is then reasonable that such extensions might also provide a solution to the problem of the origin of 
matter in the universe.
\end{abstract}

\newpage

\tableofcontents

\newpage
\section{Introduction}

The discovery of the Higgs boson at the Large Hadron Collider (LHC) \cite{higgsdiscovery} has confirmed the Brout-Englert-Higgs mechanism \cite{BEH}
for the origin of masses of  elementary particles in the standard model of particle physics and fundamental interactions (SM), providing the last missing piece for a full confirmation of its validity. 
Since then,  signs of new physics have been eagerly awaited at the LHC, in particular those that could have finally  provided (directly or indirectly) indications on the nature of the dark matter in the Universe, but so far  no compelling evidence has been found.\footnote{The most recent measurement of  the  double ratio of branching fractions in B meson decays,  $R_K= 0.846^{+0.044}_{-0.041}$,
by the LHCb experiment exhibits a $3.1\s$ deviation from the SM expected value ($R_K = 1$), hinting to a violation of lepton universality \cite{lhcb}. 
This can intriguingly be regarded as the indication of the existence of new physics at a scale $\sim 100\,{\rm TeV}$ suggesting
that the discovery of new particles might be within the reach of colliders in a near future. Moreover,  
the value of the muon anomalous magnetic moment recently measured by the  Fermilab Muon g-2 Experiment, combined with the
previous value measured at the Brookhaven National Laboratory E821 experiment,  deviates at $4.1\s$ from the SM predicted value \cite{gm2flab}. This is calculated
combining perturbative expansion in the fine-structure constant $\a$  with non-perturbative approach based on dispersive relations that require
experimental information on the hadronic cross section of $e^+ + e^-$ \cite{phenoprediction}. 
However, both anomalies require further investigation. In particular, the deviation of $R_K$ from the standard model prediction is still not statistically  significant enough considering that it represents 
just one anomaly among a multitude of other measurements not showing signs of new physics. Moreover, recent lattice calculations find an hadronic contribution 
that results into a predicted value of the muon anomalous magnetic moment in the standard model that is in good agreement with the measured value \cite{bmw}.
It is then premature to claim discovery of new physics without  first a clear understanding of the correct standard model prediction.}

However, despite null results at colliders, the existence of new physics is strongly supported  
by the necessity to reconcile the SM and the cosmological observations within a unified picture.
In addition to dark matter, another strong indication of new physics is given by the failure 
to find, within the SM, an explanation of the absence of primordial anti-matter in our observable universe, 
clearly indicating $C\!P$ violation to a macroscopic level. 

These two cosmological puzzles have been traditionally addressed separately but clearly a 
unified picture for the origin of matter in the universe is not only greatly attractive but
also quite reasonable. In the last years many ideas have been proposed. 
In this review we will primarily concentrate  on extensions of the SM also able 
to explain neutrino masses and mixing, currently the only evidence of new physics 
coming directly from particle physics.  Therefore, if the origin of matter in the universe can
be regarded as a phenomenological guidance toward a more fundamental theory, vice versa 
neutrino masses and mixing could hold the solutions to the cosmological puzzles.

This is the plan of the review. In Section 2 we briefly summarise the observational 
evidence for the matter-antimatter asymmetry of the universe and why this should be regarded as a 
strong indication of new physics. We also provide a general overview of models of baryogenesis. In Section 3 we summarise the phenomenological picture that strongly points to the existence of a dark matter component in the universe of non-baryonic nature, unaccountable within  the SM or even within a minimal extension where neutrinos are massive Dirac fermions.\footnote{This does not exclude explanation of the origin of matter in the universe where
neutrinos are Dirac particles. However, adding just Dirac neutrino masses is not a sufficient ingredient of new physics.}  In Section 4 we provide a brief review of neutrino masses and mixing and how this can be 
described within extensions of the SM based on the simplest type-I seesaw mechanism. 
In Section 5 we review the main results on leptogenesis including some recent ones.
In Section 6 we show how the role of dark matter can be played by a long-lived right-handed (RH) neutrino with  keV mass produced from the mixing with  left-handed (LH) neutrinos within the type-I seesaw mechanism, the Dodelson-Widrow mechanism. We also discuss how the active-sterile neutrino mixing can be consistently embedded within the type-I seesaw mechanism  with neutrino mass and mixing experimental results from neutrino oscillation experiments. The picture, dubbed as $\nu$MSM model, can also potentially address the matter-antimatter asymmetry puzzle with leptogenesis from RH neutrino mixing realising a unified picture based on a minimal extension of the SM. The $\nu$MSM can be tested experimentally in different ways and stringent constraints have been found.
Currently, it is not clear whether viable solutions still exist within the parameter space. In Section 7 we show how the introduction of a 5-dim operator where RH neutrinos can couple to the Higgs boson can induce a RH neutrino mixing producing an amount of long-lived `dark' RH neutrinos able to reproduce the measured dark matter abundance. In addition the matter-antimatter asymmetry can be reproduced by two RH neutrino resonant leptogenesis. 
Finally, in Section 8 we draw some conclusions, discussing the future of research of models on the origin of matter in the Universe.  

\section{Baryogenesis}

If big bang nucleosynthesis represents the first example where  knowledge of nuclear physics could be applied to cosmology
and helped understanding the existence of a hot early stage in the history of the universe, as confirmed by the discovery of the  cosmic microwave background, with baryogenesis there is an important step further: the entire universe 
is used as a unique laboratory to test new theories, a way to complement ground laboratories and go beyond their limitations.
Here we discuss the basic points and review some recent progress, mainly aiming at showing how 
the matter-antimatter asymmetry of the universe is an important motivation and investigative tool to go beyond the standard model.  

\subsection{Matter-antimatter asymmetry of the universe}

The cosmic baryon abundance is today accurately and precisely measured by an analysis of CMB temperature and polarisation
anisotropies. In particular, the ratio of the amplitude of the first to the second acoustic peak in temperature anisotropies 
is particularly sensitive  to the value of the baryon abundance.\footnote{An insightful discussion on the physics of microwave background
anisotropies can be found in \cite{hu}.}
The latest analysis of the {\em Planck} collaboration, combining {\em Planck} results on CMB anisotropies
and baryon acoustic oscillation data, finds for the baryonic contribution to the energy density parameter \cite{planck18}
\be\label{OB0}
\O_{B0}h^2 = 0.02242 \pm 0.00014 \,  .
\ee
Using the {\em Planck} result for the Hubble constant
\be\label{h}
h \equiv H_0/(100\,{\rm km}\,{\rm s}^{-1}\,{\rm Mpc}^{-1})  = 0.6766 \pm 0.0042 \, ,
\ee
one has for the baryonic contribution to the energy density parameter
\be
\O_{B0} = 0.049 \pm 0.001 \,  .
\ee
Acoustic oscillations are not sensitive to the sign of the baryon number so they actually
measure the total amount of matter and anti-matter rather than the net one. 
If we consider the possibility that at the time of recombination there is also some amount of anti-matter
in the form of anti-nucleons, CMB would be then sensitive to the sum of the two contributions. 
One can then easily derive  the total baryon-to-photon number ratio 
\be\label{etaB+}
\eta^+_{B 0} \equiv {n_{B 0} + n_{\bar{B} 0}\over n_{\gamma 0}} \simeq {\O_{B0}\,\varepsilon_{\rm c 0}\over m_p \, n_{\gamma 0}}
\simeq 273.5\,\O_{B 0}h^2\,10^{-10} \,   .
\ee
Notice that in this expression, in general, one might  have a spatial  dependence of $n_{B 0}$ and $n_{\bar{B}0}$ 
but still a homogeneous sum, as acoustic peaks require. For example, one can have  matter and antimatter spatial domains 
in the universe. However,  at the moment, let us exclude this possibility and consider the homogeneous case.
Searches of antimatter in cosmic rays, coming from space,  find  no evidence of {\em primordial} antimatter. 
The amount of antimatter that we observe in cosmic rays,  mainly positrons and anti-protons,  can be explained in terms of astrophysical processes.  This implies that at the time of recombination one can assume the amount of anti-nucleons to be negligible. The observed baryon abundance then necessarily corresponds to the  net baryon
asymmetry that had to exist prior to recombination and that  survived the 
consecutive  stages of annihilations of particles and antiparticles  
taking place  in the early universe while the temperature was dropping down. 
We can then find from Eq.~(\ref{etaB+}) the value of the net baryon-to-photon number ratio at the present time, since
$\eta^+_{B 0} \simeq \eta_{B0} \equiv (n_{B0} -  n_{\bar{B} 0})/ n_{\g 0} \simeq \eta_{B,{\rm rec}}$, so that Eq.~(\ref{OB0}) translates into
\be\label{etaB0}
\eta_{B 0} = (6.12 \pm 0.04)\times 10^{-10} \,  .
\ee
This value together with the absence of primordial antimatter in our observable universe is in agreement with
the well known result \cite{zeldovich} that starting with a vanishing asymmetry one would eventually be left 
with equal values of the relic abundances of matter and antimatter far below, 
about ten orders of magnitude, the measured value in Eq.~(\ref{etaB0}).  

However, this argument  still does not exclude the possibility, 
neglected so far,  of a universe consisting of a patchwork of matter-antimatter domains on scales 
$\lambda \gtrsim 100\,{\rm Mpc}$, i.e., larger than those  probed by the lack of primordial antimatter in cosmic rays.\footnote{Strong constraints on scales $\lambda \lesssim 20\,{\rm Mpc}$, i.e., within the
size of clusters of galaxies, also come from $\gamma$-ray observations of $X$-ray emitting clusters
not showing evidence of annihilations \cite{Steigman:2008ap}.}
If we consider such a scenario, implying the existence of some mechanism that could segregate matter from anti-matter  with the creation of matter-antimatter domains on very large scales, one could think of an overall baryon symmetric universe, with the cancellation of the contributions from matter and antimatter domains.   
However, if such large matter and anti-matter domains exist, and have comparable geometry,  then the annihilation 
radiation would contribute to the cosmic $\g$-ray diffuse background. Since such excess is not observed, this excludes
the existence of domains of matter and antimatter on scales $\lambda \gtrsim 20\,{\rm Mpc}$, so even larger than those
probed by cosmic rays and as large as the entire observable universe, thus
ruling out the possibility of a baryon symmetric universe.  Therefore, today, barring some caveats, one arrives to the conclusion 
that a matter-antimatter asymmetry had to exist prior to recombination \cite{glashow}.

Primordial nucleosynthesis is also sensitive to the amount of baryonic matter. Within a standard picture, with no matter-antimatter domains, the 
value of the baryon-to-photon ratio inferred at the time of nucleosynthesis, when Deuterium forms, is in nice agreement with the measurement from CMB.  Therefore, the matter-antimatter asymmetry, and in particular the baryon asymmetry, needs to be generated prior the onset of nucleosynthesis.  Possible deviations from the standard big bang nucleosynthesis scenario have to be such not to spoil the success of its predictions. 
For example, one could consider the existence of anti-matter domains at the time of nucleosynthesis. However, these, on the basis of the arguments discussed above, need to contain a sub-dominant component of anti-matter compared to matter and need to be sufficiently small to avoid the constraints on a baryon symmetric universe. 
In this case the synthesis of primordial anti-nuclei is expected inside these small anti-matter domains in a similar way 
to how standard big bang nucleosynthesis proceeds in matter domains. Interestingly, 
the {\em AMS-02} experiment has  recently claimed the detection  of six events compatible with being anti-$^3{\rm He}$ and two events with anti-$^4{\rm He}$ nuclei \cite{ting}. 
These events cannot be explained in terms of spallation processes, predicting an anti-$^3{\rm He}$ flux two orders of magnitude below 
{\em AMS-02} sensitivity and a anti-$^4{\rm He}$ flux approximately five orders of magnitude below.
If confirmed, this discovery would point to the existence of compact antimatter domains in the form of anti-stars or anti-clouds \cite{salati}. 
The anti-nuclei should have been created during big-bang nucleosynthesis and the measured isotopic ratio, roughly 
$\bar{^4{\rm He}}:\bar{^3{\rm He}} \simeq 1:3$, would be obtained for $\bar{\eta}_B \equiv n_{\bar{B}}/n_\g \simeq 1$--$6 \times 10^{-13}$.  
However, the creation of these compact antimatter domains in the early universe would be highly non trivial to explain. 
It would require a very specific mechanism within the very early Universe, likely related to the same mechanism that generates the asymmetry.
For example a modification of the Affleck-Dine mechanism \cite{AD} has been proposed \cite{dolgov1}.
We will not consider this kind of models but it should be clear that a confirmation of the existence of primordial antimatter 
would likely rule out the mechanisms of baryogenesis we will discuss, or in any case it would require an extension able 
to describe a mechanism of formation of antimatter domains in the universe. 

The measured value of the matter-antimatter asymmetry, in the form of baryon-to-photon ratio in Eq.~(\ref{etaB0}), 
needs to exist prior to the onset of nucleosynthesis but it  cannot be understood in terms of some 
value of the baryon charge  pre-existing the inflationary stage and that remained constant until the present time \cite{dolgovlectures}. 
This would indeed conflict with successful inflation requirements since necessarily the energy density associated to the baryon charge had to be
higher than the inflation energy density, approximately constant during inflation, before some time in past. Before this time
inflation would be then inhibited, since  it requires the inflation energy density to dominate. One can easily verify that in this way
inflation could not last more than $N= 6$--$7$ e-folds and not the required $N\simeq 60$ folds needed to solve the horizon
problem and to explain the amplitude of CMB temperature anisotropies on super-horizon scales.
For this reason the matter-antimatter asymmetry has to be explained by a mechanism of dynamical  generation, 
from processes  that occurred at the end or after inflation,  a stage commonly referred to as baryogenesis \cite{sakharov}.

\subsection{Models of baryogenesis}

There are three necessary conditions that have to be typically satisfied by a model of baryogenesis for 
the generation of a baryon asymmetry at the macroscopic level:
\begin{itemize}
\item Baryon number non-conservation;
\item $C$ and $C\!P$ violation;
\item Departure from thermal equilibrium.
\end{itemize} 
These conditions are all satisfied by the  original model proposed by Sakharov \cite{sakharov}, though not explicitly stated there,   and for this reason they are commonly referred to as {\em Sakharov's conditions}. Although there are many loopholes and there  are even examples of models that do not respect any of these conditions \cite{dolgovreport}, 
Sakharov's conditions hold in most traditional models.

There is a very long list of proposed models of baryogenesis.\footnote{There are different excellent specialistic reviews or monographs on baryogenesis \cite{dolgovreport,riottotrodden,focusissue,Garbrecht:2018mrp,buchmullerbodeker}.} 
Some of them can be grouped within the same class, this is just a partial list (as emphasised by the dots):  
\begin{itemize}
\item From heavy particle decays:
\begin{itemize}
\item GUT baryogenesis \cite{Yoshimura:1978ex,Toussaint:1978br,Ellis:1978xg,Weinberg:1979bt,Harvey:1981yk,Kolb:1983ni};
\item Leptogenesis \cite{fy} .
\end{itemize}
\item From a first order phase transition at the electroweak spontaneous symmetry breaking (electroweak baryogenesis) \cite{kuzmin}:
\begin{itemize}
\item in the SM \cite{kuzmin};
\item in the two Higgs doublet model \cite{Turok:1990zg};
\item in the minimal supersymmetric standard model (MSSM) \cite{Nelson:1991ab};
\item in the next-to-minimal supersymmetric standard model (NMSSM) \cite{Pietroni:1992in};
\item in the nearly minimal supersymmetric standard model (nMSSM)  \cite{Huber:2006wf};
\item \dots
\end{itemize}
\item Affleck-Dine baryogenesis \cite{AD};
\item Baryogenesis from black hole evaporation \cite{Hawking:1974rv,Zeldovich:1976vw,Toussaint:1978br,
Turner:1979bt,Dolgov:1980gk};
\item Spontaneous baryogenesis\footnote{Spontaneous baryogenesis is a typical example of
a baryogenesis model that does not respect Sakharov's conditions since $C\!P$ is conserved  and the
asymmetry is generated in equilibrium. In this case the crucial ingredient is a temporary, dynamical violation of $C\!P\!T$ invariance.} \cite{CK};
\item Gravitational baryogenesis \cite{Davoudiasl:2004gf};
\item Cold baryogenesis \cite{Krauss:1999ng};
\item \dots
\end{itemize}

Here we will briefly discuss the first two classes of baryogenesis models,
those that are more easily probed  by different experimental tests and, for this reason, 
have attracted more interest in recent years.

\subsection{Electroweak baryogenesis}

Electroweak baryogenesis,\footnote{For specialistic reviews on electroweak baryogenesis
see \cite{kaplannelson,rubakovshaposhnikov,trodden} or for a pedagogical introduction see \cite{gwhitebook}.}
is a popular model, or better a class of models, of baryogenesis that was initially proposed
within the SM framework. However, experimental findings  rule out this possibility and, since no other models of baryogenesis relying on known physics have been found,  the matter-antimatter asymmetry of the universe is currently regarded as evidence of new physics. 

At the classical level the baryonic current $j^\mu_B$ is conserved. However,  quantum effects can give rise to 
an anomalous current contribution \cite{adler,belljackiw} that breaks conservation and in the case of the
electroweak theory one has \cite{thooft}
\be
\partial_\mu\,j^\mu_B = n_{\rm f}\,{g^2 \over 32\,\pi^2}\,G^{\mu\nu}\,\widetilde{G}_{\mu\nu} \,   ,
\ee 
where in the chiral anomaly $n_{\rm f}$ denotes the number of families, 
$G_{\mu\nu}$ is the electroweak gauge field strength and
$\widetilde{G}_{\mu\nu}$ its dual.

The chiral anomaly can be re-expressed as a total derivative $\partial_\mu K^\mu$ of the anomalous current $K^\mu$ that does not vanish at infinite so that integrating from a time 
$t=0$ to some final time $t_{\rm f}$ between two vacuum states corresponding to $G_{\mu\nu}=0$, the change
in the baryon number is non-vanishing and one obtains
\be
\Delta B(t_{\rm f}) = \D N_{\rm CS} = n_{\rm f}\,[N_{\rm CS}(t_{\rm f}) - N_{\rm CS}(0)] \,  , 
\ee
where $N_{\rm CS} = \int \, d^3x\,K^0 $ is the Chern-Simons number, an integer number.
At zero temperature the transition rate between two vacua is tiny: $\G_{\rm sph} \sim e^{-16\pi^2/g^2} \sim 10^{-160}$.  However, at high temperatures, corresponding to energies
above the energy barrier height separating two vacua and given by the static solution of  field configuration, 
transitions over the barrier can happen. The lowest energy barrier height is found as a saddle point solution
called  {\em sphaleron} \cite{manton}. The energy barrier height is found to be a few TeV's at zero temperature
and in the case there would not be observable effects. However, at high temperatures one has to include 
finite temperature effects \cite{quiroslectures} and the energy barrier gets lower until it disappears above a critical temperature \cite{kuzmin}.
During the electroweak phase transition the universe would then pass from an unbroken phase, where 
sphaleron processes are unsuppressed, to a broken phase where they are suppressed. 
 The transition rate at finite temperatures between two electroweak vacua with $\D B \neq 0$ can then be expressed as
\be\label{sphrate}
\Gamma_{\rm sph} \propto T^4 \, \exp\left[-\kappa {v(T) \over T}\right] \,   ,
\ee
where $v \equiv \langle \Phi \rangle$ is the vacuum expectation value of the Higgs with
$v = 0$ for $T \gtrsim T_{\rm c}$ (unbroken phase) 
and $v= v(T_{\rm c})$ for $T \lesssim T_{\rm c}$ (broken phase). 

Therefore, above the critical temperature the baryon number can be violated satisfying the first Sakharov condition.
If the electroweak phase transition is first order, then during the transition broken phase bubbles nucleate. Both
inside (broken phase) and outside the bubble (unbroken phase) a macroscopic baryon number cannot be generated.
The reason is that during a sphaleron transition lepton number is also non conserved together with the baryon number,
in a way that $\D B = \D L = 3$. In this way one has $\D (B-L) = 0$, while any $(B+L)$ number is washed-out in thermal equilibrium.
However, on the bubble wall there is a strong departure from thermal equilibrium and if the phase transition is strong first order, a baryon asymmetry can survive inside the bubble as we explain in a moment. First, we need to discuss how the baryon asymmetry is produced.

Departure from thermal equilibrium and baryon number violation are still  not sufficient conditions to create an asymmetry: one also needs that $C\!P$ is violated, otherwise, on average, for any process creating a baryon number there would be a $C\!P$ conjugated  inverse process that destroys it.  The asymmetry is actually initially produced in the form of a chiral asymmetry on the bubble wall since LH and RH particles
interact with different rate with the bubble wall if $C\!P$ is violated. In this way LH particles penetrate the bubble more efficiently and this generates a chiral asymmetry  on the bubble wall.
It is then crucial that the chiral asymmetry is reprocessed into a baryon asymmetry by sphalerons inside the bubble  and this implies that electroweak baryogenesis ends when the temperature drops below the out-of-equilibrium temperature of sphalerons \cite{rummukainen}
\be\label{sphalerons}
T_{\rm sph}^{\rm out} \simeq 132\,{\rm GeV} \,   .
\ee
It should be emphasised that the sphaleron rate needs to vanish quickly inside the bubble once the baryon asymmetry has been produced, since otherwise after having reprocessed the chiral asymmetry into a baryon asymmetry, sphalerons would also destroy the baryon asymmetry if they are in equilibrium. From 
Eq.~(\ref{sphrate}) one can see that this requires the condition $v(T_{\rm c})/T_{\rm c} \gtrsim 1$\footnote{Notice that the criterium is  approximate and gauge dependent. A more sophisticate and gauge independent approach is discussed in \cite{Patel:2011th}.}
implying that the phase transition has to be a strongly first order phase transition.\footnote{Consider indeed that for a second order phase transition one has $v(T_{\rm c}) = 0$.} 
It is quite non-trivial that successful electroweak baryogenesis could, in principle, be realised within the SM.  The source of $C\!P$ violation would be in this case 
provided by the presence of phases in the quark mixing matrix, the CKM matrix. 
Moreover the Higgs potential could in principle experience a strongly first order phase transition.
Therefore, potentially, the observed baryon asymmetry could be explained in the SM, since all three Sakharov conditions 
could be satisfied. However, one finds that
the value of $v(T_{\rm c})/T_{\rm c}$  in the SM is related to the Higgs mass and it turns out that a strongly first order phase transition can occur only for $m_H \lesssim 72\,{\rm GeV}$ \cite{csikor}, clearly 
in disagreement with the measured value $m_H \simeq 125\,{\rm GeV}$. 

For this reason successful electroweak baryogenesis necessarily requires an extension of the SM. 
For example,  electroweak baryogenesis can be successfully implemented within supersymmetric models.\footnote{
Recently a different strategy based on effective theory has been proposed \cite{Ellis:2019flb}.
It has been shown that augmenting the SM by  higher-dimensional operators, the variation of the SM couplings
with temperatures is changed and this can lead to successful electroweak baryogenesis.}
In the MSSM one obtains an upper bound on the Higgs mass $m_H \lesssim 127 \,{\rm GeV}$ 
and on the stop mass $m_{\widetilde{t}} \lesssim 120\,{\rm GeV}$ \cite{quiros}. 
Current experimental constraints from the LHC have basically ruled out this scenario now \cite{curtin},
though some loopholes might still be possible \cite{profumo}. Within a non-minimal supersymmetric model,
like the NMSSM, an allowed region still survives  \cite{NMSSM,Athron:2019teq}
and also in a two Higgs doublet model only special solutions are found within the full parameter space \cite{twohiggs}.
However, the recent improved result from the ACME experiment on the electron dipole moment
of the electron \cite{ACME:2018yjb}, 
\be
|d_{\rm e}| < 1.1 \times 10^{-29}\,e \,{\rm cm} \;\; (90\% \, \mbox{\rm C.L.}) \,  ,
\ee
essentially rules out also these models  (see  \cite{buchmullerbodeker} for a recent discussion). 

These findings of course pose the legitimate question whether electroweak baryogenesis is still alive \cite{cline}.
Introducing a scalar singlet, it is possible to realise easily a strongly first order phase transition 
and at the same time avoid experimental constraints  from colliders. 
However, this seems quite an {\em ad hoc} model specifically  engineered to show that successful 
electroweak baryogenesis is still viable. Maybe, if certainly not dead, it is fair to say that 
electroweak baryogenesis is then currently is in a sleeping beauty mode, 
awaiting to get awake from some supporting experimental finding.\footnote{In this respect it is intriguing
that a strong first order phase transition might also produce a gravitational wave stochastic background
testable with gravitational wave interferometers. In the specific case of an electroweak baryogenesis
strong first order phase transition, the spectrum would peak in the millihertz range that will be tested
by an interferometer  such as LISA (see \cite{Caprini:2015zlo} and references therein).}

\subsection{Baryogenesis from heavy particle decays}

Baryogenesis from heavy particle decays is the oldest class of models of baryogenesis,
since the original proposal by Sakharov belongs to it. 
Here we review the main common features and then specialise the general discussion to 
an important specific historical example: GUT baryogenesis. 
In Section 5 we will see that leptogenesis provides another remarkable concrete realisation
within this general class of models that is tightly related to neutrino masses.

\subsubsection{Out-of-equilibrium decays}

Let us consider a self-conjugate heavy  ($M_X\gg M_{EW}$) particle $X$
whose decays are $C\!P$ asymmetric, in such a way that
the decaying rate into particles, $\Gamma_X$,  is in general different
from the decaying rate into anti-particles, $\overline{\Gamma}_X$,
and such that the single decay process into particles (anti-particles)
violate $B-L$ by a quantity $\Delta_{B-L}$ ($-\Delta_{B-L}$).
 The {\em total $C\!P$ asymmetry} is then conveniently defined as
\be\label{epsX}
\varepsilon_X={\rm sign}(\Delta_{B-L}){\Gamma_X-\overline{\Gamma}_X\over \Gamma_X+\overline{\G}_X} \, .
\ee
For example, for GUT baryogenesis one has ${\rm sign}(\Delta_{B-L}) = +1$, while for leptogenesis ${\rm sign}(\Delta_{B-L}) = -1$. At tree level $\ve_X$ vanishes. Therefore, a non-vanishing total $C\!P$ asymmetry 
relies on the interference between tree level and one-loop diagrams,
where in the loop at least a second massive particle $Y$ needs to 
be present. This means that in the limit where $M_Y \ra 0$ one necessarily has $\ve_X \ra 0$. 
Moreover $\ve_X$ also vanishes in the case of exact degeneracy, i.e., for $M_X = M_Y$. 

The {\em total decay rate} $\Gamma_X^D=\Gamma_X+\overline{\Gamma}_X$
is the product of the total decay width,
$\widetilde{\Gamma}_X^D$, times the averaged dilation factor
$\langle 1/\gamma\rangle$, explicitly:
\be
\Gamma_X^D=\widetilde{\Gamma}_X^D\,\left\langle {1\over\gamma}\right\rangle \, .
\ee
As discussed in 2.3, sphaleron processes, while inter-converting $B$ and $L$ separately,
preserve $B-L$ and for this reason the kinetic equations
are much simpler if the $B-L$ evolution is tracked instead of the separate
$B$ or $L$ evolution.
Moreover, it is convenient to use, as an independent variable, the quantity $z=M_X/T$ and
to introduce the {\em decay factor} $D_X \equiv\Gamma_X^D/(H\,z)$.  Another convenient choice is
to track the abundance of $X$ particles,
$N_X$, and  asymmetry, $N_{B-L}$, in a portion of comoving volume 
normalised in a way to contain one $X$ particle on average in ultra-relativistic thermal equilibrium 
(i.e., $N_{X}^{\rm eq}(z\ll 1)=1$).\footnote{In general the abundance of some generic quantity $X$ in a generic portion
of comoving volume $R_0^3\,a^3$ is defined as $N_X = n_X\,R_0^3\, a^3$, where $R_0$ is a normalization factor. Our normalisation corresponds to take
\be
R_0^3 = {4\over 3}\,{\pi^2 \, T_0^{-3}\over \zeta(3)\,g_X} \, {g_S^{\rm i} \over g_{S0}} \,  .
\ee 
Another popular normalisation for the abundances is to calculate them in a portion
of comoving volume per unit entropy, assumed to be constant, so that $s(T)\,R_0^3\,a^3 = 1$
and  $Y_X(T) = n_X(T) / s(T)$. If entropy is indeed conserved, then the two variables are simply proportional to each other, explicitly
\be
Y_X(T) = {135\,\zeta(3) \over 8\, \pi^4} \,  {g_X \over g_{\rm S}^{\rm i}} \, N_X(T) \,  .
\ee
In the calculation of physical observables such as $\eta_B(T)$ the normalisation factor clearly cancels out so that 
all choices are equaivalent. However, the normalisation  of the kind $N_{X}^{\rm eq}(z\ll 1)=1$ has a few advantages: 
it does not assume entropy conservation
and at $T \gg m_X$ one has conveniently $N^{\rm eq}_{N_I}(T \gg M_I) = 1$, 
while 
\be
Y_{X}^{\rm eq}(T \gg M_X) =  {135\,\zeta(3) \over 8\, \pi^4} \,  {g_X \over g_{\rm S}^{\rm i}} \,  ,
\ee
not a very convenient value. Notice moreover that often one starts from $Y_{X}$ variables 
but then quantities $Y_X/Y_{X}^{\rm eq}(T\gg M_X)$ are plotted since they are more convenient. 
However, clearly one has  $N_X = Y_X/Y_{X}^{\rm eq}(T\gg M_X)$ showing that $Y_X$ variable is an unnecessary 
intermediate auxiliary variable.}

The results for the evolution of the $X$ abundance and the $B-L$ asymmetry can be conveniently described  in terms just of the  {\em total decay parameter} defined as 
\be\label{Kdef}
K_X \equiv {\widetilde{\Gamma}_X^D \over H(z=1)} \, ,
\ee
implying $D_X=K_X\,z\,\langle 1/\gamma \rangle$.  This is proportional to the ratio of the age of the universe
when the $X$ particles become non-relativistic, for $z=1$, to the $X$ lifetime, $\tau_X=1/\widetilde{\Gamma}_X^D$.

The simplest case is realised when $\tau_X$ is much longer than the age of the Universe at $z=1$, corresponding to have $K_X \ll 1$.
In this way the bulk of decays occur when the temperature is much below the $X$ mass and
the $X$ production from inverse decays, or other possible processes, is
Boltzmann suppressed.
 In this situation decays are the only relevant processes and
the kinetic equations for the X abundance and the $B-L$ asymmetry
are particularly simple to be written,
\bea \label{keoed1}
{dN_X\over dz}& = & -D_X(z)\,N_X(z) \,  , \\ \label{keoed2}
{dN_{B-L}\over dz}& = & -\varepsilon_X\,{dN_X\over dz} \, ,
\eea
and solved,
\begin{eqnarray}\label{NBmL}
N_{B-L}(z) & = & N_{B-L}^{\rm i}+
\varepsilon_X\,\left[N_X^{\rm i}-N_X(z)\right] \\ \label{NBmL2}
N_X(z) & = & N_X^{\rm i}\,e^{-\int_{z_{\rm i}}^z\,dz' \,D(z')} \,  ,
\end{eqnarray}
where $N_{B-L}^{\rm i}$ denotes  a possible pre-existing $B-L$ asymmetry 
and $ N_X^{\rm i}$ the initial number of $X$ particles.
The dilation factor, averaged on the Boltzmann statistics, is given by
the ratio of Bessel functions,
\be\label{dilation}
\left\langle {1 \over \gamma} \right\rangle  = {{\cal K}_1(z) \over {\cal K}_2(z)} \,  ,
\ee
and can be simply approximated  by \cite{pedestrians}
\be
\left\langle{1\over\gamma}\right\rangle\simeq {z\over z+2} \,  .
\ee
This simple approximation makes possible to solve analytically
the integral in Eq. (\ref{NBmL2}), yielding the result
\be\label{outN}
N_X(z)\simeq N_{X}^{\rm i}\,
e^{-K_X\ \left[{z^2 \over 2} - {2\,z}
+ 4\ln{\left(1+{1\over 2}\, z\right)}\right]} \, .
\ee
In particular, the final $B-L$ asymmetry is given by
\be
N_{B-L}^{\rm f}=N_{B-L}^{\rm i}+\varepsilon_X\,N_X^{\rm i} \,  .
\ee
The baryon-to-photon ratio at recombination can then be obtained
dividing  by the number of photons at recombination $N_{\gamma}^{\rm rec}$
and taking into account that sphalerons will convert only a fraction
$a_{\rm sph}\simeq 1/3$ of the $B-L$ asymmetry into a baryon asymmetry.
In this way one can write:
\be\label{etaB}
\eta_B \simeq {1\over 3}\,{N_{B-L}^{\rm f}\over N_{\gamma}^{\rm rec}} \, .
\ee
\begin{figure}[t]
\centerline{\psfig{file=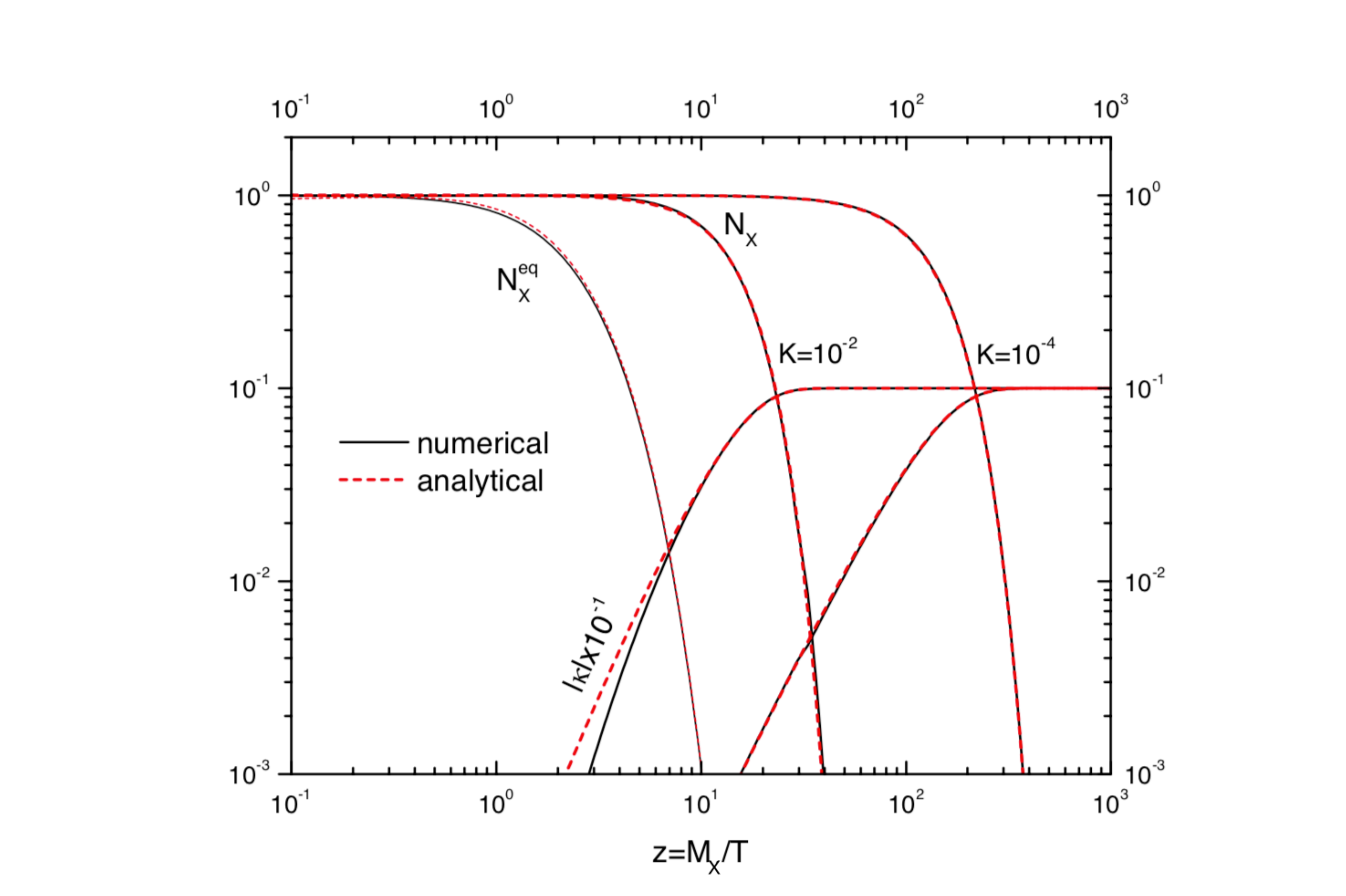,height=13cm,width=11cm,angle=90}}
\caption{Out-of-equilibrium decays for an initial thermal  $X$ abundance.}
\end{figure}
It is useful to introduce the {\em efficiency factor}, defined as
the ratio of the asymmetry produced from  $X$ decays to the $C\!P$ asymmetry:
\be
\kappa(z;K_X)\equiv {N_{B-L}(z)|_{N_{B-L}^{\rm i}=0}\over
\ve_X} \,  ,
\ee
depending uniquely on $K_X$. In the case of out-of-equilibrium decays one
has simply $\k(z)=N^{\rm i}_X-N_X(z)$ and in this way Eq. (\ref{NBmL}) can be recast as
\be
N_{B-L}(z)=N^{\rm i}_{B-L}+\ve_X\,\kappa(z) \, .
\ee
The final value of the efficiency factor is then simply given by the initial number of $X$ particles, i.e.,
$\kappa_{\rm f}\equiv \k(\infty)=N_X^{\rm i}$. In particular, it is equal to unity in the
case of an initial thermal $X$-abundance with $z_{\rm i}\ll 1$. In Fig. 1 we
show two examples of out of equilibrium decays, for $K=10^{-2}$ and
$K=10^{-4}$, assuming an initial thermal $X$-abundance ($N_X^{\rm i}=1$)
and vanishing pre-existing asymmetry ($N_{B-L}^{\rm i}=0$). The numerical results
are compared with the analytic expression Eq.~(\ref{outN}) as indicated.

 The out-of-equilibrium decay regime is an efficient way to produce an asymmetry from the decays of heavy particles.
However, it necessarily relies on the assumption that an initial $X$ abundance
was produced, either thermally or non-thermally, by some process at $t\gtrsim \tau_X$ and that
one can neglect a possible $N_{B-L}^{\rm i}$ generated during or after
inflation and prior to the the onset of $X$-decays. Therefore,
it is evident that such a scenario is plagued by a
{\em strong dependence on the initial conditions} and hence it requires
to be complemented with a picture able to specify them, for example
a detailed description of the inflationary stage.

\subsection{Inverse decays}

The case of out-of-equilibrium decays  is  valid rigorously only in the limit $K_X\rightarrow 0$.
If one defines $z_{\rm d}$ as that special value of $z$ corresponding to the time when the age of the universe
is equal to $\tau_X$, i.e., $t(z_{\rm d}) = \tau_X$, then, for $K_X\ll 1$,
one has indeed $z_{\rm d} \simeq \sqrt{2/K_X} \gg 1$. On the other hand, for $K_X\gtrsim 1$, the bulk of  
$X$ particles will decay at a temperature $T_{\rm d}=M_X/z_{\rm d} \gtrsim M_X/\sqrt{2}$ 
so that inverse decays cannot be neglected as we did so far. 
The kinetic equations (\ref{keoed1})  and (\ref{keoed2})
are then generalized in the following way \cite{kolbturner,luty,plum,bcst,cmb}
\footnote{The equations (\ref{ek1}) and (\ref{ek2}) are actually not only
accounting for decays and inverse decays but also for the
real intermediate state contribution from $2\leftrightarrow 2$
scattering processes. This term exactly cancels a $C\!P$ non conserving
term from inverse decays that would otherwise lead to an
un-physical asymmetry generation in thermal equilibrium  \cite{dolgovkolbwolfram}.}
\bea\label{ek1}
{dN_X\over dz}& = & -D_X(z)\,\left[N_X(z) - N_X^{\rm eq}(z)\right] \,  , \\ \label{ek2}
{dN_{B-L}\over dz}& = & -\varepsilon_X\,{dN_X\over dz}-
W_X^{\rm ID}(z)\,N_{B-L}(z) \, .
\eea
In the first equation, for $N_X$, the second term accounts for the inverse decays
that, importantly, can now produce the $X$'s. On the other hand, one can see that
a new term appears in the second equation for the asymmetry too, a
wash-out term that tends to destroy what is generated from decays.
This term is controlled by the  (inverse decays) wash-out factor given by
\be\label{WID}
W_X^{\rm ID}={n\over 2}\,D_X\,{N_X^{\rm eq}\over N_{b,l}^{\rm eq}} \propto K_X \, ,
\ee
where $n$ denotes the number of baryons or leptons in the $X$ decay final state
(as we will see $n=1$ in the case of leptogenesis) and $N_{b,l}^{\rm eq}$ is the equilibrium abundance
either of leptons or baryons, in any case massless, produced from $X$ decays.
Note that the decay parameter $K_X$ is still the only parameter in
the equations and thus the solutions will still depend only on $K_X$.
They can be again worked out in an integral form \cite{kolbturner}. In the
case of the $B-L$ asymmetry one can write the final asymmetry as
\be\label{NBmLf}
N_{B-L}^{\rm f}=N_{B-L}^{\rm i}\,e^{-\int_{z_{\rm i}}^{\infty}\,dz'\,W_X^{ID}(z')} +\varepsilon_X\,\k_{\rm f} \, ,
\ee
where now the efficiency factor is given by the integral
\be\label{kf}
\kappa_{\rm f}(K_X,z_{\rm i})= -\int_{z_{\rm i}}^{\infty}
\,dz'\,\left[{dN_{N_1}\over dz'}\right]\,
\,e^{-\int_{z'}^{\infty}\,dz''\,W_X^{\rm ID}(z'')} \, .
\ee
In the limit $K_X\rightarrow 0$ one has $W_X^{\rm ID} \ra 0$ in a way that the out-of-equilibrium case is recovered.
In general, one can see that the wash-out has the positive effect to
damp a pre-existing asymmetry but also the negative one to damp the same
asymmetry generated from decays, thus reducing the efficiency of the mechanism.
A quantitative analysis is crucial and it is very useful to discuss separately the
regime of strong wash out for $K_X\gtrsim 1$ and the regime  of weak wash-out for $K_X\lesssim 1$.

\subsubsection{Strong wash-out regime}

The strong wash-out regime is characterized by the existence,
for $K_X\gtrsim 3$, of an interval $[z_{\rm in},z_{\rm out}]$
such that $W_X^{\rm ID}\gtrsim 1$ and thus such that inverse decays are in equilibrium.
Practically all the asymmetry produced at $z < z_{\rm out}$
is washed-out including, noteworthy, a pre-existing one. Moreover
the calculation of the residual asymmetry is made very simple by the
possibility to use the {\em close equilibrium approximation}, i.e.,
\be\label{approx}
{dN_{X}\over dz'}\simeq {dN_{X}^{\rm eq}\over dz}=-{2\over K_X\,z}\,W_X^{\rm ID}(z) .
\ee
In this way the integral in  Eq. (\ref{kf}) can be easily evaluated for $z_{\rm i} \ll 1$ \cite{kolbturner,pedestrians}.
Indeed, this can be put in the form of a Laplace integral,
\be\label{psi}
\kappa_X^{\rm f} =\int_0^{\infty}\,dz'\,e^{-\psi(z',z)} \, ,
\ee
receiving a dominant contribution only from a small interval centered
around a special value $z_B$ such that $d\psi/dz=0$.
In this way one can use the approximation of replacing
$W_X^{\rm ID}(z'')$ with  $W_X^{\rm ID}(z'')\,z_B/z''$ in  Eq. (\ref{kf}).\footnote{It is analogous but not 
coinciding with a saddle point approximation since in this case one does not Taylor
expands the function $\psi(z',z)$ about the minimum.}
With this approximation the integral can be easily solved, obtaining
\be\label{kfth}
\k_f \simeq {2\over n\,K_X\,z_B}\,\left(1-e^{-{n\,K_X\,z_B\over 2}}\right) \,  .
\ee
For large $K_X\gg 1$ and $n=2$
this expression coincides with that one in \cite{kolbturner}.\footnote{Note, however, 
that the definition (\ref{Kdef}) for $K_X$
has to be used instead of $K_X=(1/2)\,(\Gamma_D/H)_{z=1}$.}
The calculation of $z_B$ proceeds from its definition, $(d\psi/dz)_{z_B}=0$,
approximately equivalent to the equation
\be\label{zB}
W_X^{\rm ID}(z_{\rm B}) = \left\langle {1\over\gamma}\right\rangle^{-1}(z_{\rm B})\,
-\, {3\over z_{\rm B}}\;.
\ee
This is a transcendental algebraic equation and thus one cannot find an exact
analytic solution (see \cite{pedestrians} for an approximate
procedure). However, the expression
\be\label{fit}
z_B(K)\simeq 2+4\,(n\,K_X)^{0.13}\,e^{-{2.5\over n\,K_X}} \, ,
\ee
provides quite a good fit that can be plugged into Eq.~(\ref{kfth}) thus
getting an analytic expression for the efficiency factor.
For very large $K_X$ this behaves as a power law $\kappa_{\rm f}\propto K_X^{-1.13}$.
In Fig. 2 we show a comparison of the analytic solution for $\k_{\rm f}$ Eq.~(\ref{kfth})
with the numerical solution (for $n=1$). One can see how for $K_X\gtrsim 4$
the agreement is quite good (at the percent level).
\begin{figure}[t]
\centerline{\psfig{file=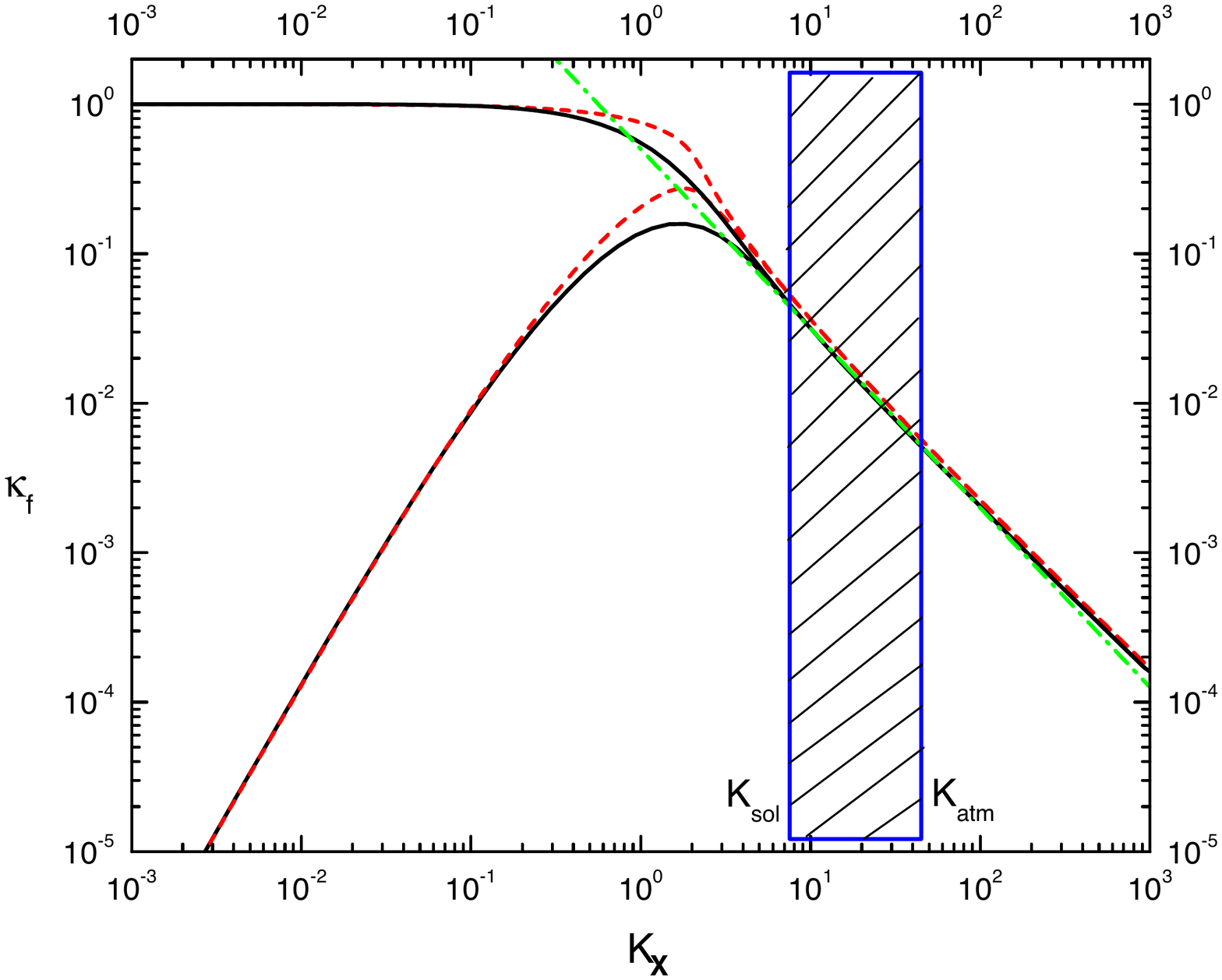,height=13cm,width=11cm,angle=90}}
\caption{Final efficiency factor  as a function of the decay parameter
$K_X$ for thermal and dynamical initial $N_X$ abundance. The solid lines
are the numerical solutions of the Eq.'s (\ref{ek1}) and (\ref{ek2}),
the short-dashed lines are the analytic results
(see Eq.~(\ref{kfth}) and Eqs.~(\ref{kf-})+(\ref{kf+})), the dot-dashed line
is the power law fit Eq. (\ref{kappaflep}). The dashed box
is the range $K_X = 8$--$45$ favoured by neutrino mixing data
in the case of leptogenesis.}
\end{figure}
Note that the Eq. (\ref{zB}) implies that  for large values
of $K_X$ one has $z_B\simeq z_{\rm out}$, that particular value of
$z$ corresponding to the last moment when inverse decays are in equilibrium
($W_X^{\rm ID}\geq 1$).  In this way almost all the asymmetry produced for
$z\lesssim z_B$ is washed-out and most of the surviving asymmetry is produced
in an interval just centred around $z_{\rm out}$, simply because the
$X$ abundance gets rapidly Boltzmann suppressed for $z > z_{\rm out}$.
An example of this picture is illustrated in Fig. 3 for $K_X=100$ (from \cite{pedestrians}).
\begin{figure}[t]
\centerline{\psfig{file=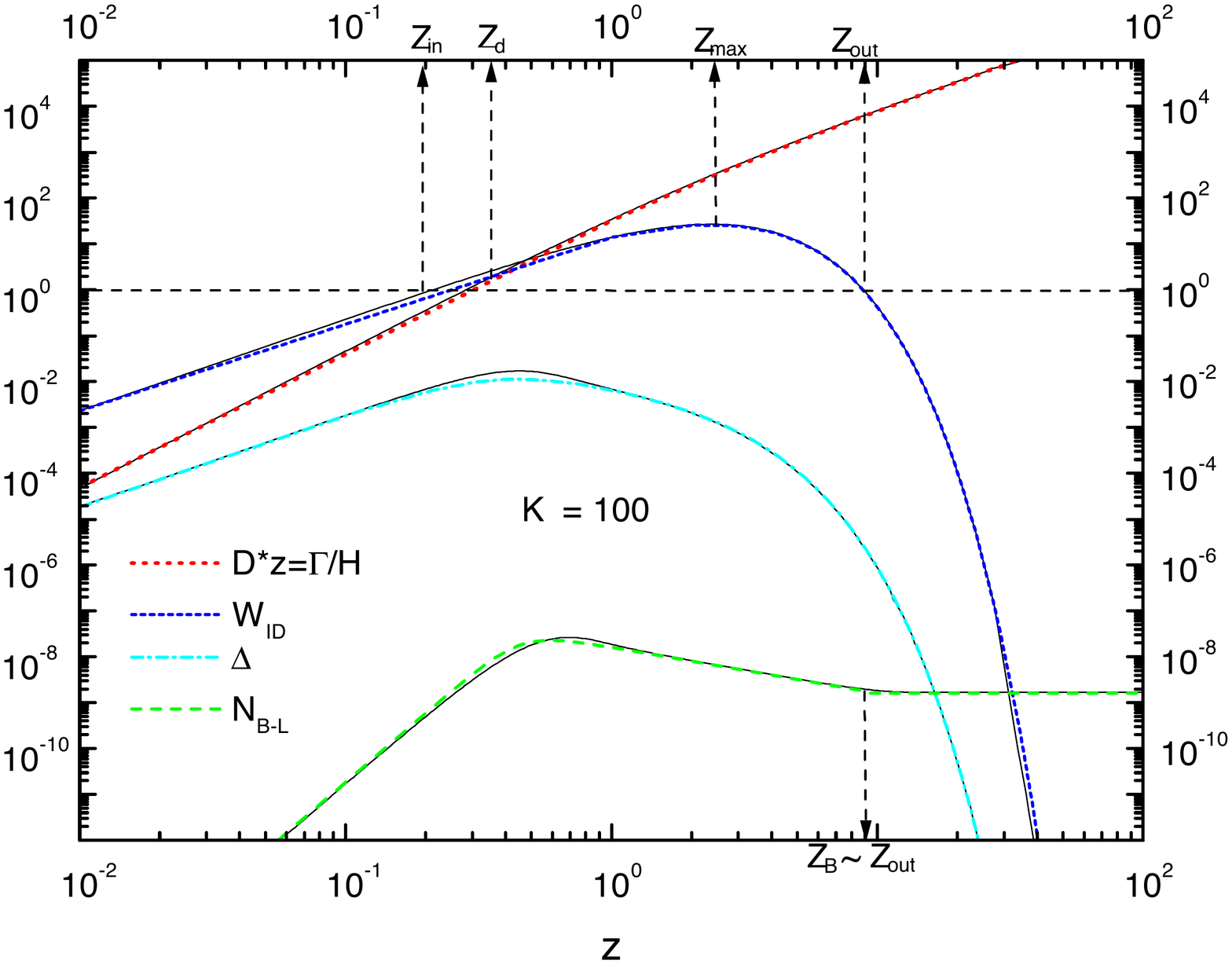,height=12cm,width=10cm,angle=90}}
\caption{\small Comparison between analytical (short-dashed lines)
and numerical (solid lines) results in the case of
strong wash-out ($K_X=100$) for $|\ve_X| = 0.75\times 10^{-6}$ (from \cite{pedestrians}).}
\label{strong}
\end{figure}
Instead of the abundance $N_X$, we plotted the deviation from the equilibrium value, the quantity
$\Delta=N_{X}-N_{X}^{\rm eq}$. The deviation grows until the $X$'s decay
at $z\simeq z_{\rm d}$, when it reaches a maximum, and decrease afterwards when
the abundance stays close to thermal equilibrium. Correspondingly, the asymmetry
grows for $z\lesssim z_{\rm d}$, reaching a maximum around $z\simeq 1$, and then
it is washed-out until it freezes at $z_B\simeq z_{\rm out}$.
The evolution of the asymmetry $N_{B-L}(z)$ can induce the
wrong impression that the residual asymmetry is some fraction
of what was generated at $z\simeq 1$ and that one cannot relax the
assumption $z_{\rm i} \ll 1$ without reducing considerably the final value of the asymmetry.
However, what is produced for $z \lesssim z_{\rm out}$ is also very quickly destroyed and as fas as $z_{\rm i} \lesssim z_B$
there is no much dependence of the final asymmetry on $z_{\rm i}$. 
A plot of the quantity $\psi(z,\infty)$, as defined in the Eq. (\ref{psi})
and shown in Fig 4 (from \cite{pedestrians}), enlightens some interesting aspects.
\begin{figure}[t]
\centerline{\psfig{file=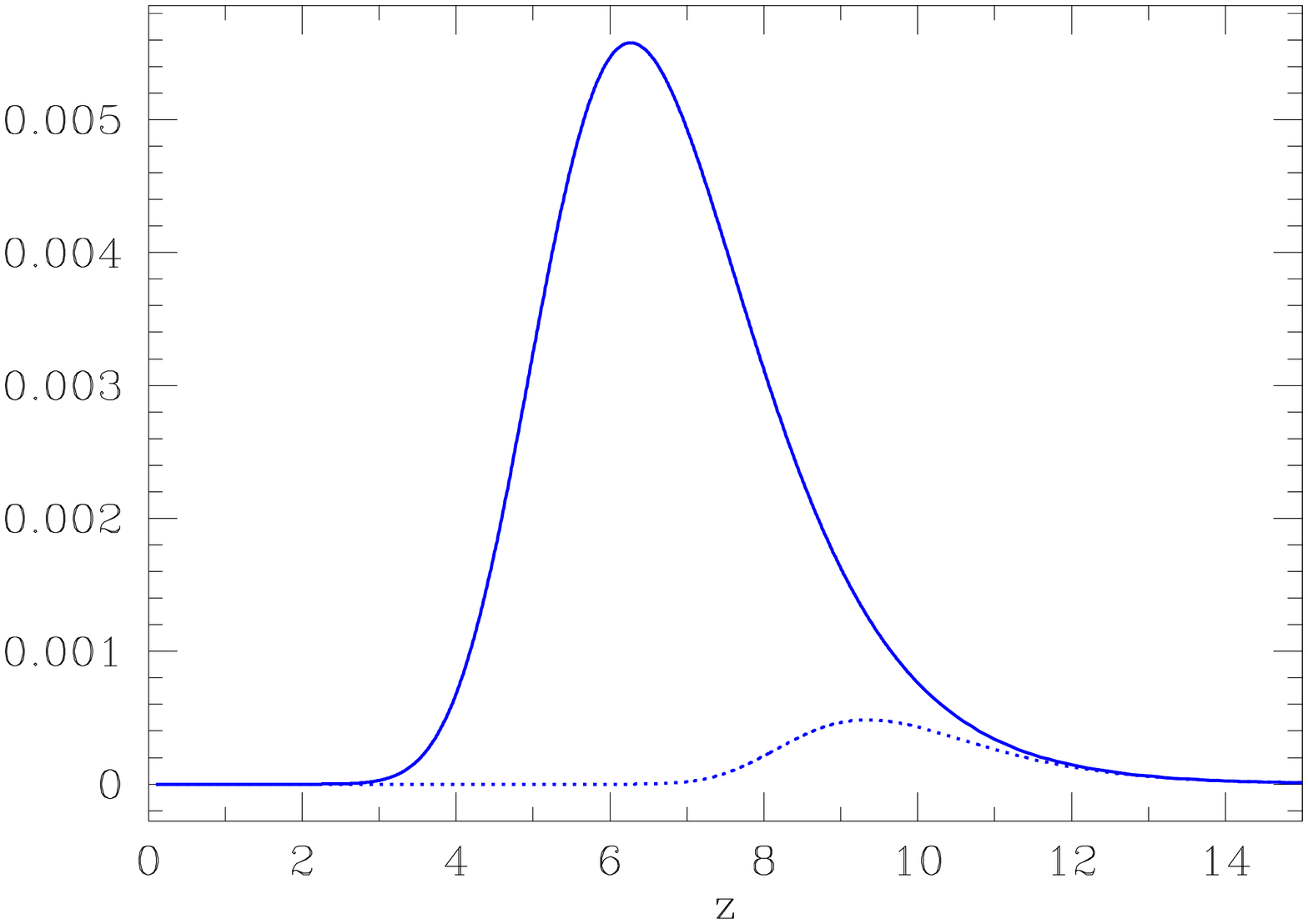,height=12cm,width=12cm}}
\vspace{-25mm}
\caption{\small The function $\psi(z',\infty)$ for $K_X=10$ (solid line) and $K_X=100$ (dashed line, from \cite{pedestrians}).}
\label{strong}
\end{figure}
This is the final asymmetry that was produced in a infinitesimal
interval around $z$. It is evident how just the asymmetry that was produced
around $z_B$ survives and, for this reason, the temperature $T_B=M_X/z_B$
can be rightly identified as the {\em temperature of baryogenesis}
for these models.
It also means that in the strong wash out regime the final
asymmetry was produced when the $X$ particles were
fully non-relativistic implying that the simple kinetic
equations  (\ref{ek1}) and (\ref{ek2}), employing
the Boltzmann approximation, give actually accurate results
and many different types of corrections, mainly from a rigorous quantum kinetic
description including thermal effects, can be safely neglected.

This is not the only nice feature of the strong wash-out regime.
Since any asymmetry generated for $z\lesssim z_B$ gets efficiently washed-out,
one can also rightly neglect any pre-existing initial asymmetry $N_{B-L}^{\rm i}$.
At the same time the final asymmetry does not depend on the initial $X$ abundance.
\begin{figure}[t]
\centerline{\psfig{file=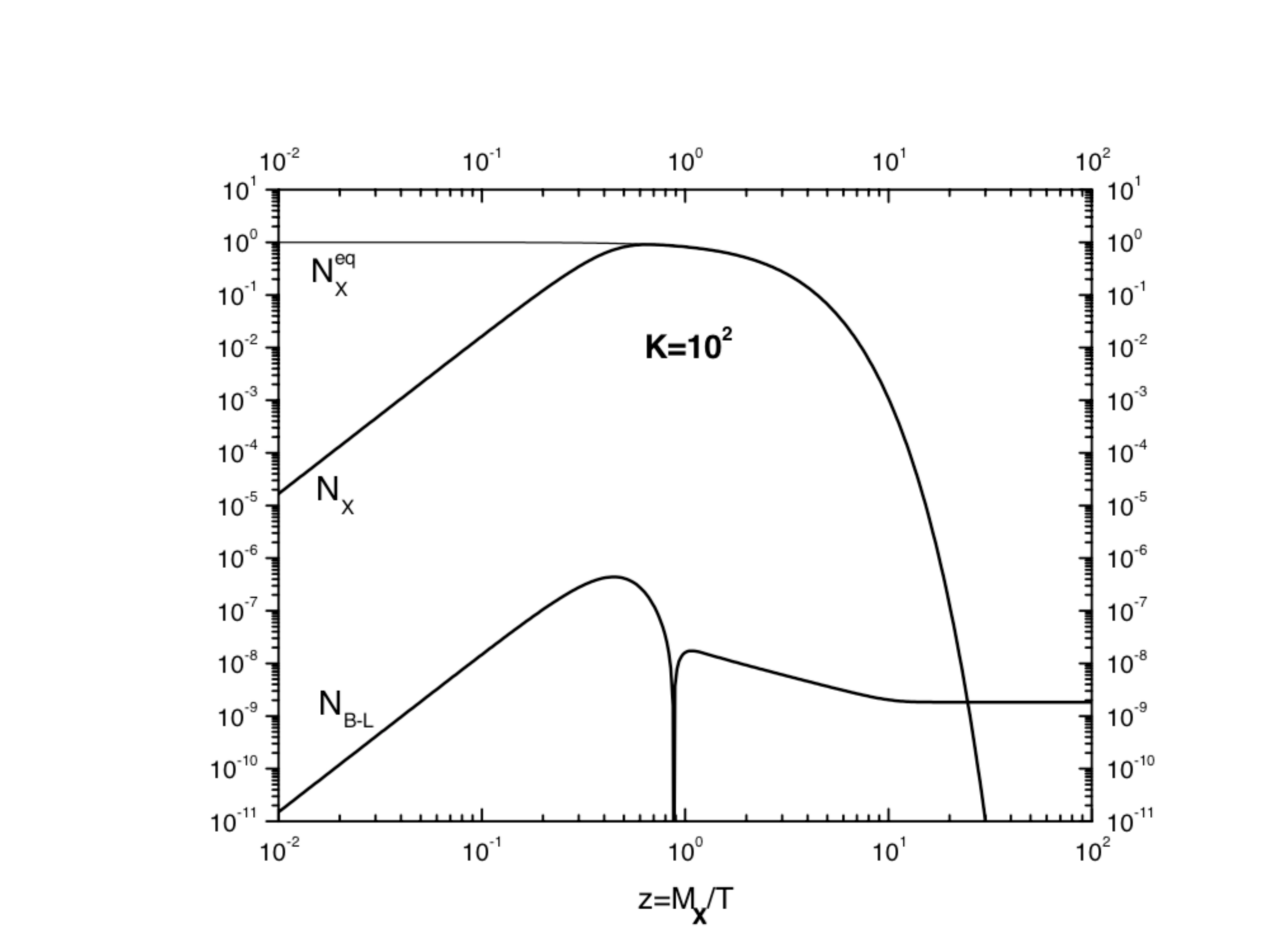,height=15cm,width=12cm,angle=90}}
\vspace{-1mm}
\caption{\small Fast thermalization of the X abundance in the strong wash-out regime.
The final $N_{B-L}$ abundance (for $\ve_X=0.75\times 10^{-6}$) is the same
as in the case of an initial thermal abundance (cf. Fig. 3) and it is
independent on the evolution at $z\ll z_B$.}
\label{strong}
\end{figure}
In Fig. 5 we show how even starting from a vanishing $X$ abundance,
the $X$'s are rapidly produced by inverse decays in a way that well before
$z_B$ the number of decaying neutrinos is always equal to the thermal number.
As we said, the final asymmetry does not even depend on the initial temperature
as far as this is higher than $\sim T_B$ and thus if one relaxes
the assumption $z_{\rm i}\ll 1$ to $z_{\rm i}\lesssim z_B-\Delta z_{B}$,
the final efficiency factor gets just slightly reduced (for example for $\Delta z_B\simeq 2$
this is reduced approximately by $10\%$).

Summarising, we can say that that in the strong wash out regime the reduced
efficiency is compensated by the remarkable fact that, for $T_{\rm i}\gtrsim T_B$,
the final asymmetry does not depend on the initial
conditions and all non relativistic approximations work very well.
These conclusions change quite drastically in the weak wash-out regime.

\subsubsection{Weak wash-out regime}

For $K_X\lesssim 1$, $z_B$ rapidly tends to unity (see Eq.~(\ref{fit})).
In Fig. 2 the analytic solution for the efficiency factor, (see Eq.~(\ref{kfth})),
is compared with the numerical solution.
It can appear surprising that, in the case of an initial thermal abundance,
the agreement is excellent not only at large $K_X\gtrsim 4$,
but also at small $K_X\lesssim 0.4$, with some appreciable deviation only in the range $0.4\lesssim K_X\lesssim 4$.
The reason is that when the wash-out processes get frozen,
the efficiency factor depends only
on the initial number of neutrinos and not on its derivative and thus
the approximation Eq. (\ref{approx}) introduces a sensible
error only in the transition regime $K_X\sim 1$.

Eq.~(\ref{kfth}) can be easily generalized to any value of the initial abundance
if one can neglect the $X$ particles produced by inverse decays.
More generally, one has to calculate such a contribution and it is convenient
to consider the limit case of an initial vanishing $X$ abundance.
The production of the $X$'s lasts until $z=z_{\rm eq}$, when its abundance is equal to the
equilibrium value, such that
\be
N_{X}(z_{\rm eq})=N_{X}^{\rm eq}(z_{\rm eq})  \, .
\ee
At this time the number of decays equals the number of inverse decays.
For $z\leq z_{\rm eq}$, decays can be neglected and Eq.~(\ref{ek1}) becomes
\be
{dN_X \over dz}=D_X(z)\,N_{X}^{\rm eq}(z)  \,.
\ee
For $z\ll 1$, one then simply finds
\be
N_{X}(z)={K_X\over 6}\,\left(z^3 - z_{\rm i}^3 \right)  \, .
\ee
In the weak wash-out regime the equilibrium is reached very late,
when neutrinos are already non relativistic and $z_{\rm eq}\gg 1$.
In this way one can see that the number of $X$ particles reaches,
at $z\simeq z_{\rm eq}$, a maximum value given by
\be
N_{X}(z_{\rm eq})\simeq N(K_X)\equiv {3\,\pi\over 4}\,K_X \, .
\ee
For $z>z_{\rm eq}$ inverse decays can be neglected
and the $X$'s decay out of equilibrium in a way that
\be
N_{X}(z>z_{\rm eq})\simeq N_X(z_{\rm eq})\,e^{-\int_{z_{\rm eq}}^z\,dz'\,D(z')} \, .
\ee
Let us now consider the evolution of the $B-L$ asymmetry calculating the
efficiency factor. Its value can be conveniently decomposed as the sum of two
contributions, a negative one, $\kappa^-_{\rm f}$, generated at $z<z_{\rm eq}$,
and a positive one, $\kappa^+_{\rm f}$, generated at $z>z_{\rm eq}$.
In the limit of no wash-out we know that the final efficiency factor
must vanish, since we have seen that in this case $\kappa_{\rm f}=N_X^{\rm i}$
and we are assuming a vanishing initial abundance. This implies
that the negative and the positive contributions cancel with each other.
The effect of the wash-out is to suppress the negative contribution more than
the positive one, in a way that the cancellation is only partial.
In the weak wash-out regime it is possible in first approximation
to neglect completely the wash-out at $z\geq z_{\rm eq}$.  In this way
it is easy to derive from the Eq. (\ref{kf}) the following expression
for the final efficiency factor:
\be\label{kfw}
\kappa_{\rm f}\simeq N(K_X)-2\left(1-e^{-{1\over 2}\,N(K_X)}\right) \, .
\ee
One can see how it vanishes at the first order in $N(K_X)\propto K_X$
and only at the second order one gets $\k_f\simeq (9\,\pi^2/64)\,K_X^2$.

\subsubsection{Final efficiency factor: global solution}

Generalising the procedure seen for the strong wash-out, it is possible to
find a global solution for $\kappa_{\rm f}(K_X)$ valid for any value of $K_X$. The calculation
proceeds separately for $\kappa^-$ and $\kappa^+$ and the final results are given by
\be\label{kf-}
\kappa^{-}_{\rm f}(K_X) = -2\ e^{-{1\over 2}\,N(K_X)}
\left(e^{{1\over 2} \overline{N}(K_X)} - 1 \right)
\ee
and
\be\label{kf+}
\kappa^{+}_{\rm f}(K_X)={2\over z_B(K_X)\,n\,K_X}
\left(1-e^{-{1\over 2} z_B(K_X)\,n\,K \overline{N}(K_X)}\right) \; .
\ee
The function $\overline{N}(K_X)$ extends, approximately, the definition of $N(K_X)$
to any value of $K_X$ and is given by
\be
\overline{N}(K_X) = {N(K_X)\over\left(1 + \sqrt{{N(K_X)}}\right)^2}\; .
\ee
The sum of the Eq.'s (\ref{kf+}) and (\ref{kf-}) is plotted, for $n=1$, in Fig. 2 (short-dashed line)
and compared with the numerical solution (solid line).

We can now draw a few conclusions about a comparison between the weak
and the strong wash-out regime. A large efficiency in the weak wash-out
regime relies on some unspecified mechanism that should have produced a large
(thermal or non-thermal) $X$ abundance before their decays.
On the other hand, the decrease of the
efficiency at large $K_X$ in the strong wash-out regime is only (approximately) linear
and not exponential \cite{kolbturner,pedestrians}. This means that for moderately large values of $K_X$
a small loss in the efficiency would be compensated by a
full thermal description  such that the predicted asymmetry
does not depend on the initial conditions, a nice situation that
resembles closely the situation in standard big bang nucleosynthesis, 
where the calculation of the primordial nuclear abundances is also independent of the initial conditions.

Another point to notice is that we discussed a very general picture taking into account just decays and inverse decays.
However, considering specific models,  one might have to consider some specific processes. 
We will discuss in Section 5 the case of leptogenesis and we will see that in that case scatterings 
also need to be taken into account
for the production of decaying particles producing the asymmetry, though in the strong wash-out regime this will
not have a great impact on the final predicted asymmetry. On the other hand, we will see that flavour effects
can dramatically change the calculation of the asymmetry in certain situations.

\subsection{GUT baryogenesis}

The advent of GUT theories in the seventies \cite{Yoshimura:1978ex,Toussaint:1978br,Ellis:1978xg,Weinberg:1979bt,Harvey:1981yk} provided a very well motivated realistic extension of the SM to embed baryogenesis from heavy particle decays. Here we want briefly to show how the general
results specialise considering a simple toy model.\footnote{For a more complete discussion we refer the reader to \cite{kolbturner}.} 

The asymmetry is in this case generated by the baryon number violating decays of super heavy bosons assumed to be complex scalars.
Since we need interference of tree level and one-loop diagrams for the total $C\!P$ asymmetries not to vanish,
one needs at least two of them that we can denote them by $X$ and $Y$ and their masses by
$M_X$ and $M_Y$ respectively. If we assume $M_Y \gg M_X$ and $K_X \gg 1$, we can neglect
the asymmetry produced by the $Y$'s in a way that the $B-L$ asymmetry will be dominantly produced
by the $X$'s and their anti-particle states $\bar{X}$. Let us assume that there are two baryon violating 
decay channels,\footnote{The existence of at least two channels is crucial since $X$ is a 
boson, and in this case the $C\!P$ conjugated state would decay exactly with the same rate because
of $CPT$ invariance. In the case of leptogenesis we will see that the decaying particle is a 
Majorana fermion and in that case one decay channel is sufficient, from this point of view
it is more economical.}
$X \ra \overline{\psi}_1 + \psi_2$ and $X \ra \overline{\psi}_3 + \psi_4$, and
its $C\!P$ conjugated, $\overline{X} \ra \psi_1 + \bar{\psi}_2$ and 
$\overline{X} \ra {\psi}_3 + \overline{\psi}_4$,
where each $\psi_i$ ($i=1,\dots,4$)  is a massive fermion with baryon number $B_i$. 

In this case the kinetic equations (\ref{ek1}) and (\ref{ek2}) still hold  
but now $N_X$ should refer to the total abundance of $X$ and $\bar{X}$ particles
and the washout factor is given by Eq.~(\ref{WID}) with $n=2$. 

In realistic models the masses of super heavy gauge bosons is close to the grand-unified scale $M_{\rm GUT}$.
Since the proton lifetime is approximately given by 
$\tau_{\rm p} \sim \a_{\rm GUT}^{-2}\, M_{X}^{4}\,m_{\rm p}^{-5}$,
the current experimental lower bound $\tau_{\rm p} \gtrsim 10^{34}\,{\rm years}$   
\cite{Super-Kamiokande:2016exg} translates into very stringent
lower bounds, $M_X \gtrsim 10^{15-16}\,{\rm GeV}$, for the masses of the superheavy gauge bosons 
that imply correspondingly very high reheat temperatures in the early universe in tension with upper
bound from CMB anisotropies. We will see that  $SO(10)$ GUT theories also predict the existence of RH neutrinos that can be much lighter. Their decays can then be responsible for the production of the 
$B-L$ asymmetry and this is at the basis of leptogenesis that we will discuss in detail in Section 5. 
Therefore, from this point of view, leptogenesis can be regarded as a way to achieve successful 
baryogenesis at lower reheat temperatures within GUT models with the addition of RH neutrinos. 
Moreover, the introduction of RH neutrinos also provides an attractive solution to the explanation of light neutrino masses and mixing and, therefore, leptogenesis somehow appears as the most likely way how the asymmetry can be produced 
within these models in a thermal way.

However, one can also consider a non-thermal production of the super heavy gauge bosons and in this 
case one can successfully reproduce the asymmetry for even much lower values of the reheat temperature. 
Various proposals have been made, such as production from inflaton decays \cite{Kumekawa:1994gx}, at preheating 
\cite{Dolgov:1989us,Traschen:1990sw,Kolb:1996jt,Giudice:2000ex} or from primordial black hole evaporation \cite{Toussaint:1978br,Turner:1979bt,Dolgov:1980gk,Hawking:1982ga,Polnarev:1985btg,Barrow:1990he,Baumann:2007yr,Hook:2014mla,Hamada:2016jnq,Hooper:2020otu}.

Non-thermal production mechanisms can of course be considered also in general, beyond GUT baryogenesis, i.e., for 
some generic heavy particles producing the asymmetry. 
In this respect recently it has been also considered the case of non-thermal heavy particle production from collisions of runaway bubbles originated during first order phase transitions \cite{Katz:2016adq,Azatov:2021irb}.


\section{Dark matter \label{sec:int}}

Very early hints of the existence of a dark matter component from anomalous motion of stars in our galaxy were already discussed 
at the beginning of  the  twentieth century  and the same name, {\em mati\`{e}re obscure} in French, was already introduced by Poincar\`{e} in 1906 
\cite{darkmatterhints}. On much large scales,  the anomalous dynamics of cluster of galaxies in 1930's \cite{zwicky} also pointed to the existence of such component. A more solid evidence was found only much later in the 1970's in galactic rotation curves \cite{freeman,rubin}.
An understanding of the large scale structure of the universe then provided the first evidence of the non-baryonic nature of the dominant component of dark matter and of its primordial origin, that had to occur prior to the matter-radiation equality time $t_{\rm eq} \simeq 52,000 \, {\rm yr}$.
 
It also became clear at the beginning of the 1980's that such component could not be constituted by massive neutrinos, since still relativistic and free-streaming  too fast at $t_{\rm eq}$  
({\em hot dark matter}), but rather by some new kind of non-standard matter that had to be already non-relativistic at $t_{\rm eq}$ ({\em cold dark matter}).
It was then only  with the discovery of CMB acoustic peaks that it was possible to firmly establish its non-baryonic nature that cannot be explained within the SM.  It is to this non-standard component that one usually  refers 
simply as {\em dark matter}.  
The cold dark matter contribution to the energy density parameter is today well determined both from dynamics of clusters of galaxies and from CMB anisotropies and from {\em Planck} +  BAO data it is found \cite{planck18}
\be\label{ODM}
\O_{DM 0} h^2 = 0.11933 \pm 0.00091 \,  
\ee
and from this, using Eq.~(\ref{h}) for the value of $h$, one finds
\be
\O_{DM 0} =  0.261 \pm 0.005 \, .
\ee
The total matter contribution,  baryonic plus cold dark matter, is then given by
\be\label{DM}
\O_{\rm M 0} h^2 = 0.14240 \pm 0.00087 
\ee
and using again Eq.~(\ref{h}) for $h$ one finds
\be
\O_{M 0} = 0.311 \pm 0.006  \,   . 
\ee
The most popular idea is that dark matter is made of new (elementary or composite) particles  
whose existence cannot be explained within the SM.\footnote{Recently the possibility that dark matter
is made of (SM) $uuddss$ hexa-quarks  has been explored in \cite{Gross:2018ivp}, showing that
a mass $\sim 1.2\,{\rm GeV}$ would be required but that on the other hand 
this is ruled out by the stability of Oxygen nuclei.}
Currently, the discovery of gravitational waves from black hole merging has drawn a lot of attention on the alternative option that dark matter is made of primordial black holes \cite{hawking,chapline}
with masses that can be potentially as large as hundred solar masses, though a host of observational constraints rules out the possibility of a population of monochromatic primordial black holes making the whole dark matter, except for a window $3.5 \times 10^{-17} \lesssim M_{\rm PBH}/M_{\bigodot} \lesssim 4 \times 10^{-12}$ 
\cite{Montero-Camacho:2019jte}
corresponding to asteroid masses. The possibility of a population of primordial black holes with a spread mass spectrum is, however, not yet completely excluded \cite{PBHreview}. Another possibility is that dark matter can be actually understood in terms of models of modified gravity rather than as an additional matter component, the most popular proposal is the
MoND theory or its covariant TeVeS theory \cite{bekenstein}. Recently mimetic dark matter
has also attracted attention since it also offers a unified solution to the dark energy puzzle and even a model for inflation \cite{mimetic}.  However, thorough statistical analyses including all cosmological observations 
disfavour such alternative models compared to the standard $\L$CDM model or variations to it where
dark matter is assumed to be made of particles. We will then focus our attention on  particle
dark matter models.

\subsection{Particle dark matter models}

In particle dark matter models, in addition to the requirement that the final relic abundance reproduces the
measured dark matter contribution to the total cosmological energy density in Eq.~(\ref{ODM}), the 
dark matter particles need to be long-lived on cosmological scales. In order to reproduce galactic rotation curves and 
the dynamics of clusters of galaxies, the life-time of dark matter particles $\tau_{\rm DM}$ 
has to be necessarily longer than the age of the  Universe $t_0$, so that one has to impose a lower bound
\be\label{tauDMlb}
\tau_{\rm DM} \gtrsim t_0 \simeq 4 \times 10^{17}\,{\rm s} \,  .
\ee
The possibility of dark matter particles decaying with $\tau_{\rm DM} \simeq t_0$ has been seriously considered to
solve well known problems of the standard CDM scenarios in reproducing observations on scales equivalent to 
that of dwarf galaxies such as the {\em missing satellite problem} and the {\em too big to fail} problem.
This {\em decaying dark matter} scenario is however currently disfavoured compared for example to other scenarios  that aims at solving the same problems, in particular models with self-interacting dark matter. It should also be said that in recent years with more accurate astronomical observations and N-body simulations that 
include baryon feedback, the shortcomings of the CDM scenario tend to be greatly ameliorated 
and currently there is no compelling motivation to abandon it.\footnote{There are of course other tensions 
in the $\L$CDM model, the strongest being the Hubble tension, but the decaying dark matter scenario still does not help addressing them. Solutions with some modification of the $\L$CDM model in the pre-recombination stage 
seem to be the most favoured from this point of view (for an extensive discussion
see \cite{Knox:2019rjx}).}

If the dark matter particle decays into SM particles, then the lower bound (\ref{tauDMlb}) becomes many orders of magnitude  more stringent and the exact value depends specifically on the mass of the dark matter particle and on its decay products. 
We will discuss specific cases.  More generally, the coupling of dark matter particles to SM particles has to be weak enough 
to escape all experimental constraints and of course also that DM particles need to decouple prior to the onset of big bang nucleosynthesis.
For this reason dark matter particles cannot interact electromagnetically and strongly (though their constituents can in case of
composite dark matter particles). 

The dark matter number density of dark matter particles at present is related to the energy density parameter simply by
\be
n_{\rm DM 0} = \ve_{\rm c0}\,h^{-2} \,  {\O_{{\rm DM}0}\,h^2 \over M_{\rm DM}} 
\simeq 1.3\,{\rm m}^{-3} \, {{\rm GeV} \over M_{\rm DM}} \,  .
\ee
From this expression we can also derive an expression for the abundance of dark matter, relatively to 
that of photons, at the production and more precisely at the freezing time $t_{\rm f}$.
Assuming that entropy is conserved between freezing and present time, one finds 
\be
\left({N_{\rm DM} \over N_\g}\right)_{\rm f} = {n_{\rm DM 0} \over n_{\g 0}} \, {g_{\rm S f} \over g_{S 0}} = 
\ve_{\rm c0}\,h^{-2} \,  {\O_{\rm DM 0}\,h^2 \over n_{\g 0} \, M_{\rm DM}} \, {g_{\rm S f} \over g_{S 0}} \simeq
8.4 \times 10^{-8}\, {{\rm GeV} \over M_{\rm DM}} \, ,
\ee
where in the last numerical expression we assumed that $T_{\rm f} \equiv T(t_{\rm f})\gtrsim 100\,{\rm GeV}$
so that we could use the SM value $g_{S}^{\rm SM} = 427/4$ for the number of ultra-relativistic degrees of freedom. 
This simple expression gives some insight on the efficiency of the production mechanism.  In particular, it shows that for higher masses
we need a less efficient mechanism since one needs a smaller number density.
If entropy is not conserved, then one has to include a further factor $S_0/S_{\rm f} \geq 1$, showing that since the relic abundance
gets additionally diluted compared to photons by an entropy increase, one needs a more efficient mechanism of dark matter production to compensate.

\subsection{The WIMP paradigm}

We mentioned that massive ordinary neutrinos play the role of dark matter but of the unwanted ({\em hot dark matter}) type.  For this reason, one gets a stringent upper bound on the sum of the neutrino masses. 
However, if one considers the case of much heavier neutral particles with weak interactions, then
a new solution appears. In the case of light active neutrinos they are produced thermally and
they decouple at a temperature $T_{\rm dec}^{\nu} \sim 1\,{\rm MeV}$
when they are still ultrarelativistic. 
If one considers considers a {\em weakly interacting massive particle} (WIMP) $X$,
its mass is assumed in the range $m_X \sim 10\,{\rm GeV}$--$1\,{\rm TeV}$,
since it is typically  associated to models where the scale of new physics is just above 
the electroweak scale to address the naturalness problem. In this way, having weak interactions,
they decouple when they are fully non-relativistic at a {\em freeze-out temperature}
$T_{\rm f}^X \simeq m_X / x_{\rm f}$, with $x_{\rm f} \simeq 25$.
This can be easily estimated  using a simple instantaneous decoupling approximation, imposing the usual criterion 
$\G_X(T_{\rm f}^X) \simeq H(T_{\rm f}^X)$  \cite{hut}.

A more accurate result can be derived solving a rate equation\footnote{It can be derived from the Boltzmann equation
for the $X$ particle distribution function integrating over momenta and within certain approximations such
as the Maxwell-Boltzmann approximation.} for the 
number density of WIMPs $n_X(t)$,  the Lee--Weinberg equation \cite{leeweinberg,dolgov}   
\be\label{leeweinberg}
{dn_X \over dt} + 3\,H\,n_X(t) = -\langle \sigma_{\rm ann} \, \b_{\rm rel} \rangle \, 
\left[ n^2_X(t) - (n_{X}^{\rm eq})^2(t) \right]  \,   ,
\ee
where $\langle \s_{\rm ann} \, \b_{\rm rel}\rangle $ is the 
thermally averaged annihilation cross section times the M\"{o}ller velocity.
Solving analytically this equation, one finds  that  the final contribution to the energy density parameter 
from WIMPs is
\be\label{DMabundance}
\Omega_{X 0} h^2 \simeq  {4 \times 10^{-10}\over \langle \s_{\rm ann} \, \beta_{\rm rel} \rangle_{\rm f}}  
\, {\rm GeV}^{-2}  \simeq {1.6 \times 10^{-37}\,{\rm cm^2} \over \langle \s_{\rm ann} \, \beta_{\rm rel}\rangle _{\rm f}} \,  ,
\ee
where $\langle \s_{\rm ann} \, \b_{\rm rel}\rangle $ is calculated at the freeze-out time.
Typical weak values of thermally averaged cross sections are given (order-of-magnitude-wise) by
 \bea\label{crosssecf}
\langle \s_{\rm ann} \, \beta_{\rm rel}\rangle ^W_{\rm f} \sim 0.1 \, {\a_W^2 \over m_X^2} \, 
& \sim  & 10^{-9}\,{\rm GeV}^{-2}  \left({100\,{\rm GeV}\over m_X} \right)^2
\\   \nonumber
& \sim & 4\times   10^{-37}\,{\rm cm}^2 \, \left({100\,{\rm GeV}\over m_X} \right)^2  \,  .
 \eea
 Here we used $\a_W \sim 0.03$, the value of the  dimensionless weak coupling constant at energies $\sim 100\,{\rm GeV}$ and for the relative velocity at the freeze-out a typical value $\beta_{\rm rel} \simeq 0.1$.
It is then intriguing that  for WIMP  masses $m_X  \sim 100\,{\rm GeV}$--$1\,{\rm TeV}$, 
the expected values from naturalness,
one obtains $\O_{X 0} h^2 \sim 0.1$, nicely reproducing the observed value of the dark matter abundance.
Because of this tantalising {\em WIMP miracle}, for long time  WIMPs 
  have been regarded as  the most attractive dark matter candidate also
  in consideration of their detectability by virtue of their weak interactions. 

Some of the proposed most popular realistic examples of WIMPs  \cite{silkbertone,feng},
are associated with specific solutions to the naturalness problem.  
Supersymmetry is certainly the most popular solution, perhaps also the most elegant one,
and for this reason the most popular  dark matter WIMP candidate
is the lightest supersymmetric neutral  spin 1/2 fermion, the so-called {\em neutralino}.    
This can be made stable by adding a discrete symmetry called
R-parity with an associate conserved quantum number with only two discrete values: +1 (even particles) or -1 (odd particles).  
Supersymmetric and standard model particles have opposite R-parity, minus and plus, respectively. 
Therefore, a supersymmetric particle cannot decay just into SM particles and this implies 
that the lightest supersymmetric particle has to be stable.
Neutralino as dark matter WIMP was proposed in the eighties \cite{goldberg,ellis} and
it became such an attractive solution that its discovery was considered just a matter of time. Moreover, as we discussed in Section 2.3,
within supersymmetry one can also have successful electroweak baryogenesis, thus realising a unified picture for the
origin of matter in the universe.\footnote{This possibility has been  investigated in particular  within
the MSSM in the presence of a light stop \cite{Balazs:2004bu,Balazs:2004ae}
and within the nMSSM \cite{Balazs:2007pf}.}
Unfortunately,  intense searches of different kinds have not fulfilled 
  expectations  and this has stimulated new ideas for models solving the naturalness 
 problem with their own associated dark matter WIMP candidate.

A popular class of theories providing a solution to the naturalness problem, alternative to supersymmetry, 
are theories with extra-dimensions.
They  also usually contain candidates for (thermally produced) dark matter WIMPs,   
the most popular one being the lightest Kaluza--Klein particle. This is
associated to some standard model particle, typically the hypercharged gauge boson B. 
Other recent examples of models addressing the naturalness problem are (in brackets their associated dark matter WIMP):  
\begin{itemize}
\item[(i)] little Higgs models (T-odd particles); 
\item[(ii)] two Higgs doublet models (neutral Higgs boson); 
\item[(iii)]  twin Higgs models (twin neutral Higgs); 
\item[(iv)] dark sector models (dark lightest particle). 
\end{itemize}

The evidence found so far for the existence of dark matter is uniquely based only on its gravitational effects. 
The attractive feature of dark matter WIMPs is that they also interact weakly
and this implies specific experimental signatures that allow to test the WIMP paradigm. 
There are three  strategies pursued to detect dark matter WIMPs 
through their weak interactions \cite{feng,roszkowski}.
If we indicate with SM some generic standard model particle, and with $X$ our WIMP, one looks
for signals of the following kind of interactions:
\begin{itemize}
\item[(i)]  ${\rm X} + {\rm SM} \rightarrow {\rm X} + {\rm SM}\;$ ({\em direct searches})  \,  ;
\item[(ii)] $ {\rm X} + {\rm X} \rightarrow  {\rm SM} + {\rm SM}\;$ ({\em indirect searches}) \,   ;
\item[(iii)]  ${\rm SM} + {\rm SM} \rightarrow {\rm X's} + {\rm SM's} \;$ ({\em collider searches}) \,   .
\end{itemize}

{\em Direct searches}. Dark matter WIMPs, though very rarely, can collide with ordinary matter through
their weak interactions. In particular, they can hit nuclei in a detector and  transfer 
them energy (nuclear recoil energy)  that can be detected. 
In the last three decades different experiments have been placing more and more 
stringent limits on the mass and on the cross section with nuclei of the dark matter WIMPs.  
However,  there are also models of WIMPs with spin dependent cross section and in this case constraints are weaker.
These constraints also depend on the evaluation of astrophysical quantities 
such as the local density and the velocity distribution of WIMPs
at our position in the Milky Way, clearly affected by some theoretical uncertainties. 

{\em Indirect searches}. A complementary strategy is to look for
the products of dark matter interactions within astronomical environments.  The interactions with
ordinary matter are in this case unhelpful. In a perfectly homogeneous universe, dark matter annihilations, though not completely  turned off,  
would be so strongly suppressed  that it would be impossible to detect any observable effect.
However, dark matter, like ordinary matter, clumps on scales smaller than 100\,{\rm Mpc}.
Since the rate of annihilations $\propto \rho^2_{DM}$, one expects that the flux of 
annihilation products is greatly enhanced in very dense dark matter regions, 
such as the galactic centre or dwarf galaxies, where the density of dark matter
particles is expected to be order of magnitudes above the average value.
First of all one can hope to detect $\g$-rays 
coming from such dense dark matter regions. These have the advantage that would travel from the production site
to us in a straight line, retaining information on the source position in the sky. 
Therefore, one can target dense dark matter regions to have better 
chances to distinguish a signal from the background.  
Ideally, one would like to observe a monochromatic line but unfortunately the rate of WIMPs 
direct annihilations  into photons is too weak and what one can realistically detect are photons produced as 
secondary particles from the primary particles produced in the annihilations. These generate an excess spread on
a wide range of energies. In the mass range $(0.1$--$1)\,{\rm TeV}$ the most stringent upper bound  on 
the WIMP annihilation cross section has been placed by the Fermi-LAT space telescope and it basically excludes  
the typical values expected for thermally produced WIMPs the value at freeze-out
in Eq.~(\ref{crosssecf}).  At values of the mass above ${\rm TeV}$, most stringent constraints 
are placed by the HESS ground based $\gamma$-ray telescope but they still allow values
$\langle \sigma_{\rm ann}\,\beta \rangle_{\rm f}$ as large as $10^{-35}\,{\rm cm}^2$.
On the other hand at values $m_X \lesssim 0.1\,{\rm TeV}$ the most stringent limits come from 
the {\em Planck} satellite observations of CMB anisotropies constraining $\langle \s_{\rm ann}\,\beta_{\rm rel} \rangle$
at the time of recombination.  These constraints strongly exclude weak cross sections for thermally produced WIMPs.

Alternatively, one  can search for dark matter products   
in {\em charged cosmic rays}. However, in this case particles are 
deflected by the galactic magnetic fields and would travel to us
along a random curvy path inside the galaxy, so that directionality is lost. 
One can simply compare the measured spectrum to the one predicted 
by the SM looking for some excess that would be the result of WIMPs annihilations. 
Typically, instruments detect electrons and protons, that can originate from many astrophysical sources,
and above all positrons \cite{Delahaye:2007fr}
and anti-protons  \cite{Donato:2003xg}, that is even more interesting for dark matter searches since these 
would be produced in equal amounts to electrons and protons in dark matter annihilations. 
Given a source for cosmic rays with a specified energy spectrum at production, 
the {\em prompt} spectrum of particles,  
the predicted energy spectrum on the Earth is the result of many complicated processes, 
with the main one being the diffusion due to the galactic magnetic fields.  This is usually calculated solving a 
{\em Fokker--Planck equation} \cite{Strong:2007nh}, that is a generalised diffusion equation.   
Cosmic rays are detected with balloon-type, ground based telescope arrays or satellite-based 
detectors. In 2009 the PAMELA satellite detector reported an excess in the positron spectrum 
in the energy range $(20$--$200)\,{\rm GeV}$. 
The Fermi-LAT and the AMS-02 satellite detectors confirmed the excess though to a lower level. 
A dark matter interpretation seems plausible but it encounters great difficulties since the 
value of the cross section required to explain the excess is much larger than the typical values 
expected for thermally produced WIMPs (see Eq.~(\ref{crosssecf})). 
Many specific models have been proposed but they are in  tension with the constraints from 
$\g$-rays so that further experiments are needed for a conclusive verdict.

Neutrinos can also be very useful for indirect searches of WIMPs. Thanks to their weak interactions,  
WIMPs would dissipate energy scattering off nuclei in central  dense regions of celestial bodies
such as stars and planets. If their velocity becomes smaller than the escape velocity, they 
get trapped accumulating gradually in the central dense regions with a capture rate $\G_{\rm capt}$. 
When density gets higher and higher,  the annihilation rate $\Gamma_{\rm ann}$ becomes 
sufficiently large to  balance the  capture rate and equilibrium is reached. WIMPs, such as neutralinos,
cannot annihilate directly into neutrinos but their annihilation products, e.g., quarks and gauge bosons, 
would produce them secondarily. Neutrinos
are the only particles able to escape the centre of stars travelling in a straight line and retaining
information on their production site. In our case the only object able to produce a detectable 
neutrino flux would be the Sun or, to a lower level, the same centre of the Earth.
Neutrino telescopes, big neutrino detectors able to track the arrival direction of neutrinos, 
 pointing toward the centre of  the Sun or the Earth  should then detect a high energy neutrino flux.  
 This strategy is sensitive both to the scattering off nuclei and annihilation cross section of WIMPs. 
 The derived upper bounds on spin independent and on annihilation cross section are less stringent than those 
 derived from direct and $\gamma$-rays searches. However, those on spin dependent cross section, derived by combining data from
 ANTARES, IceCube and SuperKamiokande neutrino telescopes, are the most stringent ones.  
 
{\em Collider searches}. Dark matter WIMPs could also be produced in colliders and currently the LHC
has the right energy to test a wide range of masses. 
Unfortunately, so far, there are no signs of new particles in the LHC  and not even  
of missing energy and momentum that cannot be explained with neutrinos.  
From LHC results one can then place upper bounds on the 
dark matter production cross sections and these are perfectly compatible with those from direct and indirect searches.  

We can fairly conclude that current experimental results rule out simple `WIMP miracle' expectations,
strongly motivating modifications or alternative models. 

The main reason why the WIMP miracle is ruled out is that the upper bounds on the cross sections 
are so stringent to point to much smaller values than those required by  Eq.~(\ref{crosssecf}):
from Eq.~(\ref{DMabundance}) one would then obtain a dark matter abundance much higher than the observed one.
However, one can still think to save the idea of WIMP dark matter relaxing
one or more assumptions on which the WIMP miracle scenario relies on. For example,
 one assumption is that the freeze-out occurs in isolation, i.e., that it is not influenced by the existence of possible 
additional new particle species that could also be WIMPs. 
Relaxing this assumption,  one can obtain the correct abundance of dark matter WIMPs,
with much lower values of $\langle \s \, \b \rangle_{\rm f}$, 
taking into account three possible effects \cite{griestB}. 
A first important one is given by so-called {\em co-annihilations},   
occurring when the mass of the dark matter WIMP is sufficiently close to the mass of a heavier unstable WIMP.  
Another possibility is that annihilations occur resonantly, at energies close to the $Z$ or Higgs boson mass, and this also tends to enhance the annihilation rate reducing the final relic abundance compared to the non-resonant case and making it compatible with the observed value. 
Both  effects can be realised within supersymmetric models, singling out particular regions
in the space of parameters (the so-called co-annihilations and funnel regions) where 
the neutralino is the lightest supersymmetric particle and its relic abundance 
reproduces the observed dark matter abundance.

\subsection{Beyond the WIMP paradigm}

The list of models beyond those of dark matter WIMPs relying on a traditional freeze-out production mechanism   
is impressively long.  This certainly shows how the {\em dark matter puzzle} is one of the greatest challenges in modern science. 
Here we mention the most popular ideas referring the reader to
more specialistic reviews \cite{feng,baer}.

\begin{itemize}

\item {\em Hidden dark matter and the WIMPless miracle}.    The WIMP miracle can be  somehow revisited  considering
dark matter particles with interactions such that
\be
\langle \sigma_{\rm ann}\,\beta_{\rm rel} \rangle_{\rm f} = {\a^2 \over m^2_X}\,{m^2_{\rm WIMP} \over \a^2_{W}}
 \langle \sigma_{\rm ann}\,\beta \rangle_{\rm f}^{\rm WIMP} \,  ,
\ee
where  $\langle \sigma_{\rm ann}\,\beta \rangle_{\rm f}^{\rm WIMP}$ denotes the
special value of $\langle \sigma_{\rm ann}\,\beta_{\rm rel} \rangle_{\rm f}$, given by 
Eq.~(\ref{crosssecf}), for $m_X = m_{\rm WIMP}$.  For $\a_X/m_X = \a_W/m_{\rm WIMP}$ one  still has 
$\langle \sigma_{\rm ann}\,\beta_{\rm rel} \rangle_{\rm f} =  \langle \sigma_{\rm ann}\,\beta_{\rm rel} \rangle_{\rm f}^{\rm WIMP}$.
This means that even though the dark matter particle is not a WIMP, 
it has the same annihilation cross section of WIMPs. 
In this way the observed dark matter abundance is still reproduced
through the usual thermal freeze-out mechanism.
For example, a dark matter particle species with 
$m_X \sim {\rm MeV}$ and $\a \sim 10^{-5}\,\a_W$
would still yield the correct relic abundance but it would
escape the tight experimental constraints we discussed.

\item {\em Feeble, extremely or super-interacting massive particles (FIMPS or SWIMPs)} 
\cite{Bernal:2017kxu,Agrawal:2021dbo}.
These dark matter particle candidates have interactions much weaker than weak interactions. 
Nevertheless, they can be still produced thermally. However, 
their abundance never reaches the thermal value before freeze-out, rather it directly 
freezes-in  to the correct value starting from an initial vanishing abundance \cite{FIMPs}.
The relic abundance is in this case proportional to the annihilation cross section instead of inversely proportional.
In this way one can obtain the correct abundance for the same values of masses as in the case of 
WIMPs but with much  smaller cross sections, thus escaping current experimental constraints.
Sterile neutrinos can be regarded as a particular important example of FIMP dark matter, in fact the first to be
proposed. We will discuss specific models both in Section 6 and Section 7. 

Alternatively,  their production can occur non-thermally from late decays of heavier particles. 
For example, one could have an unstable WIMP that, by virtue of the WIMP miracle  
has the correct relic abundance $\Omega_{\rm WIMP}$. Even though these unstable WIMPs cannot 
play the role of dark matter,  still their abundance can be transferred to a
Super-WIMP (cosmologically stable)  particle through decays, in such a way that   
\be
\Omega_{\rm SWIMP} = \Omega_{\rm WIMP}\,{m_{\rm SWIMP} \over m_{\rm WIMP}} \,  .
\ee
If $m_{\rm SWIMP}$ is not too much smaller than $m_{\rm WIMP}$, the WIMP miracle still works also for SWIMPs,
with the difference that SWIMPs escape direct and indirect search constraints. 
However, WIMPs could be still produced if some astrophysical environments and their decays
have some testable effects in cosmic rays if the decaying WIMP is charged. 
Particle colliders may also find evidence for the SWIMP scenario in this case. 
Finally decays of WIMPs into SWIMPs might have an  impact on CMB and BBN in the form for example
of extra-radiation, sometimes called {\em dark radiation} and often parameterised in terms of the 
number of effective neutrino species. In the case of BBN, for life-times between $\sim 1\,{\rm s}$ 
and $\sim 1000\,{\rm s}$, the products of decays  alter in general the values of the primordial nuclear abundances
in an unacceptable way yielding constraints on the parameters of the model.  However, for some fine-tuned choice of the parameters,  the late WIMP decays could even be beneficial, solving the long-standing 
{\em lithium problem} in BBN \cite{warmordark}. 

A popular example of SWIMP particle is the {\em gravitino}, superpartner of the graviton with spin $3/2$. 
It can play the role of  dark matter when it corresponds to the lightest supersymmetric particle, 
since in this case it can be cosmologically stable.
It can be produced both thermally, through the freeze-in mechanism, and non-thermally through the decays of the  next-to-lightest supersymmetric particle (typically the neutralino).  In order to get the correct abundance,
the reheat temperature cannot be above $\sim 10^{10}\,{\rm GeV}$. 
For gravitino masses below $\sim 1\,{\rm TeV}$, this upper bound on the 
reheat temperature, becomes a few orders of magnitude more stringent 
taking into account the constraints from BBN. Such stringent upper bound on the reheat 
temperature is incompatible with lower bounds on the reheat temperature within different
scenarios,  
thermal 
GUT baryogenesis that we discussed and, as we will see, even with the more 
relaxed analogous lower bound  arising in thermal leptogenesis: this is the well known 
{\em gravitino problem} \cite{gravitino}.
  
\item {\em Axions}.   \index{axion}
Another attractive and very popular dark matter candidate emerges naturally from 
the so-called {\em Peccei--Quinn mechanism} for the solution of the strong $C\!P$ 
violation problem in QCD \cite{Peccei:1977hh}.
This mechanism predicts the existence of a new particle, called {\em axion}, 
that can have the right features to play the role of a dark matter particle.
Axion particles would be associated to a pseudoscalar field $a$ with coupling
to gluons determined by a coupling constant $f_a^{-1}$, where $f_a$ has the dimension
of energy, determining the decay rate of axions and also the same axion mass given 
approximately by
\be
m_a \sim 10^{-5}\,{\rm eV}\, \left({10^{12}\,{\rm GeV} \over f_a} \right) \,  .
\ee
In order for axions to be stable on cosmological scales, one needs $f_a \gtrsim 10^6\,{\rm GeV}$
but actually, from astrophysical constraints, one obtains a much more stringent lower bound, 
$f_a \gtrsim 10^9 \,{\rm GeV}$ translating  into $m_a \lesssim 10\,{\rm meV}$. 
The cosmological production of axions would proceed non-thermally through a {\em vacuum
misalignment} mechanism producing an abundance
\be
\Omega_a \sim 0.3\,\theta^2 \, \left({f_a \over 10^{12}\,{\rm GeV}}\right)^{1.18} \,  ,
\ee
where $\theta$ is the initial vacuum misalignment angle. The correct dark matter density is then
achieved for $m_a \sim 10^{-3}\,\theta^2 \, {\rm meV} \sim 
(10^{-6}$--$10^{-4})\,{\rm eV}$. Despite the small mass, axions would behave similarly
to cold dark matter rather than hot dark matter, in agreement with structure formation constraints.  
This is because dark matter axions would basically form a Bose--Einstein condensate with negligible
kinetic energy. 

The axion can be detected directly since it interacts with SM particles. 
Currently, the most stringent constraints on the mass and coupling constant 
are placed by the conversion of dark matter
axions to photons through scatterings off a background magnetic field (Primakoff process). 
Within supersymmetry,  the supersymmetric partner of the axion, the {\em axino}, 
is also a viable candidate for dark matter and, contrarily to the axion, it would behave as a
SWIMP, similarly to the gravitino.

\item {\em Super massive dark matter (WIMPzillas)}. Many theories such as grand-unified theories, as we
have seen when we discussed GUT baryogenesis, predict the existence of 
super massive particles with masses as high as the grand-unified scale, 
$M_X \sim (10^{15}$--$10^{16})\,{\rm GeV}$.
Since they are above the maximum  allowed value of the reheat temperature, 
$T_{RH} \lesssim 10^{15}\,{\rm GeV}$, they cannot be produced thermally.   
However, they can be produced non-thermally at the end of inflation in different ways \cite{wimpzillas}:
\begin{itemize}
\item[(i)] At reheating, when the inflaton field energy is transferred to all other particles. 
In this case the correct dark matter abundance can be attained even for
masses that are orders of magnitude greater than the reheat temperature.
\item[(ii)] By inflaton decays if the mass of the inflaton is greater than the mass of the 
supermassive dark matter particle.
\item[(iii)] Finally, they can be produced at the so-called {\em preheating}, when the energy of the inflaton field can create supermassive particles through non-perturbative quantum effects on a  curved space leading to parametric resonant production, an effect that can be mathematically described using Bogoliubov transformation leading to a  Mathieu equation for particle production.  
Such a preheating stage at the end of inflation would also be responsible 
for the production of gravitational waves during inflation and in this way it could even be tested.
\end{itemize} 
\end{itemize}

\subsection{Asymmetric dark matter}

Asymmetric dark matter models\footnote{We just briefly discuss them here, we refer the reader to 
two recent reviews for an exhaustive discussion \cite{davoudiasl,volkas}.} 
provide a very elegant solution to circumvent the experimental stringent constraints we mentioned for
WIMPs but they also provide an attractive way to combine baryogenesis with dark matter models. 
Moreover they have the additional ambitious motivation to address the apparent 
coincidence $\O_{M0} \simeq 5\,\O_{B 0}$.

The freeze-out mechanism assumes a vanishing initial asymmetry \index{asymmetric dark matter}
between the number of dark matter WIMPs and anti-WIMPs.
However, as in the case of ordinary matter, some mechanism might have generated
an asymmetry in the dark matter sector prior to the freeze-out. 
If this asymmetry is sufficiently large, then the relic abundance
is basically given by the same initial asymmetry. This idea is quite attractive since 
one can also naturally link the problem of generation of the baryon asymmetry to the problem of dark matter, 
obtaining a combined model on the origin of matter in the universe.
The indirect detection rate from annihilation is 
of course strongly suppressed compared to the standard case (since basically one is left only with
particles or anti-particles today) and in this way the stringent  constraints 
holding in the standard freeze-out scenario are evaded. On the other hand, signals in direct detection 
searches are usually enhanced 
and many constraints have been placed excluding different regions in the parameter space. 
In particular the elegant solution of having the same asymmetry in baryonic matter and dark matter,
$|\eta_B| \simeq |\eta_{DM}|$, implying $m_{X} = (\Omega_{DM}/\Omega_B)\,m_{p} \simeq 
5\,{\rm GeV}$, where $m_{p}\simeq 1\,{\rm GeV}$ is the proton mass, 
is now ruled out by the stringent LUX constraints.

\section{Neutrino masses and mixing}

An explanation of neutrino masses and mixing necessarily requires an extension  of the SM. It is  then reasonable to search for scenarios explaining the origin of matter in the Universe that can be realised within those extensions of the SM also able to describe neutrino  masses. In this section we review the current experimental status on neutrino parameters. In the next sections we will see how extensions of the SM that explain neutrino masses and mixing can also address the origin of matter in the universe.

\subsection{Low energy neutrino experiments}

SM neutrinos interact at low energies, when the $SU(2)_L \times U(1)_Y$ electroweak symmetry is broken to $U(1)_{\rm em}$,  via neutral currents 
\be
[J_{\mu}^{NC}]_\nu = \sum_{i' = 1,\dots,N_\nu} \overline{\nu}_{L i'} \,\gamma_\m\,\nu_{L i'} \,  ,
\ee
where neutrinos couple to the $Z$ bosons, and via  the leptonic charged current
\be \label{current}
J_{- \mu}^{\rm lept}= (J_{+ \mu}^{\rm lept})^{\dagger} = \sum_{i'} \overline{\ell_{L i'}}\,\gamma_{\mu}\,\nu_{L i'} \,  ,
\ee
where neutrinos couple to the $W^{\pm}$ bosons. Like all SM lepton interactions with gauge bosons, they are flavour blind, thus respecting {\em lepton universality}. Therefore, in the previous expression leptons fields have to be meant in a generic flavour basis.  Flavour dependence originates from charged lepton Yukawa interactions. In a generic flavour basis, denoted by primed Latin indexes, these are described by the Lagrangian term
\be\label{yukawacl}
-{\cal L}_Y^{{\ell}} = \overline{L_{i'}}\,h_{i'j'}^{\ell}\,{\ell}_{R j'}\, \Phi + {\rm h.c.} \,  ,
\ee
where $\Phi =(\phi^+,\phi^0)^T$ is the Higgs doublet.
After EW symmetry breaking by a non zero vacuum expectation value (VEV) $v$
of the Higgs field, the Yukawa coupling matrix $h^{\ell}_{i'j'}$ yield the charged lepton mass matrix
\be\label{masschargedlep}
m^{\ell}_{i'j'}=v\, h^{\ell}_{i'j'} \,   
\ee
and the charged lepton Lagrangian mass term can be written as
\be
-{\cal L}_{\rm m}^{\ell} =  \overline{{\ell}_{L i'}} \, m^{\ell}_{i'j'}\,{\ell}_{R j'} \,  .
\ee
In the charged current in Eq.~(\ref{current}), the charged lepton fields are not in general mass eigenfields.
However, one can always diagonalise the charged lepton mass matrix by a bi-unitary
 transformation of the left- and RH charged lepton fields,
\bea
{\ell}_{L i'} \rightarrow \a_L \equiv {\ell}_{L \a} & = & U_{L  \, \a i'}^{\ell }\,{\ell}_{L i'} \, , \;\;\;   \\
{\ell}_{R i'} \rightarrow \a_R \equiv {\ell}_{R \a} & = & U_{R \, \a i'}^{\ell}\,{\ell}_{R i'} \, ,
\hspace{10mm}  (\a = e, \m , \t)
\eea
so that correspondingly
\be
m^{\ell}_{i'j'} = \left(U^{{\ell}\dagger}_L\,{\rm diag}(m_e,m_{\mu},m_{\tau})\,U_R^{\ell}\right)_{i'j'} \, .
\ee
Therefore, the matrices $U_L$ and $U_R$ transform the charged lepton LH and RH fields respectively 
from the generic  primed basis to the weak basis. After this transformation the leptonic charged current (see Eq.~(\ref{current})) becomes
\be
J_{- \mu}^{\rm lept}= \sum_{\a , i'} \overline{{\a}_{L}}\,\gamma_{\mu}\,U^{\ell}_{L \, \a i'}\,\nu_{L i'}  \,   .
\ee
In this way the basis where the the charged lepton mass matrix $m^{\ell}$ is diagonal
identifies a privileged flavour basis, the weak interaction basis, providing a natural lepton
basis where to describe physical processes, since it is automatically (and quickly) selected by kinematics. 
This basis is determined by a bi-unitary transformation of the left and RH charged lepton fields
\bea
{\ell}_{L i'} \rightarrow \a_L \equiv {\ell}_{L \a} & = & U_{L  \, \a i'}^{\ell }\,{\ell}_{L i'} \, , \;\;\;   \\
{\ell}_{R i'} \rightarrow \a_R \equiv {\ell}_{R \a} & = & U_{R \, \a i'}^{\ell}\,{\ell}_{R i'} \, ,
\hspace{10mm}  (\a = e, \m , \t)
\eea
in a way that 
\be
m^{\ell}_{i'j'} = \left(U^{{\ell}\dagger}_L\,{\rm diag}(m_e,m_{\mu},m_{\tau})\,U_R^{\ell}\right)_{i'j'} \,  .
\ee
In the weak interaction basis the leptonic charged current can be written as
\be
J_{- \mu}^{\rm lept}=
\sum_{\a , i'} \overline{{\a}_{L}}\,\gamma_{\mu}\,U^{\ell}_{L \, \a i'}\,\nu_{L i'}  \,  .
\ee
Defining the {\em neutrino weak interaction eigenfields}  as
\be\label{primedtoweak}
\nu_{L \a}\equiv U_{L \, \a i'}^{\ell}\,\nu_{L i'} \,   ,
\ee
the leptonic charged current can then be written as 
\be
J_{- \mu}^{\rm lept}= \sum_{\a} \overline{{\a}_L}\,\gamma_{\mu}\,\nu_{L \a} \,  ,
\ee
showing that the {\em neutrino weak interaction eigenfields} 
create the neutrino states produced in the decay 
of a $W$ boson in association with a physically charged lepton ${\a}$ of definite mass.

In the SM neutrinos are massless. If neutrinos do have a mass, as now established by neutrino oscillation 
experiments, then 
one can define a basis of $N$ neutrino mass eigenfields $\nu_1, \nu_2\, \dots, \nu_N$,  associated
respectively to masses $m_1, m_2, \dots, m_N$, that in general do not coincide with the weak eigenfields.  
The leptonic mixing matrix $U$ is defined as the matrix that operates
the transformation from the mass eigenfields $\nu_i$ to the weak eigenfields $\nu_{\a}$, 
explicitly
\be
\nu_{\a} = \sum_{i=1}^N U^{(N)}_{\a i} \, \nu_{i} \,  .
\ee
In general the number $N$ of mass eigenfields  mixing with the weak eigenfields does not have to coincide with the number of weak eigenfields (three)  but it could be higher. This can well happen since, in extensions of the SM incorporating   neutrino masses,  in addition to three active neutrino fields there can also be
sterile neutrinos mixing with the ordinary ones. In this case if one describes the mixing accounting
only for three mass eigenfields, the $(3 \times 3)$ leptonic mixing matrix $U_{a i}$ with $i=1,2,3$, also 
referred to as the PMNS matrix,  is in general non-unitary. 
We will discuss  a specific case with $N > 3$ within the type-I seesaw extension of the SM and we will see that the
active-sterile neutrino mixing can have important cosmological applications especially in connection to the origin of matter.  For the time being let us restrict ourselves to  the standard case with $N=3$ and unitary $U$. 

We can also consider a unitary transformation that brings from the neutrino mass basis to a generic primed basis:
\be
\nu_{L i'} = U_{L\, i'i}^{\nu} \,  \nu_{L i}  \,   .
\ee
In this generic primed basis the leptonic mixing matrix can be regarded as the combination of two unitary 
transformations: a transformation $U_{L}^{\nu}$ bringing from mass eigenfields basis
to the primed basis and a second transformation $U_L^{\ell}$ that, as already seen, brings
from the primed basis to the weak basis where the charged lepton mass matrix is diagonal, explicitly
\be\label{ULnu}
U= U_L^{\ell} \,  U_{L}^{\nu}  \,   .
\ee

 A generic unitary matrix would be described by nine parameters.
However, in the specific case of the leptonic mixing matrix, three
phases are unphysical since they can be absorbed in the charged lepton fields
without any observable physical consequence. In this way the leptonic
mixing matrix can be parameterised in terms of 6 parameters, 3 mixing angles
$\theta_{12},\theta_{13},\theta_{23}$ and 3 $C\!P$ violating phases $\d$, $\a_1$, $\a_2$.
A standard parametrisation is then given by 
\be\label{lmm}
U= \left( \begin{array}{ccc}
c_{12}\,c_{13} & s_{12}\,c_{13} & s_{13}\,e^{-{\rm i}\,\d} \\
-s_{12}\,c_{23}-c_{12}\,s_{23}\,s_{13}\,e^{{\rm i}\,\d} &
c_{12}\,c_{23}-s_{12}\,s_{23}\,s_{13}\,e^{{\rm i}\,\d} & s_{23}\,c_{13} \\
s_{12}\,s_{23}-c_{12}\,c_{23}\,s_{13}\,e^{{\rm i}\,\d}
& -c_{12}\,s_{23}-s_{12}\,c_{23}\,s_{13}\,e^{{\rm i}\,\d}  &
c_{23}\,c_{13}
\end{array}\right)
\cdot {\rm diag}\left(e^{i\,\rho}, 1 ,e^{i\,\sigma}
\right)\, ,
\ee
where $s_{ij}\equiv \sin\theta_{ij}$ and $c_{ij}\equiv \cos\theta_{ij}$.
This parametrisation is the same adopted also for the CKM matrix, describing mixing in the quark sector,
but with the addition of the two Majorana phases $\rho$ and $\sigma$ that cannot be reabsorbed
in the redefinition of the neutrino fields in the more general case when these are Majorana fields.
They cannot have any physical effect in low energy neutrino experiments based on kinematical processes 
since these are the same for Dirac and Majorana neutrinos. In particular Majorana phases cancel out in neutrino oscillation  probabilities and, therefore, cannot be tested in neutrino oscillation experiments.  
At the moment, the only known phenomenology that can effectively give us information on Majorana phases is neutrinoless double beta ($0\nu\b\b$) decay but since no signal has been found so far, not only we have no experimental constraints on the Majorana phases  but we do not even know whether they are physical, in the case of Majorana neutrinos, or unphysical, in the case of Dirac neutrinos.

Neutrino oscillation probabilities depend on neutrino mass squared differences and in particular they would all vanish if
the three neutrinos would be exactly mass degenerate. 
Therefore, the discovery of neutrino oscillations implies that neutrino are massive and that at least two of them  are non-vanishing  and different from each other.  Experimental results are well described in terms of two mass squared differences corresponding to
the case of three neutrino masses  and do not find evidence of a fourth mass eigenstate. 
A first mass squared  difference, $\Delta m^2_{\rm sol} \equiv m^2_2-m^2_1$, is measured in solar neutrino experiments and in the KamLAND long baseline reactor neutrino experiment. Since matter effects in the Sun depend on its sign, this is measured and found to be positive. A second mass squared
difference was initially measured in atmospheric neutrinos and now also  in long baseline neutrino
experiments. In absolute value it is much larger than the solar difference, but the sign is still unknown. For this reason
current neutrino oscillation experiments do not solve an ambiguity between two possible options: one can have either
{\em normal ordering} for $\Delta m^2_{\rm atm} \equiv  m^2_3 - m^2_1 > 0$ implying $m_1 < m_2 < m_3$, or {\em inverted ordering}, 
for $\Delta m^2_{\rm atm} \equiv m^2_3 - m^2_2  < 0$, implying $m_3 < m_1 < m_2$, where we defined 
$\Delta m^2_{\rm atm}$ as the  mass squared difference with largest
absolute value in both cases.


 The most recent global analyses find for normal ordering \cite{nufit2020}
\bea\label{normal}
\theta_{12} & = & 33.44^\circ \pm 0.75^\circ = [31.3^\circ, 35.9^\circ] \,   ,   \\ \nonumber 
\theta_{23} & = & \left. 49.2^\circ \right.^{+0.9^\circ}_{-1.2^\circ} \;\;\;\;\;\;\,   = [40.1^\circ, 51.7^\circ] \,   ,   \\ \nonumber 
\theta_{13} & = & 8.57^\circ \pm 0.12^\circ\,   \;\; = [8.20^\circ, 8.93^\circ] \,   ,   \\ \nonumber 
\delta  & = & \left. -163^\circ \right.^{+27^\circ}_{-24^\circ}   \;\;\;\;\;\; = [-240^\circ, +9^\circ]   \,  ,  \\  \nonumber 
\Delta m^2_{\rm sol}  =  (7.42 \pm 0.20) \times 10^{-3}\,{\rm eV}^2 \; 
& \Rightarrow & m_{\rm sol}  \equiv  \sqrt{\D m^2_{\rm sol}} =  (8.6 \pm 0.1) \, {\rm meV} \,  ,  \\ \nonumber
\Delta m^2_{\rm atm}  =  (2.517 \pm 0.027) \times 10^{-3}\,{\rm eV}^2 \;  
& \Rightarrow &  m_{\rm atm}  \equiv  \sqrt{\D m^2_{\rm atm}} =  (50.2 \pm 0.2) \, {\rm meV} \,  ,
\eea
and for inverted ordering
\bea\label{inverted}
\theta_{12} & = & \left. 33.45^\circ \right.^{+0.78^\circ}_{-0.75^\circ} \;\;\;\, = [31.3^\circ, 35.9^\circ] \,   ,   \\ \nonumber 
\theta_{23} & = & \left. 49.3^\circ \right.^{+0.9^\circ}_{-1.1^\circ} \;\;\;\;\;\;\,   = [40.3^\circ, 51.8^\circ] \,   ,   \\ \nonumber 
\theta_{13} & = & 8.60^\circ \pm 0.12^\circ\,   \;\; = [8.24^\circ, 8.96^\circ] \,   ,   \\ \nonumber 
\delta  & = & \left. -78^\circ \right.^{+26^\circ}_{-30^\circ}   \;\;\;\;\;\; = [-167^\circ, +8^\circ]   \,  ,  \\  \nonumber 
\Delta m^2_{\rm sol}  =  (7.42 \pm 0.20) \times 10^{-3}\,{\rm eV}^2 \;  
& \Rightarrow & \, m_{\rm sol}  \equiv   \sqrt{\D m^2_{\rm sol}} =  (8.6 \pm 0.1) \, {\rm meV} \,  ,  \\ \nonumber
\Delta m^2_{\rm atm}  =  (-2.498 \pm 0.028) \times 10^{-3}\,{\rm eV}^2 \,   
 & \Rightarrow  & \,  m_{\rm atm}  \equiv   \sqrt{-\D m^2_{\rm atm}} =  (50.00 \pm 0.25) \, {\rm meV} \,  ,
\eea
where we indicated the best fit values with the $1\s$ errors and for the three mixing angles and $\delta$ the $3\s$ ranges as well.

The measurement of two different neutrino mass scales, the atmospheric and the solar ones, implies that we can parametrise the three neutrino masses
in terms of just one unknown parameter. This can be conveniently chosen as the lightest neutrino mass that we denote by $m_{1'}$, where $1'= 1$ for normal ordering and $1' = 3$ for inverted ordering. We also 
denote by $m_{3'}$ the heaviest neutrino mass
($3' = 3$ for normal ordering and $3' = 2$ for normal ordering) and with $m_{2'}$ the intermediate neutrino
mass ($2' = 2$ for normal ordering and $2' = 1$ for inverted ordering).
If $m_{1'} \gg m_{\rm atm}$ then necessarily al three neutrino masses become quasi-degenerate. In the case of normal ordering, 
the hierarchical limit is obtained for  $m_{1'} = m_1 \ll m_{\rm sol}$. In this case in the limit $m_1 \ra 0$ 
one has $m_2 = m_{\rm sol}$ and $m_3 = m_{\rm atm}$. In the case of inverted ordering, the hierarchical limit is obtained
for $m_{1'} = m_3 \ll m_{\rm atm}$ and in this case one has $m_{3'} = m_2 = m_{m_{\rm atm}}$
and $m_{2'} = m_{1} = \sqrt{m^2_{\rm atm} - m^2_{\rm sol}} \simeq m_{\rm atm}\,[1 - m^2_{\rm sol}/(2\, m^2_{\rm atm})]$. 

In September 2019 the KATRIN collaboration released first results placing the world best 
direct neutrino mass upper bound on the electron neutrino mass \cite{KATRIN}. 
This has been recently improved combining first data taking with second data taking results, 
obtaining for the first time a sub-eV upper bound \cite{KATRIN2}, namely
\begin{equation}
m_{\nu_e} \equiv  \sqrt{\sum_i\, \left| U_{ei}\right| ^2 \, m^2_i} < 0.8\,{\rm eV}\, (90\%\mbox{C.L.}) \,  ,
\end{equation}
translating into the same upper bound on the lightest neutrino mass $m_{1'}$. If neutrinos are Majorana particles then
$0\nu\beta\beta$ decays are possible. Experiments are sensitive to  
the effective neutrinoless double beta decay neutrino mass defined as
\be\label{mee}
m_{ee} \equiv |m_{\nu ee}| = \left|m_1\,U^2_{e1}+m_2\,U^2_{e2}+m_3\,U^2_{e3} \right| \,   .
\ee
Since a  signal was not found so far,  they place an upper bound. The most stringent one comes from
the KamLAND-ZEN experiment: $m_{ee}  < 165\,{\rm meV} \, (90\%\,{\rm C.L.})$ \cite{kamlandzen}.  
  
Finally, cosmological observations place a stringent upper bound, 
$\sum_i m_i < 0.15\,(0.17)\,{\rm eV}$ ($95\%$ C.L.) \cite{hannestad}, on the sum of neutrino masses
for normally (inverted) ordered neutrino masses, translating into an upper bound 
\be\label{cosmoupperbound}
m_{1'} < 42\,{\rm meV} \;\;\;\;  (95\% \;\; {\rm C.L.}) \,  ,
\ee
 on the lightest neutrino mass, valid both for normal and inverted ordering.

\subsection{Minimally extended standard model}

We described neutrino mixing assuming that neutrino are massive but without
specifying how they acquire a mass. For this, one needs a model of neutrino masses.
In the SM mass terms arise from the Yukawa couplings with the
Higgs field $\Phi =(\phi^+,\phi^0)^T$ after electroweak spontaneous symmetry breaking. 
We have seen that in the case of charged leptons, the Yukawa coupling term
Eq.~(\ref{yukawacl}) generates the mass term Eq.~(\ref{masschargedlep}). One can then simply think 
to augment the SM adding RH neutrinos and a Yukawa interaction term analogously to the charged leptons.
In this way, in a generic lepton flavour basis, 
the total Lagrangian would become ${\cal L} = {\cal L}^{SM}+{\cal L}_Y^{\nu}$, where
\be
-{\cal L}_Y^{\nu} = \overline{L_{i'}}\,h^{\nu}_{i'j'}\,\nu_{R j'}\, \widetilde{\Phi} \,  + {\rm h.c.}  \, ,
\ee
with the dual Higgs field defined as $\widetilde{\Phi} \equiv {\rm i}\,\sigma_2\,\Phi^{\star}$.
After electroweak spontaneous symmetry breaking, a neutrino Dirac mass matrix is generated by the Higgs vev $v$
\be
m^{\nu}_{D\, i'j'} =v\,h^{\nu}_{i'j'} \, 
\ee
so that the neutrino Dirac mass term in the Lagrangian can be written as
\be\label{Diracnumass}
-{\cal L}_{{\rm Dirac \, mass}}^{\nu} = \overline{\nu_{L i'}}\, m^\nu_{D \, i'j'}\,\nu_{R j'} + {\rm h.c.} \,  .
\ee
As for the charged leptons, this can be diagonalised by means of a bi-unitary transformation
(mathematically, it corresponds to its singular value decomposition)
\be\label{biunitary}
m_{D \,i'j'}^{\nu} = \left(V^{{\nu}\dagger}_L\, D_{m_D} \, U_R^{\nu}\right)_{i'J'} \,  ,
\ee
where $D_{m_D} \equiv {\rm diag}(m_{D1},m_{D2},m_{D3})$.
In this way one can switch from the generic primed basis of LH neutrino fields 
$\nu_{L i'}$ to the basis of LH mass eigenfields $\nu_{L i}$
with the unitary transformation $\nu_{L i}=V^{\nu}_{L \, i j'}\, \nu_{L j'}$
and, analogously, for the RH neutrino fields by means of
$\nu_{R I}=U^{\nu}_{R \, I J'}\, \nu_{R J'}$. 
Since the neutrino mass basis coincides with the neutrino Yukawa basis where $m^{\nu}_D$ is diagonal,
the neutrino masses are then simply given by $(m_1,m_2,m_3)=(m_{D1},m_{D2},m_{D3})$.
The leptonic mixing matrix relating the LH mass eigenfields to the LH weak eigenfields, considering 
also the transformation from the primed to the weak basis  Eq.~(\ref{primedtoweak}), 
is then very simply given by
$U = U^{{\ell}}_L\, \left.V^{\nu}_L\right.^{\dagger}$ or, in components, 
$U_{\a i} = U^{{\ell}}_{L \, \a j'} \, V^{\nu \star}_{L \, i j'}$.
If we compare this with Eq.~(\ref{ULnu}), we can make the identification $U_L^{\nu}=V^{\nu \dagger}_L$. 
If charged lepton and neutrino Dirac mass terms were diagonal in the same 
LH neutrino basis then $\left.U^{\nu}_L\right. =
\left.U^{\ell}_L\right.^{\dagger}$  and $U=I$ and there would be no neutrino mixing. 
In the minimally extended standard model the mixing is then arising from a mismatch between the charged lepton and neutrino Yukawa bases. 
It is also important to notice  that though flavour lepton numbers are now not conserved,
as observed in neutrino oscillations, the total lepton number is still conserved perturbatively. 

The attractive features of this model is clearly that it is minimal and the addition of RH neutrinos
can somehow appear as a straightforward extension of the SM treating neutrinos 
on the same ground as the other massive fermions. On the other hand, the RH components  of the Dirac neutrino fields, 
though they share the same mass with the LH components,  are sterile degrees of freedom since they
are weak interactions singlets. 
Moreover the model presents the following limitations that suggest the necessity of a further extension:
\begin{itemize}
\item Though neutrinos also have their RH components, 
one still requires a rather special {\em ad hoc} description for neutrino masses, specifically:
\begin{itemize}
\item[-] The lightness of neutrinos compared to all other massive fermions is the result of assuming a 
tiny Yukawa couplings compared to those of the other massive fermions. For example,
the largest one, explaining the atmospheric neutrino mass scale, $m_{\rm atm}\sim 0.1\,{\rm eV}$, 
would be $h^\nu_3 \sim 10^{-12}$. 
\item[-]  The leptonic mixing matrix is the analogue of the CKM matrix in the quark sector and, therefore,
the observed largeness of neutrino mixing angles, in contrast with the smallness of quark mixing angles, also has to be 
assumed without justification.   
\end{itemize}
\item The model does not provide a solution to the cosmological puzzles and in particular to the 
origin of matter in the universe, in particular:
\begin{itemize}
\item[-] It does not provide a candidate of cold dark matter.\footnote{RH neutrino degrees of freedom 
can potentially be produced from spin-flip processes 
and behave as unwanted extra radiation at recombination and hot dark matter afterwards. However, 
with the current upper bound on neutrino masses, spin-flip processes are suppressed and the produced abundance
of RH neutrinos would be completely negligible.}
\item[-] There is no mechanism of baryogenesis provided by the minimally extended standard model and
electroweak baryogenesis would fail as in the SM, since the Higgs sector is untouched.
\end{itemize}
\end{itemize}
In addition to these limitations there is also the important observation that, having added RH neutrinos, 
nothing forbids the introduction of a Majorana mass term for the RH neutrinos. 
A possibility would be to impose lepton number conservation, at least at the perturbative level but this
would require to go beyond the minimally extended standard model we discussed, a possibility 
recently pursued in \cite{degouvea}. The model also contains a candidate for dark matter but not a clear
solution to the matter-antimatter asymmetry puzzle. More generally, one should say that 
Dirac neutrinos  can be still motivated within models involving additional ingredients
compared to the minimally extended standard model circumventing at least some of the limitations we listed
but it should be clear that in this case the model would not be minimal any more. 
For example, ways to justify  the lightness of Dirac neutrinos are found within frameworks with large \cite{arkani} 
or warped \cite{neubert} extra-dimensions. Also a mechanism of leptogenesis with Dirac
neutrinos has been proposed \cite{lindner} but it still requires some 
external source, e.g., GUT baryogenesis, for the generation of an initial $B+L$ asymmetry so that it anwyay
requires an extension of the minimally extended SM.\footnote{The presence of RH neutrinos can act in a way that part of such initial $B-L$ asymmetry 
can survive sphaleron wash-out. In this way one can have successful baryogenesis models even with $B-L =0$.}

In the following we will consider models where neutrinos are Majorana particles since,
as it will be clear, they more easily (and successfully) address the problem of the origin
of matter in the universe.

\subsection{Minimal (type I) seesaw mechanism}

As we have seen, the mere introduction of RH neutrinos and a Dirac mass term does not provide a scenario for the origin of matter in the universe. 
However, if one abandones lepton number conservation at the classical level and even $B-L$ conservation,
introducing, in addition to the Yukawa coupling term, also a bare (not generated by radiative corrections) right-right Majorana mass term $M$, 
one has, in addition to the SM Lagrangian, the term
\be\label{Y+M}
- {\cal L}_{Y+M}^{\nu}=\overline{L_{i'}}\,h^{\nu}_{i'J'}\,\nu_{R J'}\, \widetilde{\Phi} \,   +
{1\over 2}\,\overline{\nu_{R \, I'}^{\,c}}\,M_{I'\!J'}\,\nu_{R \, J'} + {\rm h.c.} \, .
\ee
Now small Latin indexes label LH fields and run from  $1$ to $3$,
while capital Latin indexes label RH neutrinos and run from $1$ to $N$ and, as before,
the primed notation indicates that we are working in a generic flavour basis.  The number of RH
neutrinos $N$ should be regarded as a free parameter within this general phenomenological framework, 
since there are no model independent theoretical arguments that can determine it. 
In particular, since RH neutrinos are SM singlets and do not carry a gauge anomaly, 
anomaly cancellation requirement does not apply. 
The Yukawa coupling matrix is then in general a $(3,N)$ matrix while the (right-right) Majorana
mass matrix is a $(N,N)$ matrix and notice that, since RH neutrinos are SM singlets, 
$M$ can be also a singlet and the term is renormalizable.

The Yukawa interaction defines the quantum number of the RH  neutrinos.
Assigning $L=+1$ to all  $\nu_R$ fields, the Dirac mass term conserves the total
lepton number  though not the individual lepton numbers or flavours.
However, though RH neutrinos carry a global charge, they do not carry 
gauge quantum numbers (colour, weak isospin, hypercharge). This is why they
can also have a Majorana mass term.
However, this now breaks $L$ by two units: therefore, the introduction of a Majorana
mass term leads to lepton number non conservation at the classical level, with processes potentially
observable even at low energy such as $0\nu\b\b$ decay. It will also prove important to notice, 
when we will discuss leptogenesis in the next section, that it also clearly violates $B-L$. 

 After spontaneous electroweak symmetry breaking, the Yukawa coupling term 
 produces the usual Dirac neutrino mass matrix  $m^\nu_{D}=v\,h^\nu$. However, this time
 this does not coincide in general with the light neutrino mass matrix $m_{\nu}$,
 unless trivially $M=0$.
 Therefore, after spontaneous symmetry breaking, we have that the neutrino
mass term can be written in a generic flavour basis (we omit primed Latin indeces) as 
\be\label{typeILag}
-{\cal L}_m^{\nu}= \overline{\nu_L}\,m_D^\nu\, \nu_R + 
{1\over 2}\,\overline{\nu^{c}_R}\,M \, \nu_{R} + {\rm h.c.}  \,  .
\ee
Using $\overline{\nu_L}\,m_D\, \nu_R = \overline{\nu^{c}_R}\,m_D^T\,\nu_L^{c}$,
we can rewrite the total neutrino mass term  as
\be
-{\cal L}^{\nu}_{\rm m}= {1\over 2}\,
\left[(\overline{\nu_L},\overline{\nu_R^{c}})
\left(
\begin{array}{cc}
                0  & m_D  \\
               m_D^T &  M    \\
\end{array}\right)
\left(
\begin{array}{c}
       \nu_L^{c}  \\
       \nu_R  \\
\end{array}\right)
\right] + {\rm h.c.} \,  .
\ee
Introducing the $3+N$ RH component neutrino field $\left.n'\right._R^T \equiv (\nu_L^{c}, \nu_R)^T$, the 
total neutrino mass term can then be written  even in a  more compact way as
\be\label{totalmass}
-{\cal L}^{\nu}_{\rm m}= {1\over 2}\,\overline{\left. n'_R\right.^{\hspace{-1.5mm}c}}\,{\cal M}_{\nu}\,n'_R  + {\rm h.c}. \,  ,
\ee
showing that this is in the form of a Majorana mass term also considering that ${\cal M}_{\nu}$
is a symmetric $(3+N,3+N)$ matrix, explicitly
\be
{\cal M}_{\nu} \equiv
\left(\begin{array}{cc}
                0  & m_D  \\
               m_D^T &  M    \\
\end{array}\right)  \,  .
\ee
Therefore, it can be Takagi diagonalised  and the mass eigenstates 
will be, in general, Majorana states.\footnote{This result is valid in general, not just in the seesaw limit.} 
Notice of course that for $M=0$ one recovers the Dirac case.
The most interesting case is the  {\em seesaw limit}, for $M\gg m_D$.
One can find a  $(3+N,3+N)$ {\em unitary} matrix ${\cal U}$ diagonalising ${\cal M}$
so that, performing the transformation
\be
n'_R \ra n_R = {\cal U}\,n'_R \,  
\ee
with $n_R^T \equiv (\nu_{L}^{c}, N_{R})^T$, one has
\be\label{M6diag}
-{\cal L}^{\nu}_{\rm m}= {1\over 2}\,\overline{n_R^c}\,{\cal D}_{{\cal M}_\nu}\,n_R  + {\rm h.c.} \,   ,
\ee
where ${\cal D}_{{\cal M}_\nu} \equiv  {\rm diag}(m_1, m_2, m_3, M_1,\dots, M_N)$.
The diagonalisation of ${\cal M}_{\nu}$ can proceed in two steps.
One can first find a block-diagonilising matrix ${\cal U}_B$ such that 
 \be
{\cal M}_{\nu} \simeq {\cal U}_B^T\,{\cal D}^B_{{\cal M}_\nu} \, {\cal U}_B   \,   ,
\ee
where 
\be
{\cal D}^B_{{\cal M}_\nu} =\left(\begin{array}{cc}
                m_{\nu} &  0  \\
                      0      &  M    \\
\end{array}\right)   
\ee
is a block diagonal matrix and  $m_\nu$ is the $(3,3)$ light neutrino mass matrix.
The block diagonalising matrix ${\cal U}_B$ is found expanding  in  powers of 
$\xi \equiv m_D\,M^{-1}$ up to terms ${\cal O}(\xi^2)$  
\be\label{multigenU}
{\cal U}_B \simeq 
\left(\begin{array}{cc}
                1 -{1\over 2}\,\xi^{\star}\,\xi^T &  - \xi^{\star} \\
               \xi^T &    1 -{1\over 2}\,\xi^T \,\xi^{\star}  \\
\end{array}\right)   \,  ,
\ee
and at the same time one finds for the light neutrino mass matrix the seesaw formula \cite{seesaw}
\be\label{seesaw}
m_{\nu} = -m_D\,\mbox{\large ${1\over M}$}\,m_D^T \,  .
\ee
The lightness of ordinary neutrinos is now explained as an algebraic byproduct 
deriving from the interplay between the electroweak and the new Majorana mass scales. 
In this way the mass spectrum splits into a set of light neutrino masses and into a set of heavy neutrinos and correspondingly
in the new basis the  expression (\ref{M6diag}) for the Lagrangian gets decomposed  
into a light and a heavy mass term 
\be\label{massbasis}
-{\cal L}^{\nu + N}_{\rm m}= {1\over 2}\,\overline{\nu_L} \, m_{\nu} \, \nu_L^{\, c} 
+{1 \over 2} \, \overline{N_R^{\, c}} \, M \, N_{R} \, . 
\ee
As a second step, since the light neutrino mass matrix $m_{\nu}$ is a symmetric matrix, it can be further diagonalised by a
unitary matrix $U_L^{\nu}$ 
\footnote{As discussed in detail in footnote 3, mathematically this is called Takagi diagonalisation of a 
complex symmetric matrix, different from the usual eigenvalue decomposition.}
in a way that  
\be
 D_m = - U_L^{\nu \dagger} \, m_{\nu}   \, U_L^{\nu \star} \,  ,
\ee
where $D_m \equiv (m_1,m_2,m_3)$. In this way, using the seesaw formula, we can write
\be\label{seesawdiagonal}
{\rm diag}(m_1,m_2,m_3) =  U_L^{\nu \dagger}\,m_D\,\mbox{\large ${1\over M}$}\,m_D^T\, U_L^{\nu \star} \, .
\ee
The leptonic mixing matrix $U$ is then given by the product
\be
U = U^{{\ell}}_L\, U^{\nu}_L  \,  .
\ee
If one starts from the basis where the charged lepton mass matrix is diagonal (the weak basis)
then $U^{{\ell}}_L = I$ and in this case one simply has $U = U^{\nu}_L$, so that 
\be\label{seesawdiagonal}
{\rm diag}(m_1,m_2,m_3) =  U^{\dagger}\,m_D \, {1\over M} \,m_D^{T} \, U^{\star} \, .
\ee
As a final step, one can also Takagi diagonalise $M$ with a transformation of the RH neutrino fields
$\nu_{R J'} = U^\dagger_{R J' I} \nu_{R I}$, 
\be\label{MRRtransform}
M_{I' J'} = U^T_{R I' I} \, D_{M I J} \, U_{R J J'} \,   ,
\ee
where $D_M \equiv {\rm diag}(M_I, M_{II},\dots,M_N)$ and we denoted by unprimed capital Latin indexes the basis where the Majorana mass is diagonal coinciding with the basis of heavy neutrino mass eigenfields. 

The  correcting term  $\propto \xi^{\star}\,\xi^T $ in the diagonal entries in Eq.~(\ref{multigenU})
produces a deviation from non-unitarity of the matrix $U_L^{\nu}$, and consequently of the 
leptonic mixing matrix $U$. Therefore, within the type-I seesaw mechanism, one can potentially have 
non-unitarity effects in neutrino oscillations within  a standard three neutrino mixing approach. 
However, in high-scale seesaw models these effects are highly suppressed but they can show up
in seesaw models at the TeV scale \cite{morisi,Antusch:2014woa,Chrzaszcz:2019inj}.

It is then finally useful to write the light and heavy mass eigenfields, that are Majorana fields, 
as a linear combination of the LH and RH neutrinos:
\bea\label{massfields}
   \nu_{i}  & \equiv &  \nu_{i L} + \nu_{i L}^c =   \left[U^\dagger_{i\a}\nu_{L\a} + (U^\dagger_{i\a}\nu_{L\a})^c \right] -  
    \left[V^\nu_{L i j'}\,\xi^\star_{j' I'} \,\nu_{R I'} + (V^\nu_{L i j'}\,\xi^\star_{j' I'} \,\nu_{RI'})^c\right]  \,  , \\  \nonumber
       N_{I} & \equiv  &  N_{I R} + N_{I R}^c   =   \left[ U_{R I I'}\,\nu_{RI'} + (U_{R I I'}\,\nu_{RI'})^c \right] + 
         \left[U_{R I I'}\, \xi^T_{I'j'} \,\nu_{L j'} + (U_{R I I'}\, \xi^T_{I'j'} \,\nu_{L j'})^c \right]   \,  .
\eea
This clearly shows how the light neutrinos are dominantly LH neutrinos but also have a subdominant RH neutrino component 
and vice-versa heavy neutrinos are dominantly RH neutrinos but also have  a small subdominant LH neutrino component 
and the mixing is  described by $\xi$.  For high-scale seesaw models $\xi$ is too tiny and the mixing does not give 
any sizeable phenomenological effects. However, for low-scale seesaw models the mixing becomes larger and this has 
a host of  potentially testable physical applications.\footnote{For a general discussion on the phenomenology of RH
neutrinos see \cite{Drewes:2013gca}.}

First of all the heavy neutrinos are not any more fully sterile and they can couple to SM
particles and can be potentially detected in colliders. Since so far no signal was found, upper bounds on the mixing parameters
that describe this active-sterile neutrino mixing are placed. However, cosmologically, even small mixing angles can be effective
in producing a heavy neutrino that can be long-lived and play the role of dark matter particle. We will discuss this 
cosmological application in Section \ref{DW}.

\subsection{Two extreme limits of type-I seesaw models}

The seesaw mechanism by itself does not specify the masses of the RH neutrinos. 
Model independently, these have to be regarded as free parameters  and, in 
general, cannot be expressed in terms of the low energy neutrino parameters. 
This interesting possibility would then require some assumption on the neutrino Dirac mass matrix.
If one considers, as an illustrative example,  a one generation toy model, the seesaw formula
reduces simply to  $M = m_D^2 /m_{\nu}$. If one assumes $m_D \sim M_{EW} \sim 100\,{\rm GeV}$,
then one obtains $M \sim 10^{15}\,{\rm GeV}$ using $m_{\nu}\sim m_{\rm sol} \sim 10\,{\rm meV}$.
This can be regarded as an encouraging exercise since the light neutrino mass scales would be obtained
from the electroweak scale and a scale very close to the grand-unified scale. However, it is clearly not compelling
and lowering the neutrino Dirac mass scale below the electroweak scale one can lower the RH neutrino mass scale
by many orders of magnitude.\footnote{As we will see, cosmologically, it will be particularly interesting to consider 
$m_D \sim 1\,{\rm eV}$, $m_{\nu} \sim 10^{-4}\,{\rm eV}$ implying $M \sim 10\,{\rm keV}$, though clearly
in this case the neutrino Dirac neutrino mass, and therefore the related Yukawa coupling, 
is again much smaller than in the case of all other massive fermions.}

Accounting for a realistic three generation case, the indetermination on the RH neutrino mass spectrum 
is even more evident since the neutrino Dirac mass matrix contains  fifteen
parameters while, on the other hand, there are only nine low energy neutrino parameters in $m_{\nu}$ and 
we do not even test all of them considering that $0\nu\b\b$ decay experiments give information on just one quantity
involving two Majorana phases.

However, using the seesaw limit and the seesaw formula Eq.~(\ref{seesawdiagonal}), 
there are some interesting limits, corresponding to definite theoretical assumptions, 
where it is easy to extract simple relations expressing the heavy neutrino masses in terms of 
the low energy neutrino parameters. 

\subsubsection{Degenerate Dirac neutrino mass spectrum}

Let us for definiteness consider the most attractive case $N=3$, motivated in different models.
A very simple assumption is to assume  $m_D = \lambda \,I$, where $\lambda$ is an overall mass scale and 
$I$ is the identity matrix in flavour space.
This form is highly symmetric and invariant in any basis and emerges quite straightforwardly from 
symmetry flavour models  \cite{flavoursymmetrylep}.

In this case one can immediately invert the diagonalised see-saw formula Eq.~(\ref{seesawdiagonal}),
since the leptonic mixing matrix $U$ cancels out, finding for the RH neutrino masses
\be
M_1 = {\lambda^2 \over m_3} \,  , \;\;
M_2 = {\lambda^2 \over m_2}  \,  , \;\;
M_3 = {\lambda^2 \over m_1}  \,   ,
\ee
simply the inverse of the light neutrino masses. 
\begin{figure}\label{SO10inspiredspectrum}
\begin{center}
\psfig{file=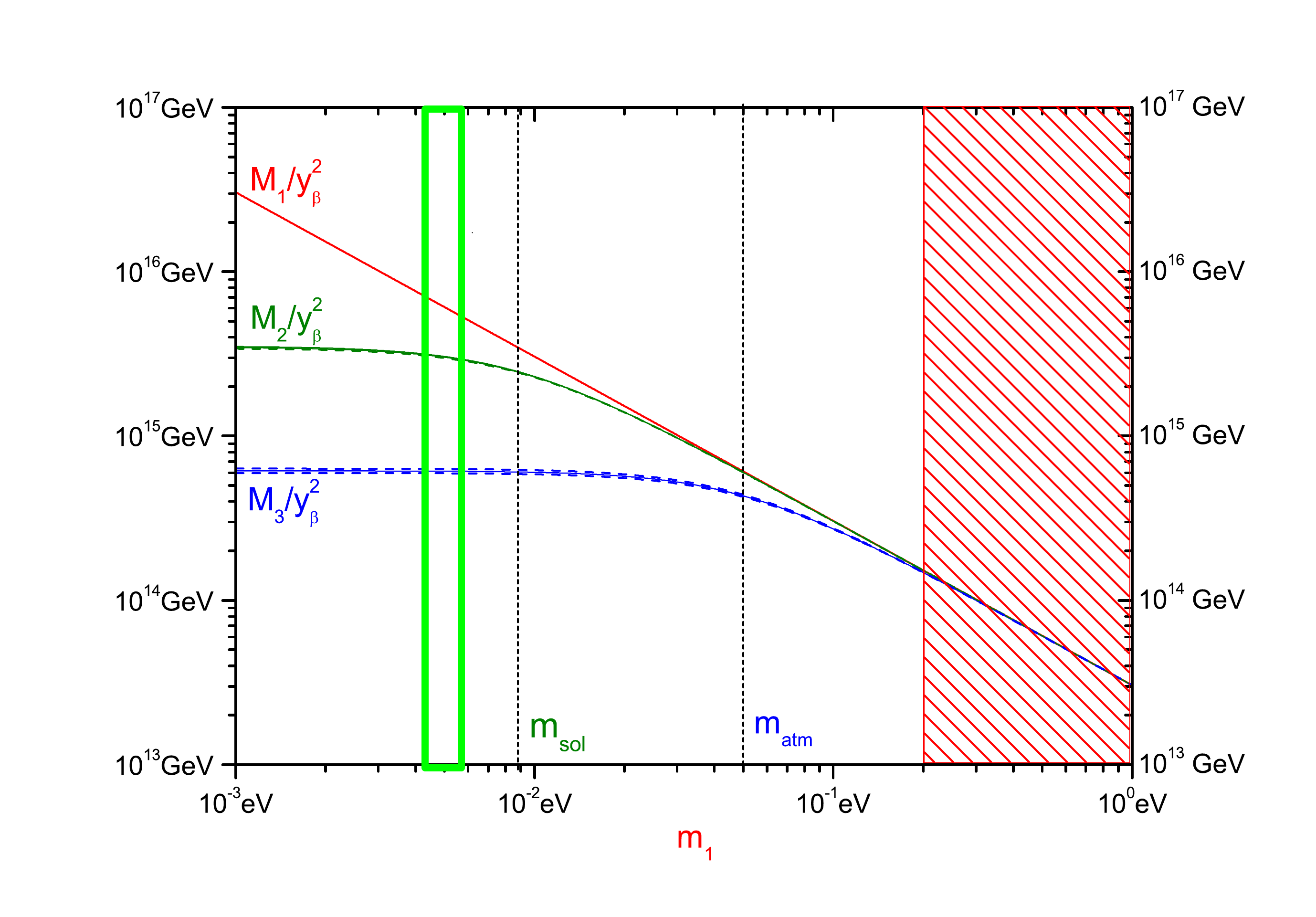,height=70mm,width=120mm}
\end{center} \vspace{-2mm}
\caption{Spectrum of RH neutrino masses versus $m_1$ in a discrete flavour symmetry model 
where $m_D = \lambda \, I$ in the case of normal ordering (from \cite{flavoursymmetrylep}). 
In the plot the parameter $y_{\beta} \equiv \lambda/v$. The vertical band indicates 
the allowed range of values for $m_1$ in the specific $A_4$ model considered in \cite{flavoursymmetrylep}. 
This fixes quite precisely the RH neutrino masses.}
\end{figure}
In Figure~6 \label{SO10inspiredspectrum} we show a plot of 
the three heavy neutrino masses versus the unknown lightest neutrino mass $m_1$ in the case
of normal ordering.

\subsubsection{$SO(10)$-inspired models}

Another important example where it is possible to express the RH neutrino mass spectrum in terms of the 
low energy neutrino parameters is the case of $SO(10)$ inspired models \cite{SO10inspired}.
In this case, in contrast with the previous limit case, one assumes that the neutrino 
Dirac mass spectrum is very hierarchical and more precisely the neutrino Dirac masses are
of the same order of magnitude as the up quark masses, explicitly 
\be\label{alphas}
m_{D1} = \a_1 \, m_{\rm up}  \,  , \;\; 
m_{D2} =\a_2 \, m_{\rm charm} \,  , \;\;
m_{D3} = \a_3 \,m_{\rm top}  \,  ,
\ee
where the three $\a_i$ are ${\cal O}$(1) factors. This implies approximately 
$m_{D1}:m_{D_2}:m_{D3} \sim m_{\rm up}: m_{\rm charm}: m_{\rm top} \sim 1 : 10^{2.5} : 10^5$.
With this assumption, one can derive the following expressions for the
three RH neutrino masses \cite{decrypting,SO10full}  
\be\label{MI}
M_1    \simeq   {m^2_{D1} \over |\widetilde{m}_{\nu 11}|} \, , \;\;
M_{2}  \simeq    {m^2_{D2} \over m_1 \, m_2 \, m_3 } \, {|\widetilde{m}_{\nu 11}| \over |(\widetilde{m}_{\nu}^{-1})_{33}|  } \,  ,  \;\;
M_{3}  \simeq   m^2_{D3}\,|(\widetilde{m}_{\nu}^{-1})_{33}|   \,  ,
\ee 
where $\widetilde{m}_{\nu}$ is the light neutrino mass in the neutrino Yukawa basis, where $m_D$ is diagonal.
Notice that these expressions are not valid in the vicinity of  $\widetilde{m}_{\nu 11} = 0$ or $(\widetilde{m}_{\nu}^{-1})_{33} =0$ or both.
In these cases the spectrum two or even all three RH neutrino masses become quasi-degenerate. The expressions we gave
work well as far as the RH neutrino masses are hierarchical. In the singular points there are so called {\em crossing level} solutions \cite{afs}.
However, it has been noticed that in these cases the light neutrino masses become highly fine-tuned and, therefore, they should be considered
as some very special cases. We will be back on this point when we discuss leptogenesis.

If we consider the neutrino Dirac mass matrix in the weak basis, where the charged lepton mass matrix is diagonal, 
and diagonalise it  by means of the bi-unitary transformation in Eq.~(\ref{biunitary}), we can write
($\a =e,\m,\t$; I=1,2,3; $k=1,2,3$)
\be\label{biunitary2}
m_{D \a I} = V^{\dagger}_{L \a k} \, D_{m_D k } \, U_{R k J}  \,   .
\ee
The unitary matrix $V_L$ now transforms the LH neutrino fields
from the weak to the neutrino Yukawa basis, where the neutrino Dirac mass matrix is diagonal.
It also transforms the  light neutrino mass matrix from the  weak to the Yukawa basis, explicitly
\be
\widetilde{m}_{\nu} = V_{L}\,m_{\nu}\,V_{L}^T \,  .
\ee 
The matrix $V_L$ has to be regarded as the analogous in the leptonic sector of the CKM matrix $V_{\rm CKM}$ in the quark sector, bringing from the basis where down quark mass matrix is diagonal  
to the basis where the up quark mass matrix is diagonal.  In $SO(10)$-inspired models
it is then assumed that the matrix $V_L \simeq V_{CKM}$. Since the largest mixing angle in $V_{\rm CKM}$ is the Cabibbo angle $\theta_{\rm c} \simeq 13^\circ$,
the dependence of the RH neutrino masses on $V_L$ is quite mild and with a good approximation
one can simply use $V_L \simeq I$.   
In Fig.~\ref{SO10inspiredspectrum} we show a typical example of RH neutrino mass spectrum
in $SO(10)$-inspired models for $V_L = I$ and $\a_1 = \a_2 = \a_3 = 1$. In particular, the three $M_I$ are plotted 
as a function of the lightest neutrino mass $m_1$ in the case of normal ordering.
\begin{figure}
\begin{center}
\psfig{file=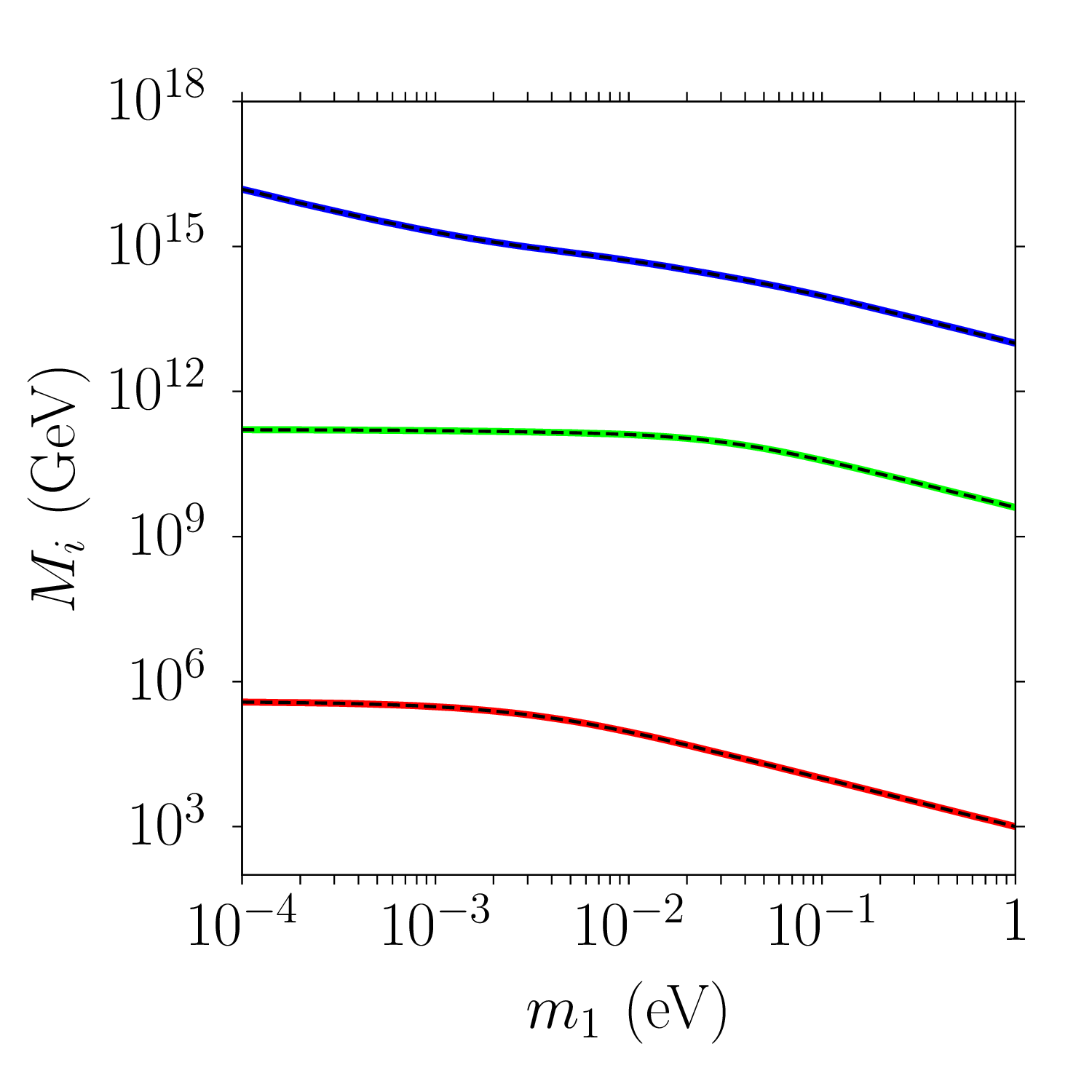,height=70mm,width=80mm}
\end{center} \vspace{-10mm}
\caption{Typical RH neutrino mass spectrum in $SO(10)$-inspired models for $V_L = I$, $\a_1 = \a_2 = \a_3 = 1$,
normal ordering, best fit values for the three mixing angles (see Eq.~(\ref{normal})) and $\rho=\sigma=\delta =0$.
The solid lines are the numerical solutions and the dashed lines are the analytical expression given in the text.
 }\label{SO10inspiredspectrum}
\end{figure}
One can notice the very strong hierarchical RH neutrino mass spectrum that is obtained. 
It is also useful, from the analytical expressions Eq.~(\ref{MI}) within the approximation $V_L = I$ and for $\a_1 = \a_2 = \a_3 = 1$, 
to derive the RH neutrino mass spectrum in the hierarchical limit for $m_1 \ra 0$
\be
M_1 \simeq {m^2_{D1} \over m_{\rm sol}\,s^2_{12}} \,   , \;\;
M_2 \simeq {m^2_{D2} \over m_{\rm atm}\,s^2_{23}} \,  ,  \;\;
M_3 \simeq  {m^2_{D3} \over m_1 } \,s^2_{23} \, s^2_{12} \,  ,
\ee
and in the quasi-degenerate limit for $m_1 \gg m_{\rm atm}$ (implying $m_1 \simeq m_2 \simeq m_3$) 
\bea
M_1 & \simeq & {m^2_{D1} \over m_1} \, \left| 1  + c^2_{12}(e^{2i\rho} - 1)\right|^{-1}  \,  , \;\;  \\ \nonumber
M_2 & \simeq & {m^2_{D2} \over m_1}  \,  {\left| 1  + c^2_{12}(e^{2i\rho} - 1)\right|} ,  \;\;         \\  \nonumber
M_3 & \simeq  & {m^2_{D3} \over m_1 } \, \left|1 + s^2_{23}(e^{2i\rho} -1) + c^2_{23} \right|  .
\eea
It can be noticed how in the hierarchical limit one has $M_3 \propto m_1^{-1}$, a necessary
consequence of the seesaw formula. In this limit the heaviest RH neutrino
decouples from the seesaw formula and effectively one obtains 
the two RH neutrino case with a considerable reduction in the number of seesaw
parameters, precisely from eighteen to eleven \cite{2RH}.  For example, for $m_1 = 10^{-4}\,{\rm eV}$
one can see from the plot that $M_1 \sim 10^6\,{\rm GeV}$,  $M_2 \sim  10^{11}\,{\rm GeV}$ and 
$M_3 \sim  10^{16}\,{\rm GeV}$.
We will see in the next section that such a RH neutrino mass spectrum will have important implication for 
the calculation of the matter-antimatter asymmetry within leptogenesis. Here we just notice that $M_3 \sim M_{\rm GUT}$ and this is indeed the most straightforward expectation for the RH neutrino mass spectrum from a grand-unified model  such as $SO(10)$.

\subsection{Beyond the minimal seesaw mechanism}

Some of the results we derived within the type-I seesaw mechanism can be  generalised within 
an effective theory approach pioneered by Weinberg \cite{weinberg}. 
 In particular, the lightness of the neutrino masses from the seesaw formula can be understood in a more general way.
 
We have seen that starting fro the type-I seesaw Lagrangian (\ref{typeILag}), after the diagonalisation of the total mass term, we could express the mass terms in the mass eigenstate basis finding that the light neutrino mass
term is given by a left-left Majorana mass term (see Eq. ~(\ref{massbasis}))
\be\label{mLL}
-{\cal L}_m^\nu = \overline{\nu_L} \, m_{\nu} \, \nu_L^{\, c} \,   .
\ee
This result is more general. First of all one could think to introduce directly  
a left-left Majorana mass term of this kind without introducing RH neutrinos and a Dirac mass term. This would seem
a very economical way not requiring apparently any extension of the SM. 
However, things are not that simple. 
First, in that case  $m_{\nu}$ would have still to be fine-tuned to reproduce the tiny neutrino masses.  
Even more importantly, the Lagrangian has to be not only Lorentz invariant but, within the SM, also 
$SU(2)_L\times U(1)_Y$ invariant not to spoil the SM renormalisability and the only way 
for the mass term in Eq.~(\ref{mLL}) to be gauge invariant is to have $m_{\nu}$ transforming 
as a $SU(2)_L$ triplet with $Y=2$.\footnote{A neutrino LH chiral field $\nu_L$ 
has isospin third component $I_3=1/2$ and hypercharge 
$Y = -1$ (see for example Table 3.5 in \cite{Giunti:2007ry}). This implies that 
$\overline{\nu_{L}}\,\nu_L^c $ has $I_3 =1$ and $Y=-2$.} However, the SM does not contain any weak isospin triplet
with $Y=2$. For this reason it is not possible to build, within the SM, a renormalizable Lagrangian term 
which can generate a left-left Majorana neutrino mass.
Therefore, in order to have a gauge invariant term without
spoiling renormalizability, one necessarily needs to extend the SM introducing new fields. 
We have seen the solution provided by the type-I seesaw mechanism.  Another 
 possibility is to introduce an isotriplet Higgs field that couples to the lepton doublets through Yukawa couplings $h_{\Delta}$. 
 After spontaneous symmetry breaking such a Yukawa term would indeed generate a left-left Majorana mass term.

However, there is a more general way, based on effective theory, 
to generate $m_{\nu}$ and  also elegantly predicting
the lightness of neutrino masses from a generalised seesaw formula.
The left-left Majorana mass term (\ref{mLL}) can be indeed regarded as generated
by  the five dimensional (lepton number violating) Weinberg operator \cite{weinberg}, 
\be\label{weinberg}
{\cal L}_5 = {g_{\a\b} \over \L} \, 
(\overline{L^c_{\a}} \,\widetilde{\Phi}^{\star})\, (\widetilde{\Phi}^{\dagger} \, L_{\b}) + {\rm h.c.} \,  , 
\ee
where $g_{\a\b}$ is a symmetric dimensionless coupling constant matrix,   and $\L$ is the  energy scale 
of new physics (e.g., the mass of a new super heavy particle).  This is the
lowest dimensional operator one can construct with the SM fields respecting SM symmetries.
After spontaneous symmetry breaking this operator does indeed generate a left-left neutrino Majorana mass
\be\label{generalseesaw}
m_{\nu\a\b} = {g_{\a\b}\,v^2 \over \L} \propto {m_D^2 \over \L}\,  .
\ee
This is quite interesting since if $\L \gg m_D$ one can naturally suppress the neutrino masses at the level we measure, 
and in this way the lightness of neutrino mass is associated to the existence of new physics at very high energy scale: 
Eq.~(\ref{generalseesaw}) can then be regarded as a generalised seesaw formula.
Since the Weinberg operator  is non renormalizable, it has to be regarded as an 
effective operator generated at energies much below the scale of new physics $\L$.
There are only three ways  to generate the Weinberg operator at tree level \cite{hirschwinter}:
\begin{enumerate}
\item Type-I seesaw mechanism: with a $SU(2)$ singlet fermion as a mediator \cite{seesaw};
\item Type-II seesaw mechanism: with a triplet scalar as a mediator \cite{typeII};
\item Type-III seesaw mechanism: with a triplet fermion as a mediator \cite{typeIII}.
\end{enumerate}

If one assumes the couplings $g_{\a\b} \sim 1$, then necessarily $\L \sim M_{\rm GUT}$ in order to
reproduce the solar and atmospheric neutrino mass scales. This is obviously quite an attractive case. The drawback
is that such high scale of new physics, and corresponding high-scale seesaw models, 
escapes direct detection at colliders. In the last years there was intense activity in considering 
alternative possibilities where neutrino masses might be generated by new physics at the $\L \sim {\rm TeV}$ scale or even below, 
directly accessible at colliders such as the LHC. In this case there are three main strategies:
\begin{itemize}
\item[(i)] Low-scale seesaw models: couplings $g_{\a\b}$ are very small; their suppression is then usually justified 
in terms of small lepton number violation (e.g., linear and inverse seesaw mechanisms, see for example \cite{morisi})
or it is not justified, small couplings are just imposed by hand;
\item[(ii)] Neutrino masses are generated radiatively and their suppression is guaranteed by a combination of
loop integrals and electroweak scale masses  entering the diagrams (see \cite{Cai:2017jrq} for a
recent review).
\item[(iii)] Neutrino masses from the Weinberg operator are forbidden ($g_{\a\b}=0$) but they
are generated by higher dimensional operators.\footnote{This case has common features
and overlap with (ii)  \cite{Cai:2017jrq}.}
\end{itemize}
We have then a picture with a very broad variety of possibilities. We will now discuss how the 
necessity to explain the origin of matter in the universe can shed light on which of these options has more
attractive features and can be regarded as more successfully. Here we just want to comment that low scale seesaw models are certainly very attractive from a phenomenological point of view but on the other hand the null results from searches of new physics at colliders and with many other different experimental strategies so far 
make these models more and more constrained experimentally. 
As we will see, cosmology will draw us to similar conclusions. 

\section{Leptogenesis}

Neutrino physics offers an elegant solution to the problem of the matter-antimatter asymmetry of the universe. 
Moreover, among the long list of baryogenesis models, has the clear advantage that relies on the 
only new physics that has been experimentally established: neutrino masses and mixing. Since the original proposal \cite{fy}, many variants have been proposed. Here we mainly focus on the minimal model of leptogenesis since it remains the most attractive scenario,
supported, as we will point out, by a few simple experimental facts and also because non-minimal extensions, usually low-energy scale scenarios, are not supported by the lack of new physics in experiments (beyond neutrino oscillations) so far.
Moreover in the minimal scenario the measured value of the baryon-to-photon number ratio
points to a RH neutrino mass scale, $M \sim 10^{10-11}\,{\rm GeV}$, that emerges quite naturally
in models.  

 \subsection{The minimal scenario of leptogenesis}

The minimal scenario of leptogenesis \footnote{For dedicated reviews see \cite{minimal,Xing:2020ald}.}
is based on two assumptions, one on  the particle physics side and one on the cosmological side:
\begin{itemize}
\item[i)] Type-I seesaw extension of the SM;
\item[ii)] Thermal production of the heavy neutrinos (thermal leptogenesis).
\end{itemize}
As discussed, the explanation of the lightness of neutrinos is well addressed by the type-I seesaw mechanism that predicts the existence
of heavy, dominantly RH, neutrinos. High-scale seesaw models are difficult to test since the parameters associated to the heavy neutrinos, like their masses,
escape the experimental investigation. However, the heavy neutrinos, though typically decaying prior to the big bang nucleosynthesis onset without affecting standard cosmology, can leave a crucial imprint at present: the matter-antimatter asymmetry of the universe in the form of the observed baryon asymmetry. 

Leptogenesis is another important example, like GUT baryogenesis, of
baryogenesis from heavy particle decays.\footnote{We can then here specialise the general formalism and results discussed in Section 2.4.}
Their decay can indeed violate $C\!P$ since in general the rate of decays into leptons
$\Gamma(N_I \ra {L}_I + \phi^\dagger)$ can be different from the decay rate into anti-leptons, $\Gamma(N_I \ra \bar{L}_I + \phi)$,
in  way that each $N_I$ decay will produce, on average, a $B-L$ asymmetry in the form of a lepton asymmetry given
by the  the total $C\!P$ asymmetry, defined as
\be
\ve_I \equiv - {\Gamma(N_I \ra {L}_I + \phi^\dagger) - \Gamma(N_I \ra \bar{L}_I + \phi) \over 
\Gamma(N_I \ra {L}_I + \phi^\dagger) + \Gamma(N_I \ra \bar{L}_I + \phi) } \,   .
\ee
Compared with the general definition Eq.~(\ref{epsX}),  we now have ${\rm sign}(\D_{B-L}) = -1$.
As we will see, when flavour effects are included there is an additional source of $C\!P$ violation.
For the time being we neglect flavour effects, we will see how these can change the results.

Even though the $B-L$ asymmetry is initially injected in the form of a lepton asymmetry, if the reheat temperature of the universe, the initial temperature at the onset of the radiation dominated regime after the end of inflation, is above the 
out-of-equilibrium temperature of sphalerons,  this will be redistributed among all SM species and in particular also among
quarks, so that a baryon asymmetry is also generated. Chemical
equilibrium \cite{harvey} ultimately enforces $L \simeq (a_{\rm sph}-1)\,(B-L)$ and $B\simeq a_{\rm sph}\, (B-L)$, with $a_{\rm sph} = 28/79 \simeq 1/3$. 

Notice that also leptogenesis, likewise electroweak baryogenesis, crucially relies on
sphalerons to generate a baryon asymmetry. In both cases the initial injected asymmetry 
is not a baryon asymmetry but
a chiral asymmetry in the case of electroweak baryogenesis and a lepton asymmetry in the case of leptogenesis.\footnote{For some
reason this aspect is always emphasized for leptogenesis but not for electroweak baryogenesis that after all could be also called
`chiralogenesis'.} The important difference is that in the case of leptogenesis $B-L$ is violated by the decays of the RH neutrinos, while it is conserved
in electroweak baryogenesis. In this respect leptogenesis is directly relying on new physics while in the case of  electroweak baryogenesis this could potentially be viable even in the SM. However, as we discussed,  electroweak baryogenesis fails to a quantitative level because 
the experimental results on the Higgs mass and $C\!P$ violation in quarks   show that departure from thermal equilibrium and $C\!P$ violation are not enough and for this reason new physics is needed. 

The final $B-L$ abundance in a portion of comoving volume will be then given by the sum of the contributions from all RH neutrino species
\be\label{NBmLf}
N_{B-L}^{\rm f} = \sum_I^{1,N} \ve_I\,\kappa^{\rm f}_I  \,   ,
\ee
where the $\kappa^{\rm f}_I$'s are the final {\em efficiency factors} taking into account the wash-out from
inverse processes, the efficiency in the RH neutrino thermal production and also contain geometrical factors depending 
on the flavour structure.  For definiteness we will consider the case $N=3$, also because this is the most attractive case
emerging in different models.  
 
From the final $B-L$ asymmetry one can then calculate the baryon-to-photon ratio at present 
\be
\eta_{B 0}^{\rm lep} = a_{\rm sph} \, {N_{B-L}^{\rm f} \over N_{\g 0}} \,  .
\ee
If we normalise the abundances in a portion of comoving volume containing one RH neutrino in ultrarelativistic thermal equilibrium,
prior to the onset of leptogenesis, at some fiducial initial time $t_{\rm i}$ 
such that $T_{\rm i}\gg M_I$, where $T_{\rm i}\equiv T(t_{\rm i})$ and $M_I$ is the mass of the heaviest RH neutrino contributing to the
final asymmetry,  then the number of photons at this stage is simply $N^{\rm i}_{\g} = 4/3$. Assuming 
a SM particle content  after RH neutrino decays and a standard cosmological expansion, implying entropy conservation,
then one has $N_{\g 0} = N_{\g}^{\rm i} \, g_{\rm S}^{\rm i}/g_{S 0}$, where the number of (entropy) degrees of freedom at 
$T\gg M_I$ is given by the number of SM degrees of freedom plus the contribution from the RH neutrino degrees of freedom, simply 
$g^{\rm i}_S = g_{\rm SM} = 427/4+7/4 = 108.5$ and the number of entropy degrees of freedom 
at the present time is $g_{S0}=43/11$ \cite{book}. In this way, one finds $N_{\g 0}= 4774/129 \simeq 37$. 
Finally, with the normalisation of abundances discussed and adopted in Section 2.4, one finds 
\be\label{etaBlep}
\eta_{B0}^{\rm lep} = {28 \over 79} \times {129\over 4774}\,N_{B-L}^{\rm f} \simeq 0.96\times 10^{-2}\,N_{B-L}^{\rm f} \,  .
\ee
This is the quantity predicted by leptogenesis that has to match the  measured value in Eq.~(\ref{etaB0}).
Clearly one needs to calculate $N_{B-L}^{\rm f}$ and from Eq.~(\ref{NBmLf}) one can see
that for this one needs to calculate the total $C\!P$ asymmetries $\ve_I$ of the RH neutrinos  and their efficiency factors. The first depend just on the particle physics and in particular on the neutrino Dirac mass matrix and, very interestingly,
on the RH neutrino mass spectrum. The efficiency factors are a mix of particle physics and cosmology, they clearly require
a solution of kinetic equations. 

A perturbative calculation from the interference of tree level with one
loop self-energy and vertex diagrams gives for the total $C\!P$ asymmetries 
\cite{flanz,Covi:1996wh,buchplumi1,Pilaftsis:1997jf}
\be\label{CPas}
\ve_I =\, {3\over 16\pi}\, \sum_{J\neq I}\,{{\rm
Im}\,\left[(h^{\dagger}\,h)^2_{IJ}\right] \over
(h^{\dagger}\,h)_{II}} \,{\xi(x_J/x_I)\over \sqrt{x_J/x_I}}\, ,
\ee
where we introduced
\be\label{xi}
\xi(x)= {2\over 3}\,x\,
\left[(1+x)\,\ln\left({1+x\over x}\right)-{2-x\over 1-x}\right] \, ,
\ee
and defined $x_J \equiv M_J^2/M_1^2$. An important property of total $C\!P$ asymmetries is that, depending
on the combination $h^\dagger\,h$, the dependence on the leptonic mixing matrix cancels out. We will see that this  will have an important consequence. 

The {\em total decay rates} are given by
\be
\Gamma^{\rm D}_{I}\equiv \Gamma(N_I \ra {L}_I + \phi^\dagger) + \Gamma(N_I \ra \bar{L}_I + \phi) \,   
\ee
and at zero temperature they correspond to decay widths given by
\be\label{decaywidth}
\widetilde{\G}^{\rm D}_{I} \equiv \Gamma^{\rm D}_{I}(T=0) = 
{M_I \over 8\,\pi}\,(h^\dagger\,h)_{II}  \,  .
\ee
The total decay rate can then be obtained multiplying the decay width by 
the averaged dilation factor,
\be\label{decayrate}
\Gamma^{\rm D}_{I} = \widetilde{\G}^{\rm D}_{I} \, 
\left\langle {1 \over \gamma_I} \right\rangle  \,  .
\ee
The averaged dilation factor is given by Eq.~(\ref{dilation}), where now
$z$ has to be replaced by $z_I \equiv M_I /T$.  The decay factor, introduced
in the general case in 2.4, can now be written as 
$D_I \equiv  \Gamma^{\rm D}_I/(H\,z_I)$. The decay term can be conveniently expressed 
in terms of the decay parameter (see Eq.~(\ref{Kdef})) as 
\be
D_I(z_I) = K_I \,  z_I \, \left\langle {1 \over \gamma_I} \right\rangle \stackrel{z\ra 0}{\longrightarrow} {1\over 2}\,K_I \, z_I^2 \,   .
\ee

The RH neutrino decay parameters are interestingly related to the light neutrino masses and can be recast as
\be\label{effectivenumass}
K_I = {\widetilde{m}_I \over m_\star} \,  , \hspace{5mm}\mbox{\rm with} 
\hspace{5mm} \widetilde{m}_I \equiv {v^2\,(h^\dagger\,h)_{II} \over M_I} 
\ee
where the $\widetilde{m}_I$'s are the {\em effective neutrino masses} \cite{plumacher} and
\be
m_{\star} \equiv  {16\,\pi^{5/2} \, \sqrt{g^{SM}_{\star}}   \over 3 \sqrt{5}} \, { v^2 \over   M_{\rm P}}  \simeq 1.07 \, {\rm meV}
\ee
is the {\em equilibrium neutrino mass} \cite{orloff}. 
The seesaw formula, combined with neutrino oscillation experiment results, typically favours $\widetilde{m}_I$ in the range $m_{\rm sol}$--$m_{\rm atm} \sim (10$--$50)\,{\rm meV}$, corresponding to
$K_I \sim 10$--$50$. More specifically, one cannot have more than one decay parameter 
$K_I \ll  m_{\rm sol}/m_{\star} \sim 10$. This point plays an important role both in leptogenesis and also
when we will discuss the possibility that one heavy seesaw neutrino could play the role of dark matter.  

As discussed in Section 4, the type-I seesaw extension of the SM introduces eleven new parameters in the case of two RH neutrinos
and eighteen new parameters in the case of three RH neutrinos. Even in the case that  
$0\nu\b\b$ decay experiments find a positive signal and measure $m_{ee}$ and cosmological observations measure the lightest neutrino mass $m_{1'}$, this will determine 
only  eight parameters in the low energy neutrino mass matrix. 
The RH neutrino masses and the complex mixing angles in the orthogonal matrix will still escape 
the experimental information from low energy neutrino experiments. Leptogenesis  can then be regarded as a special very high energy neutrino experiment that took place
once in the early universe and that is sensitive also to high energy neutrino parameters. However, in the end it just introduces one additional experimental
constraint, the successful leptogenesis condition $\eta_{B0}^{\rm lep} = \eta_{B0}^{\rm exp}$ that, in general, is still not sufficient to over constraint the seesaw parameter space
making model independent predictions, even in the case of two RH neutrinos ($N=2$). 

To this extent it is necessary to reduce the number of parameters, increasing the predictive power,
and/or to look for additional phenomenological tests. 
In the first case one can either introduce some theoretical condition within a model or a class of models or make some approximations and reasonable assumptions
in the calculation of the asymmetry that can lead to some experimental prediction on low energy neutrino parameters. In the second case one can
either lower the scale of leptogenesis, with or without going beyond a type-I seesaw mechanism, and/or try to reproduce some other phenomenological observation. 
The most remarkable ohenomenological observation one can try to combine with the matter-antimatter
asymmetry is certainly the dark matter of the universe. In this way one would obtain a unified model of neutrino masses, leptogenesis and dark matter, a model that would solve the problem of the origin of matter in the universe. 
In the following, we will discuss all these different strategies 
but of course the possibility to explain also dark matter will be our main focus.
We start from the simplest and most traditional approach that leads to a prediction on the absolute 
neutrino mass scale from leptogenesis, based on a set of assumptions and simplifications
in the calculation of the asymmetry: so-called  {\em vanilla leptogenesis} \cite{minimal}. 

\subsection{Vanilla leptogenesis}

\begin{figure}
\begin{center}
\psfig{file=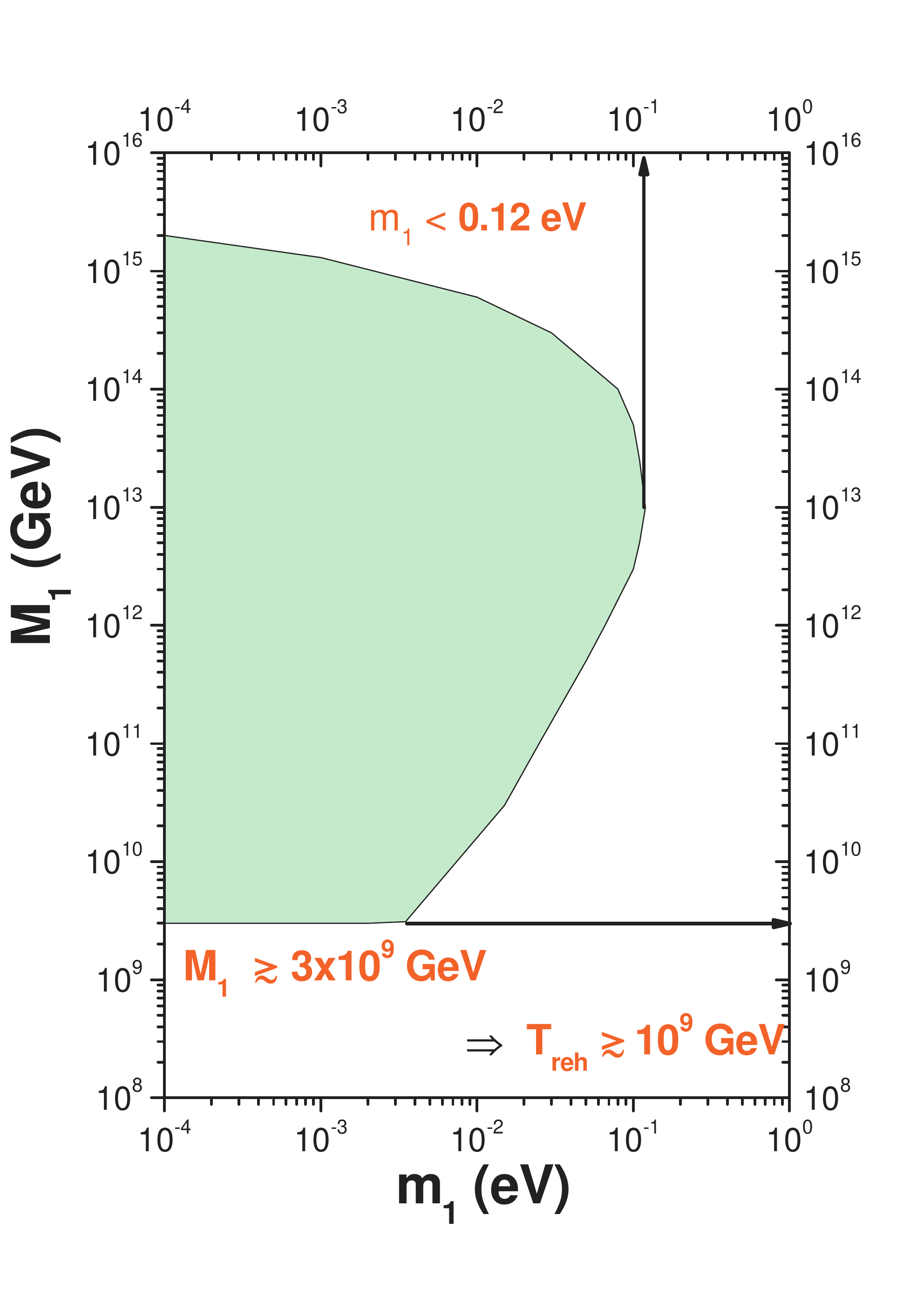,height=104mm,width=88mm}
\end{center} \vspace{-10mm}
\caption{Allowed region in the plane $M_1$ versus $m_{1}$. The lower bound on the lightest 
RH neutrino mass implies a lower bound on the reheat temperature.}\label{vanilla}
\end{figure}

Vanilla leptogenesis is the simplest scenario of leptogenesis and it is an important example of how, just imposing the successful leptogenesis condition, one can obtain predictions on physical observables, specifically an upper bound on the absolute neutrino mass scale.  It relies on the following set of assumptions:
\begin{itemize}
\item[(i)] The flavour composition of leptons and anti-leptons produced in the decays and inverse decays of RH neutrinos does not 
influence the value of the final asymmetry;
\item[(ii)] The RH neutrino mass spectrum is hierarchical and more precisely $M_2 \gtrsim 2\,M_1$;
\item[(iii)] The lightest RH neutrino decay parameter $K_1 \gg 1$;
\item[(iv)] There are no fine-tuned cancellation in the seesaw formula.
\end{itemize}
The first three assumptions imply that the 
final asymmetry is dominated by the lightest RH neutrino ($N_1$-leptogenesis), so that the
expression (\ref{etaBlep}) gets specialised into
\be
\eta_{B0}^{\rm lep} \simeq 0.96 \times 10^{-2}\,\ve_1 \, \kappa_1^{\rm fin}(K_1,m_{1'}) \,  ,
\ee
where notice that the final efficiency factor for the lightest RH neutrino decays depends just on the lightest RH neutrino
decay parameter and, quite interestingly, on the absolute neutrino mass scale expressed 
in terms of the lightest neutrino mass $m_{1'}$
(either $m_1$ for normal ordering or $m_3$ for inverted ordering).  
The lightest RH neutrino efficiency factor
can be calculated solving a simple set of just two rate equations, one for the $N_1$ abundance and one for the 
$B-L$ abundance \cite{pedestrians}
\begin{eqnarray}
{dN_{N_1}\over dz_1} & = & -D_1\,(N_{N_1}-N_{N_1}^{\rm eq}) \;, \label{lg1} \\
{dN_{B-L}\over dz_1} & = & -\ve_1\,D_1\,(N_{N_1}-N_{N_1}^{\rm eq})-W_1^{{\rm ID}}\,N_{B-L} \;,\label{lg2}
\end{eqnarray} 
where, analogously to the decay term $D_1$, we defined the inverse decay wash-out term as
\be
W^{{\rm ID}}_1 \equiv {1 \over 2} \, {\Gamma^{{\rm ID}}_1 \over H(z_1)\,z_1} = {1\over 4} \,K_1 \, z_1^3 \, {\cal K}_1(z_1) \,   .
\ee
In the last expression we used that the inverse decay rate is related to the decay rate (see Eq.~(\ref{decayrate})) simply by
\be
\Gamma^{{\rm ID}}_1(z_1) = \Gamma^{{\rm D}}_1(z_1) \, {N_{N_1}^{\rm eq}(z_1) \over N_{\ell}^{\rm eq}(z_1)} = 
{1\over 2}\, \Gamma^{{\rm D}}_1 \, z_1^2 \, {\cal K}_2(z_1) \,  .
\ee
A simple solution for the final efficiency factor can be written in the form \cite{pedestrians,predictions}
\be
\k_1^{\rm f}(K_1,m_1') = \k_1^{\rm f}(K_1) \, \exp\left[-{{\o\over z_B}\,{M_1 \over 10^{10}\,{\rm GeV}}\,
\left({\overline{m}^2 \over {\rm eV^2}}\right)} \right]  \,   ,
\ee
where 
\be
\k_1^{\rm f}(K_1) = {2 \over z_B(K_1)\,K_1}\, \left(1 -e^{-{1\over 2}\,z_B(K_1)\, K_1}\right) \,   ,
\ee
\be
z_B(K_1) \simeq 2 + 4\,K_1^{0.13}\,e^{-{2.5 \over K_1}} \,  
\ee
and 
\be
\overline{m}^2 \equiv \sum_i  m_i^2  \,   .
\ee 
The exponential is due to wash-out from $\D L =2$ scatterings and the constant $\o$  is given by
\be
\o={9\sqrt{5}\,M_{\rm P}\,10^{-8}\,{\rm GeV}^3 \over
4\pi^{9/2}\,g_l\,\sqrt{g_{\star}}\,v^4}\simeq 0.186 \,   .
\ee
 This value implies that for $M_1 \lesssim 10^{13}\,{\rm GeV}$ the exponent is negligible
 but for $M_1 \gtrsim 10^{13}\,{\rm GeV}$ it becomes large, inducing an exponential suppression.
Finally, using the assumption (iv), one can derive an upper bound on the  total $C\! P$ asymmetry \cite{di}
\be
\ve_1 \lesssim \overline{\ve}(M_1) \equiv {3 \over 16\,\pi}\, {M_1 \, m_{\rm atm} \over v^2} \simeq 10^{-6}\,
\left({M_1 \over 10^{10}\,{\rm GeV}} \right)  \, {m_{\rm atm} \over m_1 + m_3} \,  .
\ee
In this way one has an upper bound on the asymmetry depending just on three parameters: $K_1$, $m_{1'}$, $M_1$, explicitly:
\be\label{etaB0ub}
\eta_{B0}^{\rm lep} \lesssim 0.96 \times 10^{-2}\,\overline{\ve}_1(M_1) \, \kappa_1^{\rm fin}(K_1,m_{1'}) \,  .
\ee
Imposing the successful leptogenesis condition,  from the upper bound (\ref{etaB0ub})
one finds the allowed region shown in Fig.~\ref{vanilla} in the plane $M_1$ versus $m_{1}$.\footnote{The figure is obtained for normal ordering but a very similar result is also found for inverted ordering.}
One can first of all notice the lower bound on the lightest RH neutrino mass $M_1 \gtrsim 10^{9}\,{\rm GeV}$ \cite{di,cmb}, coming from the upper bound on $\ve_1$ and also, interestingly, the existence of an upper bound on the 
neutrino masses $m_1 \lesssim 0.1 \, {\rm eV}$ \cite{pedestrians}, 
now confirmed by the cosmological upper bound
obtained within the $\Lambda$CDM model. 
This nice confirmation is certainly an encouraging result supporting vanilla leptogenesis
but it also  provide a clear simple example of how, under certain assumptions, 
one can get predictions on experimental quantities. 

Another encouraging feature of vanilla leptogenesis 
is that the assumption of strong washout, $K_1 \gg 1$, is 
supported by neutrino oscillation experiments. As we said, 
the seesaw formula typically favours 
$\widetilde{m}_1$ in the range $m_{\rm sol}$--$m_{\rm atm} \sim (10$--$50)\,{\rm meV}$ corresponding to
$K_1 \sim 10$--$50$.\footnote{In this range a simple and very good analytical fit for the final efficiency factor is given by: 
\be\label{kappaflep}
\kappa_1^{\rm f}(K_1) \simeq {0.5 \over K_1^{1.2}}\,  .
\ee}
For this reason having $K_1 \lesssim 1$ is a very special situation.\footnote{Recently it was shown that for 
randomly uniformly (in flavour space) generated  seesaw models respecting neutrino oscillation experimental results, 
the probability to have $K_1\lesssim 1 $ is less than $0.1\%$ \cite{seesawmotion}.} This range of values of $K_1$ corresponds to
efficiency factors $\k^{\rm f} \sim 10^{-3}$--$10^{-2}$ that still allows successful leptogenesis if $M_1$ is above the  lower bound. 
On the other hand, for such values of $K_1$ the efficiency factor is independent of the initial $N_1$ abundance and moreover any pre-existing asymmetry is very efficiently washed-out, since its final value is exponentially suppressed 
\be\label{preexisting}
N^{\rm p,f}_{B-L} = N^{\rm p,i}_{B-L} \, e^{-{3\pi \over 8}\, K_1} \,  . 
\ee
Therefore, neutrino oscillation experimental results and the upper bound on neutrino masses from cosmology greatly 
support the vanilla leptogenesis scenario. However, there is an unpleasant point in these results: 
the lower bound on $M_1$ rules out $SO(10)$-inspired models we discussed in the previous section 
where, as we have seen, one has $M_1 \sim 10^6 \, {\rm GeV}$ 
unless one considers special crossing level solutions \cite{afs}. However, these solutions 
imply a large fine-tuning in neutrino masses from the seesaw formula \cite{decrypting,SO10full}. We will see that the inclusion of flavour effects dramatically changes
this conclusion opening up a new interesting scenario of leptogenesis where the
asymmetry is produced from the next-to-lightest RH neutrinos ($N_2$-leptogenesis).

The existence of a lower bound on the lightest RH neutrino mass also translates into a lower bound on the reheat temperature
of the universe that is slightly more relaxed since the asymmetry is produced quite sharply around a temperature 
$T_B \sim M_1/z_B \sim M_1/(2$--$5)$.  The existence of such a lower bound on the reheat temperature can be a problem
in natural gravity mediated supersymmetric models because of the gravitino problem \cite{gravitino} whose solution typically requires
reheat temperatures below $10^9\,{\rm GeV}$. However, the current experimental status simply does not support natural supersymmetric models
and points to a situation where in any case some level of fine-tuning in the 
stabilization of the electroweak scale is unavoidable and the upper bound from the gravitino problem gets either relaxed or simply does not hold, in any case it is certainly less compelling than initially thought.

The existence of the lower bounds on $M_1$ and on the reheat temperature have motivated an intense research on how to evade them departing from the simple vanilla leptogenesis scenario. 
One of the motivation, as we mentioned was the gravitino problem, 
but in recent years the possibility to have successful leptogenesis at the TeV scale gained a lot of attention for two reasons. A first reason  is that the presence of RH neutrinos  tends to destabilise the electroweak scale unless some degree of fine-tuning is requested. Imposing that the fine-tuning is not higher than percent level, one obtains the upper bound \cite{vissani} $M_1 \lesssim 10^8 \, {\rm GeV}$.

At the same time,  the possibility to have leptogenesis at the TeV scale opens the opportunity to test 
the existence of RH neutrinos at colliders. Of course this second reason is justified from the first one, essentially naturalness. Since the LHC is placing severe lower bounds on the scale of new physics, this has, however, risen doubts on naturalness driven arguments and  in the last years these two motivations became less compelling. 

A first way to lower the scale of leptogenesis is {\em resonant leptogenesis} 
\cite{resonant,Dev:2017wwc,Klaric:2021cpi}.  In this case it just relies on relaxing the assumption of
a hierarchical RH neutrino spectrum since the $C\!P$ asymmetries get enhanced when the lightest and  
next-to-lightest RH neutrino, or all of them, get quasi-degenerate. 
However, barring fine-tuning in the type-I seesaw, Yukawa couplings get
necessarily too small at the TeV scale to make the RH neutrinos detectable. Therefore, resonant leptogenesis needs to work in combination
with some additional extension of the seesaw Lagrangian. One possibility is that RH neutrinos have extra gauge interactions and a typical example comes from gauging a $U_{B-L}(1)$ symmetry getting broken at low scale \cite{Blanchet:2009bu,Iso:2009nw}.
In this case one gets a rich phenomenology at colliders with $Z'$ detection. Currently stringent lower bounds have been placed at the LHC \cite{Atre:2009rg,Drewes:2013gca,Deppisch:2015qwa,Cai:2017mow}.
Another popular way to lower the scale of leptogenesis and at the same time to have chances to detect RH neutrinos at the LHC relies on going beyond a pure type-I seesaw mechanism, for example considering leptogenesis with both
type-3 and type-1 contributions to the seesaw formula \cite{Albright:2003xb}
or in recent years leptogenesis within linear and inverse seesaw mechanism has been also  explored \cite{inverselinear}.  In this case the one can justify large Yukawa couplings 
that, as we said, in a type-I seesaw mechanism would be highly fine-tuned.
Finally, another popular low-scale scenario of leptogenesis relies on a production of the asymmetry from RH neutrino mixing rather than from decays, the so-called ARS scenario \cite{ARS,Abada:2018oly,Drewes:2021nqr}.

\subsection{Flavour effects}

Within the minimal scenario defined in Section 5.1, the most important generalisation of the vanilla leptogenesis scenario 
comes from considering flavour effects \cite{nardi1,riottolosada}.\footnote{For a recent review on flavour effects see \cite{reviewflavour}. Before the the general discussion in \cite{nardi1,riottolosada}, 
flavour effects were specifically considered within resonant leptogenesis 
\cite{Pilaftsis:2004xx,Pilaftsis:2005rv} and $N_2$-leptogenesis  \cite{vives}.} 
The flavour composition of lepton in final states is neglected in vanilla leptogenesis.
However, in certain situations this can dramatically change the calculation of the final asymmetry.  
If the mass of the decaying RH neutrino is much above $10^{12}\,{\rm GeV}$, then active processes in the early universe 
are flavour blind. In this case  the unflavoured approximation and the simple Boltzmann equation Eq.~(\ref{lg2}) for the 
calculation of the asymmetry provides a fair description. However, for lighter RH neutrinos with 
masses  in the range 
$10^9\,{\rm GeV} \ll M_1 \ll 10^{12}\,{\rm GeV}$, Yukawa interactions of the lepton quantum states $|{\ell}_1\rangle$
produced in the $N_1$-decays with the RH tauons in the thermal bath, will measure, on average,  the tauon component 
before they inverse decay. In this way the inverse decay wash-out will now act independently on 
the tauon component  and the electron plus muon coherent component of the $|{\ell}_1\rangle$'s. 
One has then to track separately
the tauon asymmetry $\D_\tau \equiv B/3 - L_\tau$ and the electron plus muon asymmetry 
$\Delta_{e+\m} \equiv 2B/3 - L_e - L_\mu$. In this two fully flavoured regime 
the set of Boltzmann equations (\ref{lg1}) and (\ref{lg2}) is then replaced by 
\bea\label{flke}
{dN_{N_1}\over dz} & = & -D_1\,(N_{N_1}-N_{N_1}^{\rm eq}) \,  , \\ \nonumber
{dN_{\D_{e}}\over dz} & = & \ve_{1e}\,D_1\,(N_{N_1}-N_{N_1}^{\rm eq}) -N_{\D_{e}} \,p_{1 e}^{0}\,W_1^{\rm ID} \,  ,  \\ \nonumber
{dN_{\D_{\mu}}\over dz} & = & \ve_{1\mu}\,D_1\,(N_{N_1}-N_{N_1}^{\rm eq}) -N_{\D_{\mu}} \,p_{1 e+\mu}^{0}\,W_1^{\rm ID}  \,   .
\eea
Here $p_{1 e}^{0}$ and $p_{1 \mu}^{0}$ are the probabilities, at tree level, that a quantum lepton state 
$|\ell_1 \rangle$ is measured by tauon Yukawa interactions respectively as a tauon or an electron plus muon lepton state
and in terms of the neutrino Dirac mass matrix they can be simply calculated as
$p^0_{1\a} = |m_{D1\a}|^2 / (m_D^\dagger\,m_D)_{11}$. From these one can also introduce the 
flavoured decay parameters $K_{1\a} \equiv p^0_{1\a} \, K_1$. 
In Eqs.~(\ref{flke}) the flavoured $C\!P$ asymmetries $\ve_{1\a}$ are defined as
\be
\ve_{1\a}\equiv
\frac{\Gamma_{1\a} - \overline{\Gamma}_{1\a}}{\Gamma_1 + \overline{\Gamma}_1} \,  .
\ee
The flavoured decay rates are related to the total decay rates by
$\Gamma_{1\a} \equiv p_{1\a} \, \Gamma_1 $ and
$\overline{\Gamma}_{1\a} \equiv \overline{p}_{1\a} \, \overline{\Gamma}_1$,
where $p_{1\a}$ and $\overline{p}_{1\a}$ are the probabilities that the 
lepton and anti-lepton quantum state $|\ell_1 \rangle$ and $|\overline{\ell}_1 \rangle$
are measured in the flavour $\a$ by the interaction with a RH charged lepton $\a$. 
Notice that in general $\Delta p_{1\a} \equiv p_{1\a} - \overline{p}_{1\a} \neq 0$ and this introduces
an additional source of $C\!P$ violation. For this reason one can have the generation of
an asymmetry even if the total $C\!P$ asymmetry vanishes \cite{nardi1}.\footnote{
The tree level probabilities  $p^0_{1\a} \simeq (p_{1\a} + \overline{p}_{1\a})/2$.} 

Solving the set of Boltzmann equations one finds the final flavoured asymmetries and from these
the final $B-L$ asymmetry is simply given by their sum
\be
N^{\rm f}_{B-L} = N^{\rm f}_{\D_e} + N^{\rm f}_{\D_{e+\mu}}  \,   . 
\ee
Notice that in the limit of vanishing wash-out, for $K_1 \ll 1$, the sum of the two equations for the flavoured
asymmetries simply reduces to the Eq.~(\ref{lg2}) and, therefore, one obtains exactly the same solution as in the unflavoured regime.
Therefore, flavour effects can have an impact on leptogenesis predictions only if wash-out is non negligible, for  $K_1 \gtrsim 1$ \cite{predictions}.  
In this case one can arrive to the following approximate expression for the final $B-L$ asymmetry \cite{bounds}:
\be\label{NBmLapprox}
N^{\rm f}_{B-L} \simeq 2\,\ve_1 \, k^{\rm f}(K_1) + {\D p_{1\tau} \over 2} \,  
\left[\kappa_1^{\rm f}(K_{1\tau}) -  \kappa_1^{\rm f}(K_{1e+\mu})\right] \,  .
\ee
The first term is just twice what one would obtain in the unflavoured case, an effect that was already spotted
in early works studying flavour effects \cite{bcst}, but the second term is something genuinely arising 
from the account of flavour effects and can generate dramatic changes in the result compared to the unflavoured case.
The quantities $\D p_{1\tau}$ and $\D p_{1e+\mu} = - \D p_{1\tau}$ represent the difference in the flavour composition
of lepton quantum states and anti-lepton quantum states $C\!P$ conjugated and rise from radiative corrections. 
If the flavoured decay parameters in the two flavours the difference in the second term is non-vanishing and 
the $\D p_{1\a}$ can, under certain conditions, be comparable or even higher than the total $C\!P$ asymmetry.
In this situation the final $B-L$ asymmetry can be strongly enhanced compared to the unflavoured case. 
In this respect it can be shown how $C\!P$ violating phases, and in particular the Majorana phases, play
an important role in making in a way that difference between the two final efficiency factors  in the two flavours does not vanish. 
Therefore, flavour effects introduce a dependence of the final asymmetry on the parameters in the leptonic mixing matrix and in particular
on the Majorana phases, a dependence that completely cancels out in vanilla leptogenesis. In this way flavour effects can potentially improve the predictive power of leptogenesis, though one still needs some parameter reduction as discussed in the unflavoured case.

An extreme case occurs when the first term in Eq.~(\ref{NBmLapprox}) vanishes for $\ve_1 =0$, corresponding to 
total lepton number conservation in $N_1$-decays. All the asymmetry has then to be generated
by the second term \cite{nardi1}. In this case one can show easily that necessarily the $\D p_{1\a}$'s stem uniquely from
the $C\!P$ violating low energy neutrino phases, mainly from the Majorana phases, 
and that successful leptogenesis is attainable \cite{predictions,pascoliriotto,branco}.

\begin{figure}
\begin{center}
\psfig{file=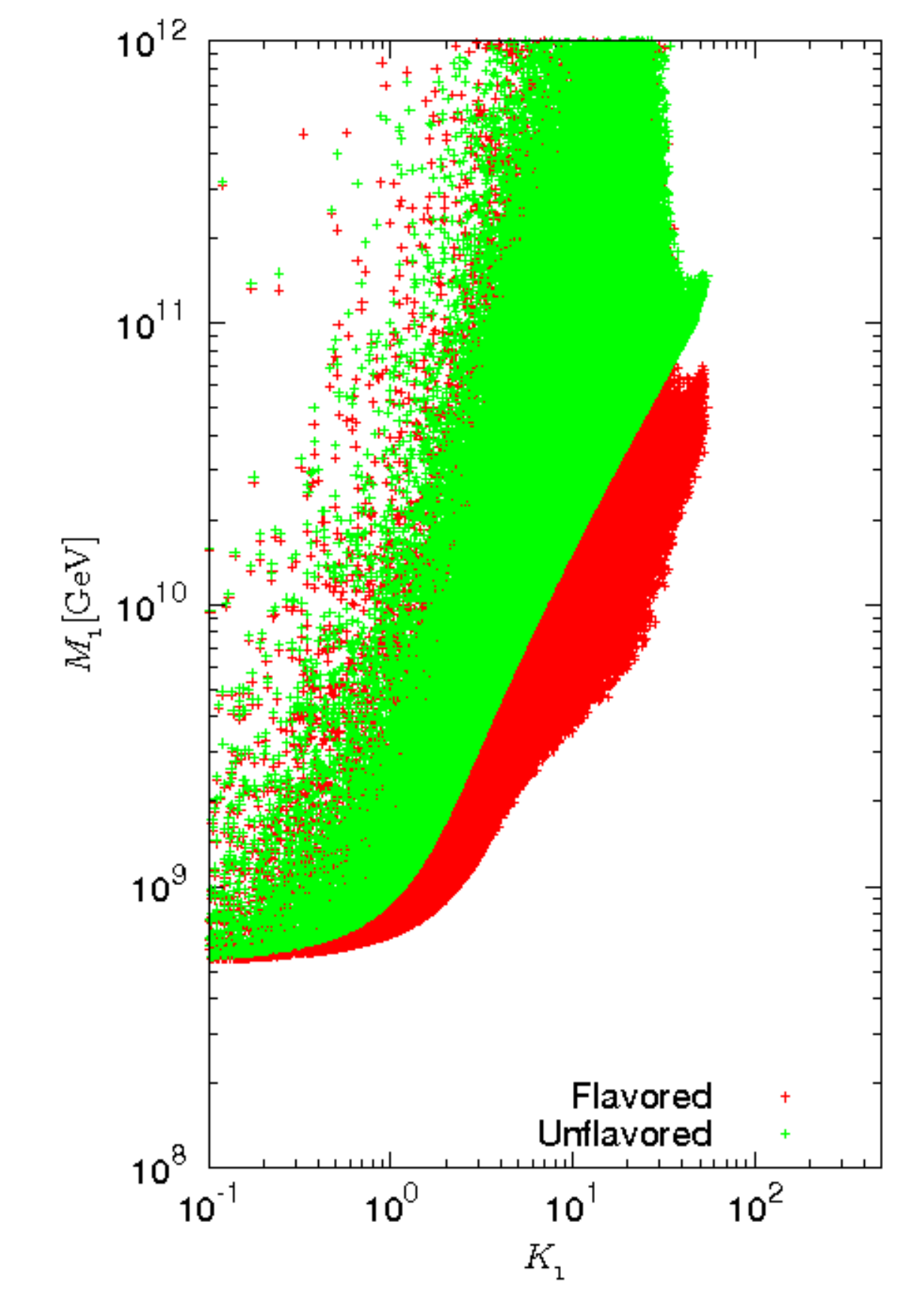,height=70mm,width=70mm} \hspace{5mm}
\psfig{file=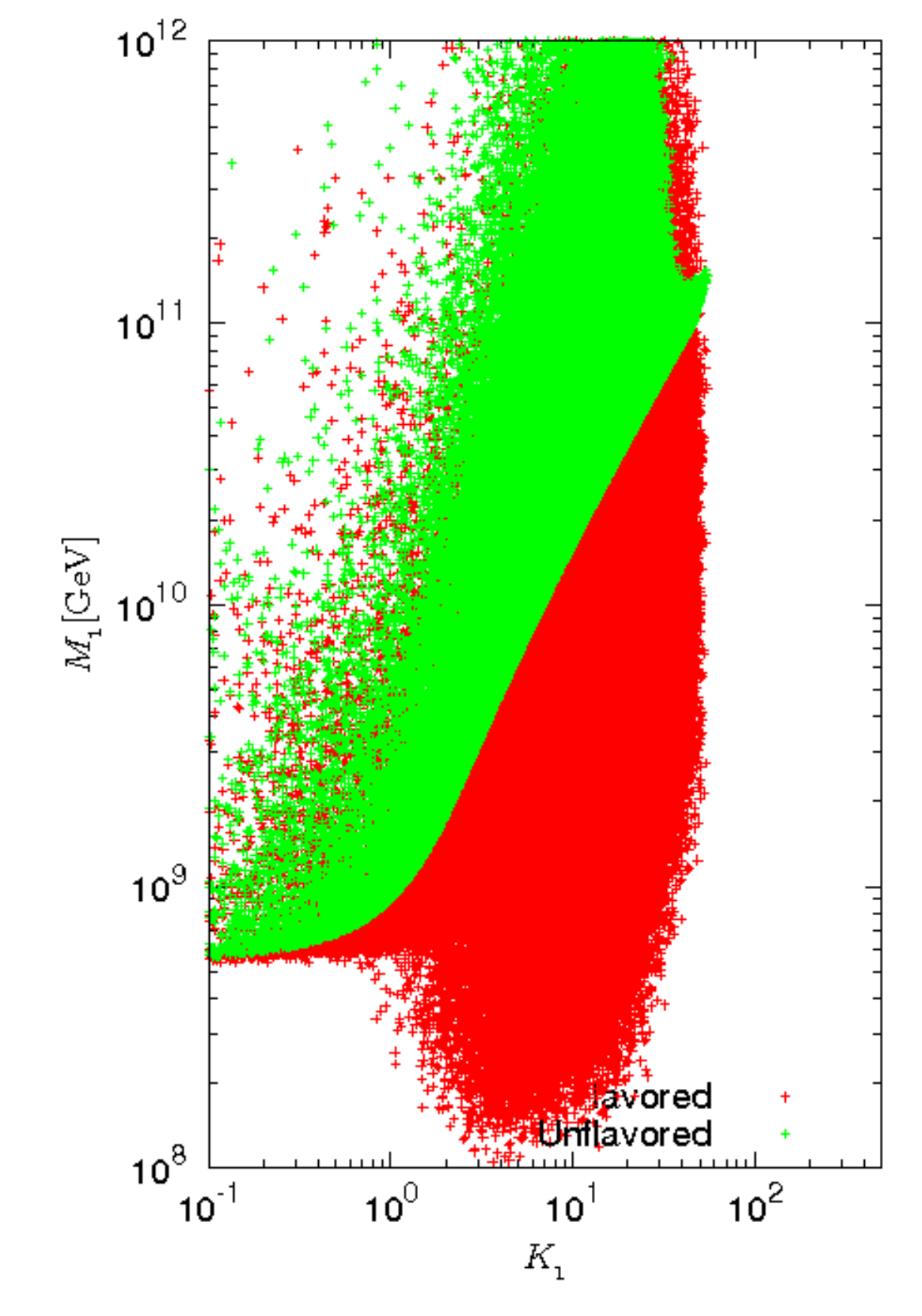,height=70mm,width=70mm}
\end{center} \vspace{-3mm}
\caption{Allowed region in the plane $M_1$ versus $m_{1}$ when flavour effects are included (from \cite{bounds}).
In the left panel no fine-tuning in the seesaw formula is allowed, in the right panel a mild fine-tuning is allowed.}\label{flbounds}
\end{figure}

For this reason, calculating the final asymmetry within a certain model including flavour effects is absolutely necessary.
However, despite this, the absolute bounds on $M_1$ and $m_{1'}$ we discussed within vanilla leptogenesis and shown in Fig.~8 are not dramatically relaxed in the hierarchical case. 
The reason is that in the absence of wash-out, as we mentioned, one recovers the unflavoured regime and,
in the presence of wash out the relaxation is large but, unless some fine-tuning in the seesaw formula is introduced, is not at the level to relax the lower bound on $M_1$ considerably and similarly for the upper bound on $m_{1'}$. However, if some fine-tuning in the seesaw formula is allowed then,
even for hierarchical RH neutrinos, one can lower the scale of leptogenesis.
In Fig.~(\ref{flbounds}) we show a comparison between the lower bound on $M_1$ when flavour effects are included compared to 
vanilla leptogenesis. In the left panel no fine-tuning in the seesaw formula is allowed while in the right panel a mild fine-tuning is allowed
and one can see how the lower bound is relaxed by one order of magnitude.
This possibility of lowering the scale of leptogenesis has been also implemented in a model \cite{antuschblanchet}
and it has been also recently thoroughly studied in \cite{jturner}.

Notice that if $M_1 \ll 10^{9} \, {\rm GeV}$  interactions of leptons produced by RH neutrino decays
with the RH muons in the thermal bath are effective in decohering the $e+\mu$ component of the $|{\ell}_1 \rangle$'s as well.
In this situation one has to consider a three fully flavoured regime where now one has to track the flavoured asymmetries
in all three flavours. Notice that for values of $M_1 \sim 10^{12}\,{\rm GeV}$ and $M_1 \sim 10^9 \, {\rm GeV}$ one is in a sort
of transition between the unflavoured and the two fully flavoured regime and between the two fully flavoured regime and the 
three fully flavoured regime respectively.  In these transition regimes the validity of Boltzmann equations breaks down and
one needs to use a density matrix formalism to describe the transition \cite{bcst,riottolosada,desimone,bdraffelt,jturner}. 

{\bf $N_2$-leptogenesis}. We have so far assumed that the asymmetry is dominated by the 
contribution of $N_1$-decays. 
As far as flavour effects are neglected, in the case of vanilla leptogenesis, this assumption works in 
most cases with just some special exception relying on having $K_1 \lesssim 1$, in which case one can
have that the asymmetry produced by next-to-lightest RH neutrinos can dominate   ($N_2$-leptogenesis) \cite{geometry}.

However, when flavour effects
are taken into account, the asymmetry produced from the next-to-lightest ($N_2$) RH neutrinos can find its way to survive the $N_1$ wash-out much more easily and $N_2$-leptogenesis cannot be regarded as  a special case any more. Therefore, the contribution to the final asymmetry from $N_2$-decays needs in general to be systematically taken into account, unless special assumptions are made, like for example that $M_2 \gg T_{\rm RH} \gg M_1$. 

There are two reasons why the asymmetry from $N_2$-decays can survive more easily.
The first reason is that for $M_1 \ll 10^{9}\,{\rm GeV}$  the lightest RH neutrino wash-out occurs in the
fully three flavoured regime. At the lightest RH neutrino wash-out stage,
the asymmetry produced from $N_2$-decays
 can be regarded as a pre-existing initial asymmetry.
However, now the Eq.~(\ref{preexisting}) holding in the unflavoured case,
has to be written as the sum of three contributions, one  from each flavour \cite{vives} 
\be\label{flpreexisting}
N_{B-L}^{\rm p,f} =   
N_{\D_e}^{\rm p,i}\, e^{-{3\pi\over 8}\,K_{1 e}} +
N_{\D_\mu}^{\rm p,i}\, e^{-{3\pi\over 8}\,K_{1 \mu}}+
N_{\D_\tau}^{\rm p,i}\,e^{-{3\pi\over 8}\,K_{1 \tau}} \, .
\ee
In our  case the three flavoured pre-existing asymmetries are produced from $N_2$-decays.
Now, having included flavour, it is sufficient that just one $K_{1\a} \lesssim 1$ for the asymmetry in that flavour to survive. 
In the same random generation of models performed in \cite{seesawmotion}, where it was found that 
the probability to have $K_1 \lesssim 0.1$ is approximately $0.1 \%$, it was also found that the
probability for $K_{1e} \lesssim 1$ is $26\%$ and the probabilities for $K_{1\mu}, K_{1\tau} \lesssim 1$ are $\sim 7\%$.
Therefore, it is clear that in this case the survival of an asymmetry produced from $N_2$-decays 
is not special  but it becomes a very plausible situation. As we will discuss in the next subsection, this situation is crucial for $SO(10)$-inspired models. 

A second reason why the asymmetry from $N_2$-decays, or at least some fraction of it,  can survive $N_1$-decays is 
simply due to  {\em flavour projection}: considering the full three-dimensional structure of flavour space,  the 
wash-out from $N_1$ inverse processes acts only on a component that is parallel to the ${\ell}_1$ flavour,
i.e., the flavour of leptons produced in $N_1$-decays and that can induce $N_1$ production from $N_1$ inverse decays.
The orthogonal component will completely escape the wash-out \cite{bcst,nardi2}. This  effect is interesting since it  
holds even if the $N_1$ and $N_2$ masses fall in the same flavour regime and, in particular, even in the unflavoured regime, for 
$M_1, M_2 \gg 10^{12}\,{\rm GeV}$.  An interesting application is that a new allowed region, where the 
$N_2$ asymmetry dominates, opens up in the space of parameters in the  two RH neutrino case \cite{2RHNlep}.

\subsection{$SO(10)$-inspired leptogenesis}

One of the most intriguing application of flavour effects is that they rescue $SO(10)$-inspired leptogenesis \cite{riotto1}.
Given the typical very hierarchical RH neutrino mass spectrum rising in $SO(10)$-inspired models, see Fig.~\ref{SO10inspiredspectrum}, one has 
$M_1 \ll 10^9\,{\rm GeV}$ and, as we already pointed out,
the asymmetry produced from the lightest RH neutrinos would be orders of magnitude below the observed value.  Moreover, in $SO(10)$-inspired models
one has $K_1 \gg 1$. Considering the expression Eq.~(\ref{preexisting}) for the wash-out of an asymmetry generated prior to the
lightest RH neutrino wash-out assumed to occur before sphalerons go out-of-equilibrium, 
the asymmetry produced from $N_2$-decays would be efficiently washed out and $N_2$-leptogenesis would not be viable.

However, when flavour effects are taken into account, the expression Eq.~(\ref{preexisting}) for the
final value of a pre-existing asymmetry gets modified into Eq.~(\ref{flpreexisting}) and, specifically
in the case of $N_2$-leptogenesis, when $N_2$-decays take place in the two fully flavoured regime for 
$10^9\,{\rm GeV} \ll M_2 \ll 10^{12}\,{\rm GeV}$, one has
\cite{riotto1,riotto2,decrypting}
\bea\label{twofl} \nonumber
N_{B-L}^{\rm lep, f} & \simeq &
\left[{K_{2e}\over K_{2\tau_2^{\bot}}}\,\ve_{2 \tau_2^{\bot}}\kappa(K_{2 \tau_2^{\bot}}) 
+ \left(\ve_{2e} - {K_{2e}\over K_{2\tau_2^{\bot}}}\, \ve_{2 \tau_2^{\bot}} \right)\,\kappa(K_{2 \tau_2^{\bot}}/2)\right]\,
\, e^{-{3\pi\over 8}\,K_{1 e}}+ \\ \nonumber
& + &\left[{K_{2\mu}\over K_{2 \tau_2^{\bot}}}\,
\ve_{2 \tau_2^{\bot}}\,\kappa(K_{2 \tau_2^{\bot}}) +
\left(\ve_{2\mu} - {K_{2\mu}\over K_{2\tau_2^{\bot}}}\, \ve_{2 \tau_2^{\bot}} \right)\,
\kappa(K_{2 \tau_2^{\bot}}/2) \right]
\, e^{-{3\pi\over 8}\,K_{1 \mu}}+ \\
& + &\ve_{2 \tau}\,\kappa(K_{2 \tau})\,e^{-{3\pi\over 8}\,K_{1 \tau}} \,  ,
\eea
where $K_{2\tau_2^{\bot}} \equiv K_{2e} + K_{2\mu}$ and 
$\ve_{2\tau_2^{\bot}} \equiv \ve_{2e} + \ve_{2\mu}$. In this expression we have
also included so called phantom terms \cite{fuller} that, however, in the case of $SO(10)$-inspired
leptogenesis give a very small contribution and can be neglected.

\begin{figure}
\begin{center}
\psfig{file=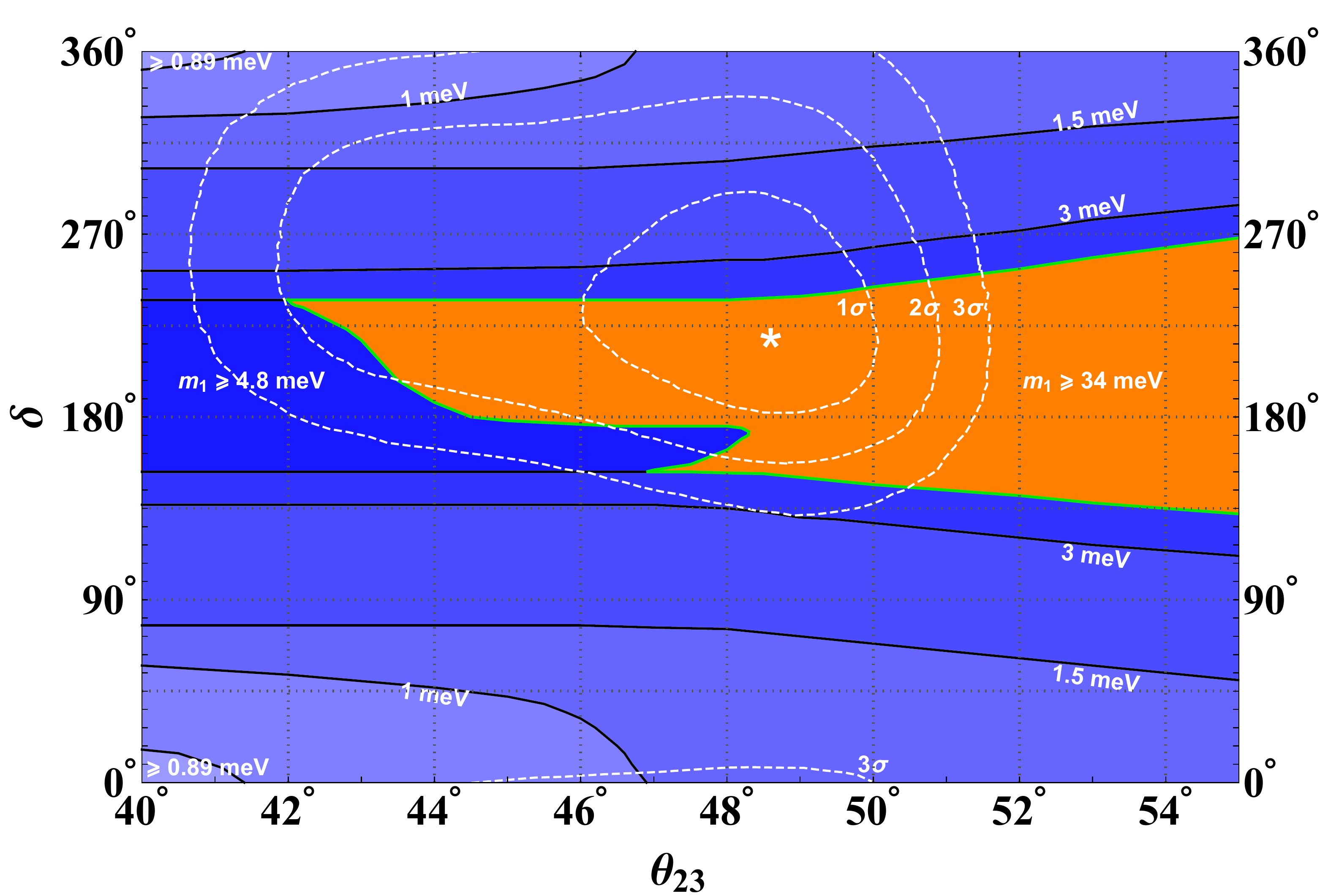,height=70mm,width=77mm} \hspace{3mm}
\psfig{file=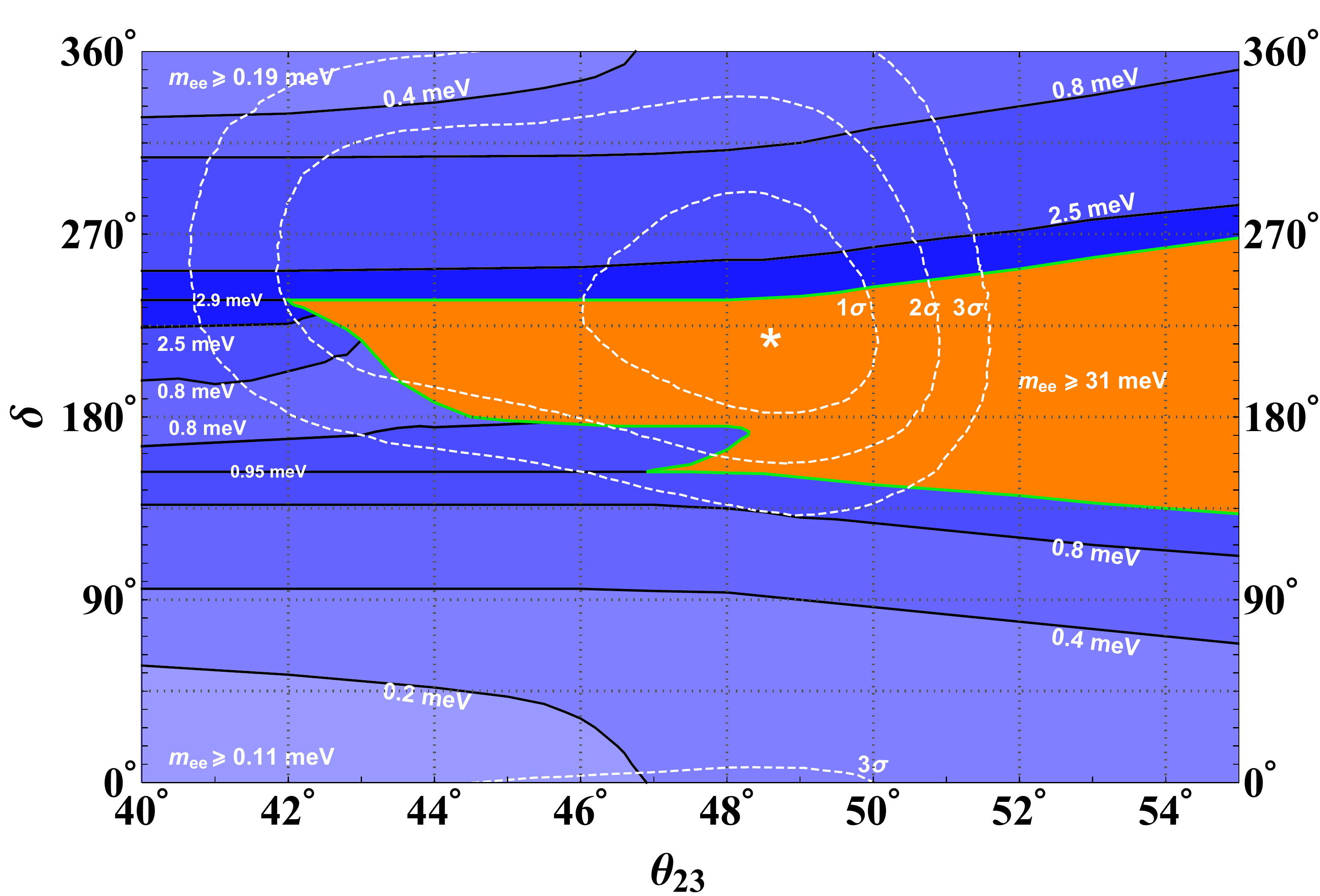,height=70mm,width=77mm}
\end{center} \vspace{-1mm}
\caption{Lower bound on $m_1$ (left panel) and $m_{ee}$ (right panel) from $SO(10)$-inspired leptogenesis 
(from \cite{opportunity}). }\label{boundsSO10lep}
\end{figure}

The interesting property  is that, analogously to vanilla leptogenesis where the final asymmetry depends only on the high energy parameters associated to the
lightest RH neutrino $N_1$, the final asymmetry in $SO(10)$-inspired leptogenesis depends only on the high energy parameters 
associated to the next-to-lightest RH neutrino  $N_2$. The flavoured $C\!P$ asymmetries and the flavoured decay parameters
can be expressed through the parameterisation we discussed in 4.4.2 (see Eqs.~(\ref{alphas}) and (\ref{biunitary2})). Out of the three $\alpha_i$'s,
the asymmetry depends only on $\alpha_2$. Moreover, since the matrix $V_L \simeq V_{CKM} \simeq I$, the dependence
on the six parameters in the matrix $V_L$ is mild. For this reason, from the successful leptogenesis condition, one is able to obtain predictions
on the low energy neutrino parameters, and to place interesting experimental constraints, 
in particular \cite{riotto1,riotto2,fuller}:
\begin{itemize}
\item Inverted ordering is now in strong tension with the cosmological upper bound on neutrino masses (see (Eq.~\ref{cosmoupperbound}));
\item There is a lower bound $\theta_{13} \gtrsim 2^\circ$, now nicely confirmed by the $\theta_{13}$ discovery;
\item There is a lower bound on the lightest neutrino mass $m_1 \gtrsim 1\,{\rm meV}$;
\item  Majorana phases are constrained within narrow ranges around special values and for this reason the effective
           $0\nu\b\b$ mass $m_{ee}$ cannot be  arbitrarily small and bulk solutions are found for values above ${\rm meV}$. 
\end{itemize}
Recently it has been shown \cite{opportunity} how the lower bound on the absolute neutrino mass scale, both on $m_1$ and $m_{ee}$, becomes
much more stringent for values of $\delta$ and $\theta_{23}$ in a region (in orange in Fig.~\ref{boundsSO10lep})
including current best fit values of $\delta$ and $\theta_{23}$  (see Eq.~(\ref{normal})),
and from Fig.~\ref{boundsSO10lep} one can see that one obtains
\be
m_1 \gtrsim 34 \, {\rm meV} \,  \hspace{5mm} \mbox{\rm and} \hspace{5mm} m_{ee} \gtrsim 31 \, {\rm meV} \,  .
\ee
This is quite interesting since future experiments should be able to fully test these lower bounds.
It is in particular interesting that, despite these solutions are found for normally ordered neutrino masses,
there is such a stringent lower bound  on $m_{ee}$. These results clearly show that even high-scale 
leptogenesis models are testable. From this point of view the discovery of $0\nu\b\b$ decay would be a 
clear (necessary) breakthrough discovery supporting leptogenesis.\footnote{While on the other hand
notice from Fig.~\ref{boundsSO10lep} how a (low energy) $C\!P$ conserving value $\d = \pi$
is perfectly compatible with successful leptogenesis. This is a clear example showing how observing $C\!P$ violation in neutrino oscillation experiments is neither a necessary nor sufficient condition for successful leptogenesis. However, of course, one would in general expect to observe it, while ruling it out with great precision would represent quite a special 
situation casting some doubts, also because there are no compelling models where this is predicted to occur.} 


\section{Dark matter from active-sterile neutrino mixing and the $\nu$MSM model} \label{DW}

As we discussed in the previous section, leptogenesis provides an elegant way to the solution of the matter-antimatter asymmetry of the universe.  In its minimal, and original, version it just relies on the type-I seesaw extension of the SM Lagrangian Eq.~(\ref{Y+M}) that also successfully explains the observed neutrino masses and mixing.
It is then legitimate to wonder whether the same extension might also address the dark matter of the universe. It is quite intriguing that there is indeed a very simple solution \cite{dw} relying on the 
active-sterile neutrino mixing, expressed by Eq.~(\ref{massfields}),
predicted by the type-I seesaw Lagrangian. 

The possibility for the heavy seesaw neutrinos to provide a viable dark matter candidate 
necessarily relies on special regions of parameter space where they are stable on cosmological scales.
Their lifetime needs necessarily to be longer than the age of the Universe to be able to play the 
gravitational role of dark matter as {\em cosmic glue}, explaining the dynamics of 
clusters of galaxies and galactic rotation curves.
However, depending on their mass, the lower bound on the life-time is orders of magnitude more stringent, in order to avoid experimental
constraints deriving from indirect searches and also, at some energies, 
from cosmic microwave background and 21 cm global cosmological signal. 

If the mass of the heavy neutrino $N_I$ is higher than the Higgs boson mass, then its lifetime, 
taking the inverse of the decay width in Eq.~(\ref{decaywidth}), is simply given by
 \be\label{tiKi}
\t_I= {8\,\pi \over (h^\dagger\,h)_{II}\,M_I} \simeq {5\over K_I}\,
\left({{\rm TeV}\over M_I}\right)^2\times 10^{-13}\,{\rm sec} \,  ,
\ee
that we wrote in terms of the decay parameters  defined in Eq.~(\ref{Kdef}).
Therefore, one can see that the stability condition $\t_I \gtrsim  t_0 \simeq 4\times 10^{17}\,{\rm sec}$,
would translate into an upper bound \cite{ad}
\be\label{lifetime}
K_{I}\, \lesssim {10^{-30}}\,\,\left({\rm TeV \over M_I} \right)^2 \, .
\ee
In this limit, the RH neutrino would decouple from the seesaw formula and one would have simply
$m_1 = \widetilde{m}_I \propto 1 / M_I$. 
 Within this minimal extension of the SM there cannot be any production mechanism relying on such tiny Yukawa couplings, even barring fine-tuning considerations on the feasibility of building such a model. One could resort to some  gravitational production mechanism but clearly this would lack predictive power being disconnected from neutrino parameters.

\subsection{The Dodelson-Widrow mechanism}

There is, however, a second possibility to consider. One RH neutrino could be lighter than gauge bosons 
and in this case it would not directly decay through their Yukawa couplings into gauge bosons and leptons. 
However, it would still decay into SM particles because of the active-sterile neutrino mixing since, as expressed by Eq.~(\ref{massfields}),
heavy seesaw neutrinos have a LH component. The dominant decay would be the 
decay into three leptons.  The possibility to have $\tau_I > t_0$ can be realised if $M_I < 2\, m_e$. 
One would still need small Yukawa couplings with $K_I \ll 1$ but not as tiny as in Eq.~(\ref{lifetime}). 
In this case,  the dark matter heavy neutrino has to be necessarily $N_1$,  the  lightest of the three, in order for the two heavier ones   to reproduce the 
solar and atmospheric neutrino mass scales from the seesaw formula \cite{asakablanchet}. 
The lightest  seesaw neutrino mass would then be inversely proportional to the lightest ordinary
neutrino mass $m_{1'}$. The general expression Eq.~(\ref{massfields}) for the mixing can then be specialised to the case when one has a $\nu_1-N_1$ dominant mixing: 
\bea
\nu_1 & = & \left[U^\dagger_{1\a}\nu_{L\a} + (U^\dagger_{1\a}\nu_{L\a})^c \right]  -  
\left[V^\nu_{L 1\a}\,\xi^\star_{\a 1} \,\nu_{R 1} + (V^\nu_{L 1\a}\,\xi^\star_{\a 1} \,\nu_{R1})^c\right]  \\
N_1  & =&  \left[ \nu_{R 1} + \nu_{R1}^c \right] + 
         \left[\xi^T_{1\a} \,\nu_{L \a} + (\xi^T_{1\a} \,\nu_{L \a})^c \right]  \,   .
\eea
This mixing would be then responsible for its decay dominantly into three neutrinos with a lifetime 
\be
\tau_1 \simeq 5 \times 10^{28}\,{\rm s}\,\left({M_1 \over {\rm keV}} \right)^{-5} \, \left({10^{-4} \over \theta}\right)^2  \,   ,
 \ee
where $\theta \equiv \sum_\a \, |m_{D\a 1} / M_1 |^2$ is an effective active-sterile neutrino mixing angle and where
one can notice the strong inverse-fifth power dependence on the mass typical of three particle decays (Sargent's law).
The expression is written in a way to highlight that values $\theta \sim 10^{-4}$ are necessary 
in order to have such long lifetimes that would escape stringent experimental constraints from X-ray observations  in galaxy clusters.  The striking coincidence is that this order-of-magnitude for the value of the  mixing angle is also the correct one able to reproduce the observed dark matter abundance (see Eq.~(\ref{ODM})). Indeed, solving 
a simple Boltzmann equation for the production of the $N_1$'s, one finds for its
contribution to the energy density parameter 
\be
\O_{N_1} h^2 \sim 0.1 \, \left({\theta \over 10^{-4} }\right)^2 \,  \left({M_1 \over {\rm keV}} \right)^{2} \,  ,
\ee
that can be indeed satisfied for $\theta\sim 10^{-4}$ and $M_1 \sim {\rm keV}$.  This is often referred to as the Dodelson-Widrow mechanism for the production of a sterile neutrino abundance able to reproduce the observed dark matter abundance \cite{dw}.  
Interestingly, for keV masses, the $N_1$'s would behave as warm dark matter, 
implying that compared to cold dark matter there would be a reduced power in the large scale
structure at scales corresponding to dwarf galaxies ($\sim 0.1\,{\rm Mpc}$ in comoving length). This would also
help in solving some potential issues in pure cold dark matter N-body simulations that seem to predict too
many satellite galaxies, in a galaxy like ours, compare to what is observed astronomically. At the same time warm dark matter
would also smooth the cusp profile in galaxies that is predicted by cold dark matter N-body simulations but that is
in tension with different observations.  

Sub-dominantly the $N_1$'s can also radiatively decay and this offers a very beautiful experimental signature to test them with X-ray observations that place an upper bound 
$M_1 \lesssim 10\,{\rm keV}$ for $\theta \sim 10^{-4}$ 
(see \cite{Drewes:2016upu,Boyarsky:2018tvu}). 

Moreover large scale structure N-body simulations place a lower bound $M_1 \gtrsim {\rm keV}$, since otherwise it would be too warm.
These constraints have progressively ruled out the minimal scenario relying on the assumption of a non-resonant
active-sterile neutrino mixing \cite{horiuchi,Boyarsky:2018tvu}, holding in the reasonable case that the initial lepton asymmetry is of the order of the baryon asymmetry
(as for example it occurs in the minimal scenario of leptogenesis). However, if a large lepton asymmetry ($L \gtrsim 10^{-5}$) is
present at the time of mixing, then the mixing would be resonant and the production is enhanced for the same mixing angle 
compared to a non-resonant production \cite{shifuller,dolgovhansen}. 
In this way one can reproduce the correct abundance and at the same time having longer life-times, reconciling predictions
with the experimental constraints.   

Interestingly, the observation of a new  $3.5\,{\rm keV}$ line in the X-ray observations of different 
clusters of galaxies \cite{bulbul,boyarsky} could be explained by such a mechanism for a decaying  
$7\,{\rm keV}$ sterile neutrino with a mixing angle $\theta \simeq 4 \times 10^{-6}$ and its 
relic abundance can explain the observed dark matter density parameter for a 
lepton asymmetry  at the resonance $L \simeq 4.6 \times 10^{-4}$ \cite{abazajian}.
Recently, observations from blank-sky observations testing 
emission from the dark matter Milky Way halo seems to have excluded
the interpretation of the $3.5\,{\rm keV}$ anomaly in terms of a 7 keV decaying sterile
neutrino dark matter \cite{dessert}. However, such a conclusion has 
very recently been  questioned \cite{replies} and 
more observations will be likely necessary to definitively test 
this exciting possibility.

\subsection{The $\nu$MSM model}

The non-resonant production of sterile neutrino dark matter can be nicely embedded within the seesaw mechanism
in a way that neutrino masses and mixing parameters are also explained \cite{asakablanchet}.
Even more ambitiously, one can also explain the matter-antimatter asymmetry of the universe via leptogenesis
from the mixing of the two heavier RH neutrinos, the so-called ARS scenario \cite{ARS,Abada:2018oly,Drewes:2021nqr}, 
obtaining a unified model of neutrino masses and mixing, dark matter and baryogenesis: 
the so-called $\nu$MSM model \cite{nuMSM}.

The necessity of a resonant production with the generation of a large lepton asymmetry prior to the 
sterile neutrino production is clearly more challenging to reproduce but it can also be accommodated within
the $\nu$MSM, assuming just a type-I seesaw extension of the SM with three RH neutrinos. This is possible if 
the two heavier RH neutrinos $N_2$ and $N_3$ produce the large lepton asymmetry after sphaleron processes go out-of-equilibrium 
(by mixing or decays) and clearly prior to the 
lightest RH neutrino production.  This late leptogenesis would then  be instrumental in the 
 sterile neutrino dark matter resonant production \cite{shap,cdshap1,cdshap2}.

The possibility to generate a large lepton asymmetry prior to the dark matter imposes  very stringent constraints.
First of all the two heavier RH neutrinos have to be extremely degenerate with a mass difference that has to be 
in an extremely narrow  range of values around $M_3 - M_2 \sim 10^{-13}\, M_2$, a condition rising from
a necessary strong fine-tuning between light and heavy neutrino
mass differences~\cite{royshap}. Moreover, 
in order to have sufficient  $C\!P$ violation one also obtains a  lower bound (holding 
for normal ordering) $M_2 \gtrsim 2\,{\rm GeV}$. Despite the strong necessary fine-tuning, 
this  scenario  is quite interesting since it can be tested either from the decays of the
dark matter particles that, again, should produce a line in X-ray observations, 
and/or also from B meson decays, that might produce
the two heavier RH neutrinos if these are lighter than
$\sim 5\,{\rm GeV}$, in experiments such as SHiP~\cite{ship},
MATHUSLA~\cite{mathusla} and the approved
FASER~\cite{FASER:2019aik}.
Clearly the observation of the 3.5 keV line
has triggered great excitement. It should be said however that a recent numerical analysis where  the mass  of $N_1$
has been fixed to $7\,{\rm keV}$, the value that would explain the anomalous 3.5 keV line, has found no solutions in the space of parameters
able to reproduce both the baryon asymmetry and the observed dark
matter energy density parameter~\cite{laine}.
The main problem is the difficulty to produce, after sphaleron out-of-equilibrium and prior to dark matter production, 
the large lepton asymmetry needed to enhance resonantly the dark matter production in active--sterile neutrino mixing.
In order to achieve dark matter resonant 
production, solutions where such large 
asymmetry is produced from the $N_{2,3}$ late decays have been recently found \cite{Ghiglieri:2020ulj} but an exquisite degree of fine-tuning of the involved parameters is required. Indeed, the necessary lepton asymmetry for the resonant production is generated if the physical splitting $(M_3 - M_2) \sim 10^{-16}\,M_2$ with $M_2 \sim {\rm GeV}$ and if the real part of the complex angle in the orthogonal matrix is close to $\pm \pi/2$ with a relative precision of $\sim 10^{-8}$ for normal hierarchy
and $\sim 10^{-7}$ for inverted hierarchy (see Fig.~2 in \cite{Ghiglieri:2020ulj}). These solutions are also compatible  with successful leptogenesis from RH neutrino mixing. If such fine-tunings are not imposed, then a generation of just $(5$--$10)\%$ of the observed dark matter abundance is generically obtained for
$M_{2,3} \sim {\rm GeV}$.

\section{Beyond the minimal  seesaw Lagrangian}

The difficulty, if not impossibility, to obtain a scenario on the origin of matter in the universe within a minimal type-I seesaw extension of the SM, motivates an extension of the scenario. Moreover, unless an origin of the
3.5 keV anomaly in terms of sterile neutrino decays is confirmed,\footnote{The launch in 2022 of the XRSM satellite should definitively solve the 
controversial situation \cite{XRISM}.}  the strong  existing constraints from X-ray observations and structure formation seem to disfavour
a solution of the dark matter puzzle in terms of a $\sim {\rm keV}$ seesaw neutrino. 
This encourages the investigation of an extension of the SM beyond the minimal type-I seesaw. 
We briefly discuss a few recent proposals.

\subsection{Higgs induced RH neutrino mixing (RHiNo) model}

The success of minimal leptogenesis in explaining the matter-antimatter asymmetry of the universe in relation to the observed neutrino masses and mixing, motivates to retain the type-I seesaw Lagrangian as a successful starting point 
and considering an extension to it. 

Even more, one can think to repeat, in an upgraded version, the same successful strategy. We saw that the type-I seesaw 
is the simplest way to realise a UV-completion of the SM that corresponds to a low energy effective theory
described by the 5-dim Weinberg operator (\ref{weinberg}). Having now introduced RH neutrinos, leading to an
explanation of neutrino masses and mixing via seesaw mechanism and matter-antimatter asymmetry via leptogenesis,
one can form a similar new 5-dim operator, the Anisimov operator \cite{anisimov,ad} ($I, J = 1,2,3$)
\be\label{anisimov}
{\cal O}_A = {\la_{IJ} \over \L} \, \Phi^\dagger \, \Phi \, \overline{N_I^c} \, N_J \,  ,
\ee
a variant of the Higgs portal operator. Having introduced this operator, one can go back to the idea of having one 
heavy RH neutrino,  with vanishing Yukawa coupling as a candidate of cold dark matter. 
The Anisimov operator interestingly would induce, via the Higgs, a mixing between a seesaw heavy neutrino
and the dark matter RH neutrino. This mixing would be
responsible both for the dark matter RH neutrino production at high temperatures 
but also for its decays at zero temperature. This is the Higgs induced RH-RH neutrino mixing  (RHiNo) model.  
The interesting feature of the model is that at high temperatures the 
mixing is enhanced $\propto T^3$ by thermal effects. In this way the effect of the Anisimov operator is 
such to have an enhanced production at high temperatures but would fade out today, still however inducing dark matter
decays, with the possibility to test them in indirect searches. Clearly the possibility to satisfy stability while still 
reproducing the observed dark matter energy density parameter has to be verified quantitatively and goes through
an accurate calculation both of the dark matter production and of all decay channels contributing to the total decay width of the dark matter RH neutrino and to indirect search signals \cite{DiBari:2016guw}.

Initially, calculations of the final relic dark matter abundance have been done imposing a resonance condition, leading to the condition that the dark matter is heavier than the source RH neutrino, and then using a Landau-Zener formula for the description of the non-adiabatic production of dark matter RH neutrinos at the resonance.  However, recent results within a density matrix formalism \cite{densitym} show that 
 the Landau-Zener formula fails to reproduce the correct dark matter abundance unless there is a strong degeneracy between dark and source RH neutrino. The problem is that the Landau-Zener assumes that oscillations have already developed at the resonance. However, this  is true only if 
 $\D M/M_{\rm DM} \ll 10^{-12} \, M_{\rm DM}/{\rm GeV}$, where $\D M$ is the mass difference 
 between the dark matter and the source RH neutrinos.  This condition clearly requires the two RH neutrinos to
 be quasi-degenerate.
For hierarchical RH neutrinos, density matrix calculations show that a non-resonant production is still occurring but the final abundance
is order of magnitudes lower than within a Landau-Zener approach. This non-resonant production has analogies with the  extremely
non-adiabatic limit in solar neutrinos. In this case essentially thermal effects generated by the Anisimov operator are still important but those generated by the Yukawa coupling of the source RH neutrino, corresponding to medium effects in the case of  solar neutrinos, are negligible.  

The results is that for $M_{\rm DM} > M_{\rm S} > M_{\rm Higgs} \simeq 125\,{\rm GeV}$ one can reproduce the correct
measured dark matter energy density parameter only assuming the thermalisation of the source RH neutrino at the time
of the dark matter production from the mixing. This thermalisation however cannot be achieved by Yukawa couplings so somehow
one needs to resort to some external mechanism. However, for dark matter masses below the Higgs and gauge boson masses 
a two body decay channel, $N_{\rm DM} \ra N_{\rm S} \ra A + {\ell}$, is kinematically forbidden and this allows to have larger source RH neutrino Yukawa couplings still satisfying dark matter stability on cosmological scales. 
In this case however, the usual scenario of leptogenesis from decays cannot be successful for the generation of the matter-antimatter asymmetry and one needs to resort to the ARS scenario of leptogenesis from mixing \cite{ARS}, as in the $\nu$MSM model.

On the other hand, since the production is non-resonant anyway, a condition $M_{\rm DM} > M_{\rm S}$ is not needed any more 
and one can explore the case $M_{\rm S} > M_{\rm DM}$. This might open a new interesting solution where the dark matter cam be
the lightest RH neutrino species $N_1$ and the two heavier ones, $N_2$ and $N_3$, can produce the matter-antimatter asymmetry of the Universe  just within the standard leptogenesis scenario and generate neutrino masses in two RH neutrino seesaw models. 

It is interesting that the mechanism can also be combined with a phase transition driven by a scalar that is also responsible
for the generation of the source RH neutrino mass. During the phase transition both dark matter, neutrino masses and 
gravitational waves are generated\cite{gwaves}.

\subsection{Frogatt-Nielsen model and dynamical Yukawa couplings}

The idea that a heavy RH neutrino can play the role of dark matter and be produced through couplings
that were much stronger at high temperatures but almost vanish at present time to ensure stability, 
can be exported to other models. Recently, it was proposed that Yukawa couplings of dark matter RH neutrino
in the type-I seesaw Lagrangian could dramatically change during a phase transition of a new scalar $\Sigma$
in a Frogatt-Nielsen  model, passing from ${\cal O}(1)$ couplings to tiny values consistent with cosmological stability \cite{lindnerrode}.
Before the $\S$ phase transition the large Yukawa couplings thermalise the dark matter RH neutrino. 
At the phase transition the dark matter RH neutrino Yukawa coupling is strongly suppressed and this induces the freeze-out of the dark matter RH neutrino abundance that has to be such that to yield the  
observed dark matter energy density parameters. In addition to the dark matter RH neutrino, the presence of two additional RH neutrinos can reproduce the neutrino masses and mixing via type-I seesaw mechanism and the matter-antimatter asymmetry via leptogenesis.

\subsection{Seesaw and dark side sector}

The RHiNo model \cite{ad} is the first example of a model connecting neutrino masses from type-I seesaw mechanism to dark matter. Since it introduces a non-renormalisable operator, it is not UV-completed. 
Notice that the dark sector is represented by the RH neutrino $N_{\rm DM}$.

Along similar lines one can connect the seesaw physics to dark matter introducing a dark sector made of a new singlet complex scalar
$\phi$ and a singlet fermion $\chi$ and imposing a $Z_4$ symmetry \cite{Bhattacharya:2018ljs}.
The SM Lagrangian is then extended by 
\be
{\cal L} -{\cal L}_{\rm SM} = {1 \over \Lambda}\,\overline{L}\,\widetilde{H}\,\phi \, N +
{1 \over 2}\, M \,\overline{N^c} \, N + y\, \phi\,\overline{\chi}\, \chi + V(H,\phi) \, .
\ee
Prior to the $Z_4$ symmetry breaking the dark matter remains massless and does not couple to the SM but after
symmetry breaking it can be shown that $\phi$ mixes with the SM Higgs and dark matter can be produced
and annihilate through processes ${\rm SM} + {\rm SM} \leftrightarrow \chi + \overline{\chi}$. The abundance
can then be calculated via standard freeze-out solution.  The dark matter is stable due to a relic $Z_2$ symmetry. 

With the same dark sector a variation of the RHiNo models that allows to obtain a UV-completed model can be obtained  introducing a {\em neutrino portal}  term \cite{Chianese:2018dsz}
\be
{\cal L}_{\rm portal} = y_{DS}\,\phi \,\overline{\chi}\,N_I + {\rm h.c.}
\ee
In this case there is no mixing of $\phi$ with the standard model Higgs and the dark matter production occurs via freeze-in solution.  In both models the DM particle is stable so that there are no indirect search signatures and this can be seen as a phenomenological drawback.  

\subsection{Scotogenic model}

In the scotogenic model  \cite{Ma:2006km} the SM is extended by adding three copies of 
SM singlet fermions plus an additional Higgs doublet that is inert since it does not develop 
a vacuum expectation value. The new sector is odd under an unbroken $Z_2$ symmetry so that  
the lightest $Z_2$ odd particle, that turns out the lightest neutral scalar of the inert doublet, 
can play the role of a stable dark matter particle candidate. The relic abundance is
obtained via the standard freeze-out mechanism. 
The $Z_2$ symmetry and the corresponding charges of the fields prevent the inert Higgs doublet to
couple with SM fermions at tree level so that light neutrino masses are forbidden at tree level (otherwise 
one would have a seesaw generation) but can be generated radiatively at one loop.

Interestingly, the out-of-equilibrium decays of the three SM singlet $Z_2$ odd fermions can also 
generate the baryon asymmetry via leptogenesis  \cite{Borah:2018rca}. As in the case of vanilla leptogenesis,
the asymmetry would be dominated by the decays of the lightest $Z_2$ odd fermion $N_1$ with mass
$M_1$ and generated at temperatures $T \sim M_1$. It is found that  successful leptogenesis can be obtained for $M_1 \sim 10\,{\rm TeV}$.

\section{Final remarks}

Understanding the origin of matter in the universe poses a challenge to our current knowledge of fundamental physics, 
since within our current description of fundamental interactions based on general relativity and the SM 
there is no model able to account for the dark matter and matter-antimatter asymmetry of the universe.  This is extremely interesting since a model of the origin of matter in the universe necessarily implies the existence of new physics. 
We mainly discussed solutions relying on an extension of the SM
and we have particularly focussed on those solutions that can also address neutrino masses and mixing, since in the absence
of clear new physics from other phenomenologies, this is an attractive path. From this point of view, the continuation of the
already well established programme of low energy neutrino experiments might provide an important insight in future, especially in connection with leptogenesis. In this respect, the discovery of $0\nu\b\b$ decays would represent a clear breakthrough, establishing the Majorana nature of neutrinos and lepton violation at tree level. 
At the same time, the measurement of the $0\nu\b\b$ effective neutrino mass would 
provide a way to test specific models and we have seen in particular the case of $SO(10)$-inspired models. 
When $0\nu\b\b$ experiments are combined together with the cosmological information on the absolute neutrino mass, one obtains a powerful 
test that might even lead, in optimistic scenarios, to pin down specific scenarios of leptogenesis. 
We have also seen how the 3.5 keV anomaly  would in case point also to a solution of the dark matter puzzle based on neutrino physics.   
On this front, planned laboratory experiments, such as SHiP, MATHUSLA and  FASER 
will test the existence of RH neutrinos lighter 
than $\simeq 5 \, {\rm GeV}$ that are predicted in scenarios on the origin of matter such as the $\nu$MSM.  
However, in the last years a wide variety of different experimental discoveries 
open new ways to investigate models of the origin of matter in the universe.  The discovery of very high energy neutrinos
at IceCube provides a new investigative tool on scales much above the TeV scale, unaccessible to ground laboratory experiments. 
The first experimental information on 21 cm cosmological global signal   allows already to place 
new constraints on dark matter models probing a so far unexplored stage in the Universe history, the so-called dark ages at redshifts
$z \gtrsim 10$.  Finally, the discovery of gravitational waves opens an incredible variety of new tests on models associated to the production
of stochastic backgrounds and/or primordial black holes. 

Clearly history teaches us that one should not be overly optimistic about getting close  to the solution to the problem of the origin of matter in the universe: too many times great announcements have then been followed by great disappointments. However, it is fair to say that we are certainly entering a new era where the variety of theoretical models, often seen as a sign of a perilous navigation in uncharted waters, is at least counterbalanced  by an incredible variety of new phenomenological tools that can guide us through the journey, making everything extremely exciting. 
For these reasons, finding a solution to the origin of matter in the universe should certainly inspire new explorers eager to embark in such a journey toward new lands.

\vspace{-4mm}
\subsection*{Acknowledgments}

I wish to thank Wilfried Buchm\"uller, Marco Drewes, Jacopo Ghiglieri and Graham White for useful comments and suggestions.
I also wish to thank Enrico Bertuzzo, Steve Blanchet, Wilfried Buchm\"uller, Marco Chianese, Ferruccio Feruglio, David Jones, Sophie King, Steve King, 
Danny Marfatia, Luca Marzola,  Enrico Nardi, Michael Pl\"{u}macher, Michele Re Fiorentin, Tony Riotto, Sergio Palomares-Ruiz, Rome Samanta, Ye-Ling Zhou  for fruitful collaborations
on topics discussed in this review. We are all indebted to Steven
Weinberg for his  scientific breakthroughs. Personally, I was
greatly inspired, as a teenager,  by the reading of the
\textit{The first three minutes}~\cite{first}, the unsurpassed
popular science book masterpiece. 
I acknowledge financial support from the STFC UK Consolidated Grant ST/T000775/1.


\begin{thebibliography}{99}
\itemsep -2pt 
\bibitem{higgsdiscovery}
S.~Chatrchyan \textit{et al.} [CMS],
{\em Observation of a New Boson at a Mass of 125 GeV with the CMS Experiment at the LHC},
Phys. Lett. B \textbf{716} (2012), 30-61
[arXiv:1207.7235 [hep-ex]];
G.~Aad \textit{et al.} [ATLAS],
{\em Observation of a new particle in the search for the Standard Model Higgs boson with the ATLAS detector at the LHC},
Phys. Lett. B \textbf{716} (2012), 1-29
[arXiv:1207.7214 [hep-ex]].

\bibitem{BEH}
F.~Englert and R.~Brout,
{\em Broken Symmetry and the Mass of Gauge Vector Mesons},
Phys. Rev. Lett. \textbf{13} (1964), 321-323;
P.~W.~Higgs,
{\em Broken symmetries, massless particles and gauge fields},
Phys. Lett. \textbf{12} (1964), 132-133;
P.~W.~Higgs,
{\em Broken Symmetries and the Masses of Gauge Bosons},
Phys. Rev. Lett. \textbf{13} (1964), 508-509.

\bibitem{lhcb}
R.~Aaij \textit{et al.} [LHCb],
{\em Test of lepton universality in beauty-quark decays},
[arXiv:2103.11769 [hep-ex]].

\bibitem{gm2flab}
B.~Abi \textit{et al.} [Muon g-2],
{\em Measurement of the Positive Muon Anomalous Magnetic Moment to 0.46 ppm},
Phys. Rev. Lett. \textbf{126} (2021), 141801
[arXiv:2104.03281 [hep-ex]].

\bibitem{phenoprediction}
T.~Aoyama, N.~Asmussen, M.~Benayoun, J.~Bijnens, T.~Blum, M.~Bruno, I.~Caprini, C.~M.~Carloni Calame, M.~C\`e and G.~Colangelo, \textit{et al.},
{\em The anomalous magnetic moment of the muon in the Standard Model},
Phys. Rept. \textbf{887} (2020), 1-166
doi:10.1016/j.physrep.2020.07.006
[arXiv:2006.04822 [hep-ph]].

\bibitem{bmw}
S.~Borsanyi, Z.~Fodor, J.~N.~Guenther, C.~Hoelbling, S.~D.~Katz, L.~Lellouch, T.~Lippert, K.~Miura, L.~Parato and K.~K.~Szabo, \textit{et al.}
{\em Leading hadronic contribution to the muon 2 magnetic moment from lattice QCD},
[arXiv:2002.12347 [hep-lat]].

\bibitem{hu}
W.~Hu, N.~Sugiyama and J.~Silk,
{\em The Physics of microwave background anisotropies},
Nature \textbf{386} (1997), 37-43
[arXiv:astro-ph/9604166 [astro-ph]].

\bibitem{planck18}
Y.~Akrami {\it et al.} [Planck Collaboration],
  {\em Planck 2018 results. I. Overview and the cosmological legacy of Planck},
  arXiv:1807.06205 [astro-ph.CO].

\bibitem{zeldovich}
Y.~B.~Zeldovich and I.~D.~Novikov,
{\em RELATIVISTIC ASTROPHYSICS. VOL. 2. THE STRUCTURE AND EVOLUTION OF THE UNIVERSE}.

\bibitem{Steigman:2008ap}
G.~Steigman,
{\em When Clusters Collide: Constraints On Antimatter On The Largest Scales},
JCAP \textbf{10} (2008), 001
[arXiv:0808.1122 [astro-ph]].

\bibitem{glashow}
A.~G.~Cohen, A.~De Rujula and S.~L.~Glashow,
{\em A Matter - antimatter universe?},
Astrophys. J. \textbf{495} (1998), 539-549
[arXiv:astro-ph/9707087 [astro-ph]].

\bibitem{ting}
S. Ting, {\em The First Five Years of the Alpha Magnetic Spectrometer} on the ISS (2016).

\bibitem{salati}
V.~Poulin, P.~Salati, I.~Cholis, M.~Kamionkowski and J.~Silk,
{\em Where do the AMS-02 antihelium events come from?},
Phys. Rev. D \textbf{99} (2019) no.2, 023016
[arXiv:1808.08961 [astro-ph.HE]].

\bibitem{AD}
I.~Affleck and M.~Dine,
{\em A New Mechanism for Baryogenesis},
Nucl. Phys. B \textbf{249} (1985), 361-380.

\bibitem{dolgov1}
S.~I.~Blinnikov, A.~D.~Dolgov and K.~A.~Postnov,
{\em Antimatter and antistars in the universe and in the Galaxy},
Phys. Rev. D \textbf{92} (2015) no.2, 023516
[arXiv:1409.5736 [astro-ph.HE]].

\bibitem{dolgovlectures}
A.~D.~Dolgov,
{\em Baryogenesis, 30 years after},
[arXiv:hep-ph/9707419 [hep-ph]].

\bibitem{sakharov}
A.~D.~Sakharov,
{\em Violation of CP Invariance, C asymmetry, and baryon asymmetry of the universe},
Sov. Phys. Usp. \textbf{34} (1991) no.5, 392-393.

\bibitem{dolgovreport}
A.~D.~Dolgov, {\em NonGUT baryogenesis},
Phys. Rept. \textbf{222} (1992), 309-386.

\bibitem{riottotrodden}
A.~Riotto and M.~Trodden,
{\em Recent progress in baryogenesis},
Ann. Rev. Nucl. Part. Sci. \textbf{49} (1999), 35-75
[arXiv:hep-ph/9901362 [hep-ph]].

\bibitem{focusissue}
P.~Di Bari, A.~Masiero and R.~Mohapatra,
{\em Focus on the origin of matter},
New J. Phys. \textbf{15} (2013), 035030.


\bibitem{Garbrecht:2018mrp}
B.~Garbrecht,
{\em Why is there more matter than antimatter? Calculational methods for leptogenesis and electroweak baryogenesis},
Prog. Part. Nucl. Phys. \textbf{110} (2020), 103727
[arXiv:1812.02651 [hep-ph]].


\bibitem{buchmullerbodeker}
D.~Bodeker and W.~Buchmuller,
{\em Baryogenesis from the weak scale to the GUT scale},
[arXiv:2009.07294 [hep-ph]].



\bibitem{Yoshimura:1978ex}
M.~Yoshimura,
{\em Unified Gauge Theories and the Baryon Number of the Universe},
Phys. Rev. Lett. \textbf{41} (1978), 281-284
[erratum: Phys. Rev. Lett. \textbf{42} (1979), 746].

\bibitem{Toussaint:1978br}
D.~Toussaint, S.~B.~Treiman, F.~Wilczek and A.~Zee,
{\em Matter - Antimatter Accounting, Thermodynamics, and Black Hole Radiation},
Phys. Rev. D \textbf{19} (1979), 1036-1045.

\bibitem{Ellis:1978xg}
J.~R.~Ellis, M.~K.~Gaillard and D.~V.~Nanopoulos,
{\em Baryon Number Generation in Grand Unified Theories},
Phys. Lett. B \textbf{80} (1979), 360
[erratum: Phys. Lett. B \textbf{82} (1979), 464].

\bibitem{Weinberg:1979bt}
S.~Weinberg, {\em Cosmological Production of Baryons},
Phys. Rev. Lett. \textbf{42} (1979), 850-853.

\bibitem{Harvey:1981yk}
J.~A.~Harvey, E.~W.~Kolb, D.~B.~Reiss and S.~Wolfram,
{\em Calculation of Cosmological Baryon Asymmetry in Grand Unified Gauge Models},
Nucl. Phys. B \textbf{201} (1982), 16-100

\bibitem{Kolb:1983ni}
E.~W.~Kolb and M.~S.~Turner,
{\em Grand Unified Theories and the Origin of the Baryon Asymmetry},
Ann. Rev. Nucl. Part. Sci. \textbf{33} (1983), 645-696

\bibitem{fy}
M.~Fukugita and T.~Yanagida,
  {\em Baryogenesis Without Grand Unification},
  Phys.\ Lett.\ B {\bf 174} (1986) 45.

\bibitem{kuzmin}
V.~A.~Kuzmin, V.~A.~Rubakov and M.~E.~Shaposhnikov,
{\em On the Anomalous Electroweak Baryon Number Nonconservation in the Early Universe},
Phys. Lett. B \textbf{155} (1985), 36.


\bibitem{Turok:1990zg}
N.~Turok and J.~Zadrozny,
{\em Electroweak baryogenesis in the two doublet model},
Nucl. Phys. B \textbf{358} (1991), 471-493.


\bibitem{Nelson:1991ab}
A.~E.~Nelson, D.~B.~Kaplan and A.~G.~Cohen,
{\em Why there is something rather than nothing: Matter from weak interactions}
Nucl. Phys. B \textbf{373} (1992), 453-478.

\bibitem{Pietroni:1992in}
M.~Pietroni,
{\em The Electroweak phase transition in a nonminimal supersymmetric model},
Nucl. Phys. B \textbf{402} (1993), 27-45
[arXiv:hep-ph/9207227 [hep-ph]].

\bibitem{Huber:2006wf}
S.~J.~Huber, T.~Konstandin, T.~Prokopec and M.~G.~Schmidt,
{\em Electroweak Phase Transition and Baryogenesis in the nMSSM},
Nucl. Phys. B \textbf{757} (2006), 172-196
[arXiv:hep-ph/0606298 [hep-ph]].


\bibitem{Hawking:1974rv}
S.~W.~Hawking, {\em Black hole explosions}, Nature \textbf{248} (1974), 30-31.

\bibitem{Zeldovich:1976vw}
Y.~B.~Zeldovich,
{\em Charge Asymmetry of the Universe Due to Black Hole Evaporation and Weak Interaction Asymmetry},
Pisma Zh. Eksp. Teor. Fiz. \textbf{24} (1976), 29-32.

\bibitem{Turner:1979bt}
M.~S.~Turner, {\em BARYON PRODUCTION BY PRIMORDIAL BLACK HOLES},
Phys. Lett. B \textbf{89} (1979), 155-159.

\bibitem{Dolgov:1980gk}
A.~D.~Dolgov,
{\em HIDING OF THE CONSERVED (ANTI)-BARYONIC CHARGE INTO BLACK HOLES},
Phys. Rev. D \textbf{24} (1981), 1042.


\bibitem{CK}
A.~G.~Cohen and D.~B.~Kaplan,
{\em Thermodynamic Generation of the Baryon Asymmetry},
Phys. Lett. B \textbf{199} (1987), 251-258.

\bibitem{Davoudiasl:2004gf}
H.~Davoudiasl, R.~Kitano, G.~D.~Kribs, H.~Murayama and P.~J.~Steinhardt,
{\em Gravitational baryogenesis},
Phys. Rev. Lett. \textbf{93} (2004), 201301
[arXiv:hep-ph/0403019 [hep-ph]].


\bibitem{Krauss:1999ng}
L.~M.~Krauss and M.~Trodden,
{\em Baryogenesis below the electroweak scale},
Phys. Rev. Lett. \textbf{83} (1999), 1502-1505
[arXiv:hep-ph/9902420 [hep-ph]].

\bibitem{kaplannelson}
A.~G.~Cohen, D.~B.~Kaplan and A.~E.~Nelson,
{\em Progress in electroweak baryogenesis},
Ann. Rev. Nucl. Part. Sci. \textbf{43} (1993), 27-70
[arXiv:hep-ph/9302210 [hep-ph]].

\bibitem{rubakovshaposhnikov}
V.~A.~Rubakov and M.~E.~Shaposhnikov,
{\em Electroweak baryon number nonconservation in the early universe and in high-energy collisions},
Usp. Fiz. Nauk \textbf{166} (1996), 493-537
[arXiv:hep-ph/9603208 [hep-ph]].

\bibitem{trodden}
M.~Trodden,
{\em Electroweak baryogenesis},
Rev. Mod. Phys. \textbf{71} (1999), 1463-1500
[arXiv:hep-ph/9803479 [hep-ph]].

\bibitem{gwhitebook}
G.~A.~White,
{\em A Pedagogical Introduction to Electroweak Baryogenesis},
2016 Morgan \& Claypool Publishers, doi:10.1088/978-1-6817-4457-5.



\bibitem{adler}
S.~L.~Adler,
{\em Axial vector vertex in spinor electrodynamics},
Phys. Rev. \textbf{177} (1969), 2426-2438.

\bibitem{belljackiw}
J.~S.~Bell and R.~Jackiw,
{\em A PCAC puzzle: $\pi^0 \to \gamma \gamma$ in the $\sigma$ model},
Nuovo Cim. A \textbf{60} (1969), 47-61.

\bibitem{thooft}
G.~'t Hooft, {\em Symmetry Breaking Through Bell-Jackiw Anomalies},
Phys. Rev. Lett. \textbf{37} (1976), 8-11.

\bibitem{manton}
F.~R.~Klinkhamer and N.~S.~Manton,
{\em A Saddle Point Solution in the Weinberg-Salam Theory},
Phys. Rev. D \textbf{30} (1984), 2212.

\bibitem{quiroslectures}
For a pedagogical discussion of finite temperature effects in electroweak baryogenesis see: 
M.~Quiros, {\em Finite temperature field theory and phase transitions},
[arXiv:hep-ph/9901312 [hep-ph]].


\bibitem{rummukainen}
M.~D'Onofrio, K.~Rummukainen and A.~Tranberg,
  {\em Sphaleron Rate in the Minimal Standard Model},
  Phys.\ Rev.\ Lett.\  {\bf 113} (2014) no.14,  141602
  [arXiv:1404.3565].

\bibitem{Patel:2011th}
H.~H.~Patel and M.~J.~Ramsey-Musolf,
{\em Baryon Washout, Electroweak Phase Transition, and Perturbation Theory},
JHEP \textbf{07} (2011), 029
[arXiv:1101.4665 [hep-ph]].




\bibitem{csikor}
F.~Csikor, Z.~Fodor and J.~Heitger,
{\em Endpoint of the hot electroweak phase transition},
Phys. Rev. Lett. \textbf{82} (1999), 21-24
[arXiv:hep-ph/9809291 [hep-ph]].

\bibitem{Ellis:2019flb}
S.~A.~R.~Ellis, S.~Ipek and G.~White,
{\em Electroweak Baryogenesis from Temperature-Varying Couplings},
JHEP \textbf{08} (2019), 002
[arXiv:1905.11994 [hep-ph]].

\bibitem{quiros}
M.~Carena, G.~Nardini, M.~Quiros and C.~E.~M.~Wagner,
{\em The Baryogenesis Window in the MSSM},
Nucl. Phys. B \textbf{812} (2009), 243-263
[arXiv:0809.3760 [hep-ph]].

\bibitem{curtin}
D.~Curtin, P.~Jaiswal and P.~Meade,
{\em Excluding Electroweak Baryogenesis in the MSSM},
JHEP \textbf{08} (2012), 005
[arXiv:1203.2932 [hep-ph]].

\bibitem{profumo}
S.~Liebler, S.~Profumo and T.~Stefaniak,
{\em Light Stop Mass Limits from Higgs Rate Measurements in the MSSM: Is MSSM Electroweak Baryogenesis Still Alive After All?},
JHEP \textbf{04} (2016), 143
[arXiv:1512.09172 [hep-ph]].

\bibitem{NMSSM}
S.~V.~Demidov, D.~S.~Gorbunov and D.~V.~Kirpichnikov,
{\em Split NMSSM with electroweak baryogenesis},
JHEP \textbf{11} (2016), 148
[erratum: JHEP \textbf{08} (2017), 080]
[arXiv:1608.01985 [hep-ph]].

\bibitem{Athron:2019teq}
For a recent review on electroweak baryogenesis in the NMSSM see: 
P.~Athron, C.~Balazs, A.~Fowlie, G.~Pozzo, G.~White and Y.~Zhang,
{\em Strong first-order phase transitions in the NMSSM \textemdash{} a comprehensive survey},
JHEP \textbf{11} (2019), 151
[arXiv:1908.11847 [hep-ph]].

\bibitem{twohiggs}
J.~M.~Cline, K.~Kainulainen and M.~Trott,
{\em Electroweak Baryogenesis in Two Higgs Doublet Models and B meson anomalies},
JHEP \textbf{11} (2011), 089
[arXiv:1107.3559 [hep-ph]].

%
\bibitem{ACME:2018yjb}
V.~Andreev \textit{et al.} [ACME],
{\em Improved limit on the electric dipole moment of the electron},
Nature \textbf{562} (2018) no.7727, 355-360.


\bibitem{cline}
J.~M.~Cline,
{\em Is electroweak baryogenesis dead?},
Phil. Trans. Roy. Soc. Lond. A \textbf{376} (2018) no.2114, 20170116
[arXiv:1704.08911 [hep-ph]].

\bibitem{Caprini:2015zlo}
C.~Caprini, M.~Hindmarsh, S.~Huber, T.~Konstandin, J.~Kozaczuk, G.~Nardini, J.~M.~No, A.~Petiteau, P.~Schwaller and G.~Servant, \textit{et al.}
{\em Science with the space-based interferometer eLISA. II: Gravitational waves from cosmological phase transitions}, JCAP \textbf{04} (2016), 001
[arXiv:1512.06239 [astro-ph.CO]].

\bibitem{pedestrians}
W.~Buchmuller, P.~Di Bari and M.~Plumacher,
{\em Leptogenesis for pedestrians},
Annals Phys. \textbf{315} (2005), 305-351
[arXiv:hep-ph/0401240 [hep-ph]].

\bibitem{kolbturner}
 E.~W.~Kolb and M.~S.~Turner, {\em The Early universe},
  Front.\ Phys.\  {\bf 69} (1990) 1.
   
\bibitem{luty}
M.~A.~Luty,
{\em Baryogenesis via leptogenesis},
Phys. Rev. D \textbf{45} (1992), 455-465;
M.~Plumacher,
{\em Baryogenesis and lepton number violation},
Z. Phys. C \textbf{74} (1997), 549-559
[arXiv:hep-ph/9604229 [hep-ph]].

\bibitem{bcst}
R.~Barbieri, P.~Creminelli, A.~Strumia and N.~Tetradis,
{\em Baryogenesis through leptogenesis},
Nucl.\ Phys.\ B {\bf 575} (2000) 61
[arXiv:hep-ph/9911315].

\bibitem{cmb}
W.~Buchmuller, P.~Di Bari and M.~Plumacher,
{\em Cosmic microwave background, matter - antimatter asymmetry and neutrino masses},
Nucl. Phys. B \textbf{643} (2002), 367-390
[erratum: Nucl. Phys. B \textbf{793} (2008), 362]
[arXiv:hep-ph/0205349 [hep-ph]].

\bibitem{dolgovkolbwolfram}
A.D.~Dolgov, Sov. J. Nucl. Phys. {\bf 32}, 831 (1980);
E.~W.~Kolb, S.~Wolfram, \np{172}{1980}{224}.


\bibitem{Super-Kamiokande:2016exg}
K.~Abe \textit{et al.} [Super-Kamiokande],
{\em Search for proton decay via $p \to e^+\pi^0$ and $p \to \mu^+\pi^0$ in 0.31  megaton 
years exposure of the Super-Kamiokande water Cherenkov detector},
Phys. Rev. D \textbf{95} (2017) no.1, 012004
[arXiv:1610.03597 [hep-ex]].

\bibitem{Kumekawa:1994gx}
K.~Kumekawa, T.~Moroi and T.~Yanagida,
{\em Flat potential for inflaton with a discrete R invariance in supergravity},
Prog. Theor. Phys. \textbf{92} (1994), 437-448
[arXiv:hep-ph/9405337 [hep-ph]].

\bibitem{Dolgov:1989us}
A.~D.~Dolgov and D.~P.~Kirilova,
{\em ON PARTICLE CREATION BY A TIME DEPENDENT SCALAR FIELD},
Sov. J. Nucl. Phys. \textbf{51} (1990), 172-177 JINR-E2-89-321.


\bibitem{Traschen:1990sw}
J.~H.~Traschen and R.~H.~Brandenberger,
{\em Particle Production During Out-of-equilibrium Phase Transitions},
Phys. Rev. D \textbf{42} (1990), 2491-2504.

\bibitem{Kolb:1996jt}
E.~W.~Kolb, A.~D.~Linde and A.~Riotto,
{\em GUT baryogenesis after preheating},
Phys. Rev. Lett. \textbf{77} (1996), 4290-4293
[arXiv:hep-ph/9606260 [hep-ph]].

\bibitem{Giudice:2000ex}
G.~F.~Giudice, E.~W.~Kolb and A.~Riotto,
{\em Largest temperature of the radiation era and its cosmological implications},
Phys. Rev. D \textbf{64} (2001), 023508
[arXiv:hep-ph/0005123 [hep-ph]].

\bibitem{Hawking:1982ga}
S.~W.~Hawking, I.~G.~Moss and J.~M.~Stewart,
{\em Bubble Collisions in the Very Early Universe},
Phys. Rev. D \textbf{26} (1982), 2681.

\bibitem{Polnarev:1985btg}
A.~G.~Polnarev and M.~Y.~Khlopov,
{\em COSMOLOGY, PRIMORDIAL BLACK HOLES, AND SUPERMASSIVE PARTICLES},
Sov. Phys. Usp. \textbf{28} (1985), 213-232.

\bibitem{Barrow:1990he}
J.~D.~Barrow, E.~J.~Copeland, E.~W.~Kolb and A.~R.~Liddle,
{\em Baryogenesis in extended inflation. 2. Baryogenesis via primordial black holes},
Phys. Rev. D \textbf{43} (1991), 984-994.

\bibitem{Baumann:2007yr}
D.~Baumann, P.~J.~Steinhardt and N.~Turok,
{\em Primordial Black Hole Baryogenesis},
[arXiv:hep-th/0703250 [hep-th]].

\bibitem{Hook:2014mla}
A.~Hook, {\em Baryogenesis from Hawking Radiation},
Phys. Rev. D \textbf{90} (2014) no.8, 083535
[arXiv:1404.0113 [hep-ph]].


\bibitem{Hamada:2016jnq}
Y.~Hamada and S.~Iso,
{\em Baryon asymmetry from primordial black holes},
PTEP \textbf{2017} (2017) no.3, 033B02
[arXiv:1610.02586 [hep-ph]].

\bibitem{Hooper:2020otu}
D.~Hooper and G.~Krnjaic,
{\em GUT Baryogenesis With Primordial Black Holes},
Phys. Rev. D \textbf{103} (2021) no.4, 043504
[arXiv:2010.01134 [hep-ph]].

\bibitem{Katz:2016adq}
A.~Katz and A.~Riotto,
{\em Baryogenesis and Gravitational Waves from Runaway Bubble Collisions},
JCAP \textbf{11} (2016), 011
[arXiv:1608.00583 [hep-ph]].

\bibitem{Azatov:2021irb}
A.~Azatov, M.~Vanvlasselaer and W.~Yin,
{\em Baryogenesis via relativistic bubble walls},
[arXiv:2106.14913 [hep-ph]].

\bibitem{darkmatterhints}
For an exhaustive history of dark matter and references see 
G.~Bertone and D.~Hooper, {\em History of dark matter},
Rev. Mod. Phys. \textbf{90} (2018) no.4, 045002
[arXiv:1605.04909 [astro-ph.CO]].

\bibitem{zwicky}
F.~Zwicky, {\em Die Rotverschiebung von extragalaktischen Nebeln},
Helv. Phys. Acta \textbf{6} (1933), 110-127.

\bibitem{freeman}
K.~C.~Freeman, {\em On the disks of spiral and SO Galaxies},
Astrophys. J. \textbf{160} (1970), 811.

\bibitem{rubin}
V.~C.~Rubin and W.~K.~Ford, Jr.,
{\em Rotation of the Andromeda Nebula from a Spectroscopic Survey of Emission Regions},
Astrophys. J. \textbf{159} (1970), 379-403.

\bibitem{Gross:2018ivp}
C.~Gross, A.~Polosa, A.~Strumia, A.~Urbano and W.~Xue,
{\em Dark Matter in the Standard Model?},
Phys. Rev. D \textbf{98} (2018) no.6, 063005
[arXiv:1803.10242 [hep-ph]].

\bibitem{hawking}
S.~Hawking, {\em Gravitationally collapsed objects of very low mass},
Mon. Not. Roy. Astron. Soc. \textbf{152} (1971), 75.

\bibitem{chapline}
G.~F.~Chapline,
{\em Cosmological effects of primordial black holes},
Nature \textbf{253} (1975) no.5489, 251-252.

\bibitem{Montero-Camacho:2019jte}
P.~Montero-Camacho, X.~Fang, G.~Vasquez, M.~Silva and C.~M.~Hirata,
{\em Revisiting constraints on asteroid-mass primordial black holes as dark matter candidates},
JCAP \textbf{08} (2019), 031
[arXiv:1906.05950 [astro-ph.CO]].


\bibitem{PBHreview}
For a recent review see A.~M.~Green and B.~J.~Kavanagh,
{\em Primordial Black Holes as a dark matter candidate},
[arXiv:2007.10722 [astro-ph.CO]].

\bibitem{bekenstein}
J.~D.~Bekenstein,
{\em Alternatives to Dark Matter: Modified Gravity as an Alternative to dark Matter},
[arXiv:1001.3876 [astro-ph.CO]].

\bibitem{mimetic}
For a review on how mimetic gravity addresses dark matter see L.~Sebastiani, S.~Vagnozzi and R.~Myrzakulov,
{\em Mimetic gravity: a review of recent developments and applications to cosmology and astrophysics},
Adv. High Energy Phys. \textbf{2017} (2017), 3156915.
[arXiv:1612.08661 [gr-qc]].

\bibitem{Knox:2019rjx}
L.~Knox and M.~Millea,
{\em Hubble constant hunter\textquoteright{}s guide},
Phys. Rev. D \textbf{101} (2020) no.4, 043533
[arXiv:1908.03663 [astro-ph.CO]].

 \bibitem{hut}
P.~Hut,
  {\em Limits on masses and number of neutral weakly interacting particles},
  Phys.\ Lett.\  {\bf 69B} (1977) 85;
  
\bibitem{leeweinberg}
B.~W.~Lee and S.~Weinberg,
  {\em Cosmological lower bound on heavy neutrino masses},
  Phys.\ Rev.\ Lett.\  {\bf 39} (1977) 165.
 
 
 \bibitem{dolgov}
M.~I.~Vysotsky, A.~D.~Dolgov and Y.~B.~Zeldovich,
  {\em Cosmological restriction on neutral lepton masses},
  JETP Lett.\  {\bf 26} (1977) 188
   [Pisma Zh.\ Eksp.\ Teor.\ Fiz.\  {\bf 26} (1977) 200].
  
   
 \bibitem{silkbertone}
 G.~Bertone, D.~Hooper and J.~Silk,
  {\em Particle dark matter: Evidence, candidates and constraints},
  Phys.\ Rept.\  {\bf 405} (2005) 279
  [hep-ph/0404175].
 
 \bibitem{feng} 
  J.~L.~Feng,
  {\em Dark matter candidates from particle physics and methods of detection},
  Ann.\ Rev.\ Astron.\ Astrophys.\  {\bf 48} (2010) 495
  [arXiv:1003.0904 [astro-ph.CO]].
  
  \bibitem{goldberg}
 H.~Goldberg,
  {\em Constraint on the photino mass from cosmology},
  Phys.\ Rev.\ Lett.\  {\bf 50} (1983) 1419
   Erratum: [Phys.\ Rev.\ Lett.\  {\bf 103} (2009) 099905].
  
 \bibitem{ellis}
J.~R.~Ellis, J.~S.~Hagelin, D.~V.~Nanopoulos, K.~A.~Olive and M.~Srednicki,
  {\em Supersymmetric relics from the Big Bang},
  Nucl.\ Phys.\ B {\bf 238} (1984) 453.
  
\bibitem{Balazs:2004bu}
C.~Balazs, M.~Carena and C.~E.~M.~Wagner,
{\em Dark matter, light stops and electroweak baryogenesis},
Phys. Rev. D \textbf{70} (2004), 015007
[arXiv:hep-ph/0403224 [hep-ph]].
 
\bibitem{Balazs:2004ae}
C.~Balazs, M.~Carena, A.~Menon, D.~E.~Morrissey and C.~E.~M.~Wagner,
{\em The Supersymmetric origin of matter},
Phys. Rev. D \textbf{71} (2005), 075002
[arXiv:hep-ph/0412264 [hep-ph]].
 
\bibitem{Balazs:2007pf}
C.~Balazs, M.~Carena, A.~Freitas and C.~E.~M.~Wagner,
{\em Phenomenology of the nMSSM from Colliders to Cosmology},
JHEP \textbf{06} (2007), 066
[arXiv:0705.0431 [hep-ph]].
  
\bibitem{roszkowski}
L.~Roszkowski, E.~M.~Sessolo and S.~Trojanowski,
  {\em WIMP dark matter candidates and searches - current issues and future prospects},
   Rep.\ Prog.\ Phys., (2018),  doi.org/10.1088/1361-6633/aab913 [arXiv:1707.06277 [hep-ph]].
  
  
  
\bibitem{Delahaye:2007fr}
T.~Delahaye, R.~Lineros, F.~Donato, N.~Fornengo and P.~Salati,
{\em Positrons from dark matter annihilation in the galactic halo: Theoretical uncertainties},
Phys. Rev. D \textbf{77} (2008), 063527
[arXiv:0712.2312 [astro-ph]].
  
  
\bibitem{Donato:2003xg}
F.~Donato, N.~Fornengo, D.~Maurin and P.~Salati,
{\em Antiprotons in cosmic rays from neutralino annihilation},
Phys. Rev. D \textbf{69} (2004), 063501
[arXiv:astro-ph/0306207 [astro-ph]].
 
\bibitem{Strong:2007nh}
A.~W.~Strong, I.~V.~Moskalenko and V.~S.~Ptuskin,
{\em Cosmic-ray propagation and interactions in the Galaxy},
Ann. Rev. Nucl. Part. Sci. \textbf{57} (2007), 285-327
[arXiv:astro-ph/0701517 [astro-ph]].
  
 \bibitem{griestB}
K.~Griest and D.~Seckel,
  {\em Three exceptions in the calculation of relic abundances},
  Phys.\ Rev.\ D {\bf 43} (1991) 3191.
  
  \bibitem{baer}
H.~Baer, K.~Y.~Choi, J.~E.~Kim and L.~Roszkowski,
  {\em Dark matter production in the early Universe: beyond the thermal WIMP paradigm},
  Phys.\ Rept.\  {\bf 555} (2015) 1
  [arXiv:1407.0017 [hep-ph]].
  
\bibitem{Bernal:2017kxu}
N.~Bernal, M.~Heikinheimo, T.~Tenkanen, K.~Tuominen and V.~Vaskonen,
{\em The Dawn of FIMP Dark Matter: A Review of Models and Constraints},
Int. J. Mod. Phys. A \textbf{32} (2017) no.27, 1730023
[arXiv:1706.07442 [hep-ph]].
  
\bibitem{Agrawal:2021dbo}
P.~Agrawal, M.~Bauer, J.~Beacham, A.~Berlin, A.~Boyarsky, S.~Cebrian, X.~Cid-Vidal, 
D.~d'Enterria, A.~De Roeck and M.~Drewes, \textit{et al.}
{\em Feebly-Interacting Particles:FIPs 2020 Workshop Report},
[arXiv:2102.12143 [hep-ph]].
  
\bibitem{FIMPs}
L.~J.~Hall, K.~Jedamzik, J.~March-Russell and S.~M.~West,
 {\em Freeze-In Production of FIMP Dark Matter},
  JHEP {\bf 1003} (2010) 080
  [arXiv:0911.1120 [hep-ph]].
  
\bibitem{warmordark}
P.~Di Bari, S.~F.~King and A.~Merle,
{\em Dark Radiation or Warm Dark Matter from long lived particle decays in the light of Planck},
Phys. Lett. B \textbf{724} (2013), 77-83
[arXiv:1303.6267 [hep-ph]].
  
\bibitem{gravitino}
M.~Y.~Khlopov and A.~D.~Linde, Phys.\ Lett.\ B {\bf 138} (1984) 265;
J.~R.~Ellis, J.~E.~Kim and D.~V.~Nanopoulos, Phys.\ Lett.\ B {\bf 145} (1984) 181;
K.~Kohri, T.~Moroi and A.~Yotsuyanagi, Phys.\ Rev.\ D {\bf 73} (2006) 123511;
 M.~Kawasaki, K.~Kohri, T.~Moroi and A.~Yotsuyanagi,
  {\em Big-Bang Nucleosynthesis and Gravitino},
  Phys.\ Rev.\ D {\bf 78} (2008) 065011
  [arXiv:0804.3745 [hep-ph]].

\bibitem{Peccei:1977hh}
R.~D.~Peccei and H.~R.~Quinn,
{\em CP Conservation in the Presence of Instantons},
Phys. Rev. Lett. \textbf{38} (1977), 1440-1443.
  
  
 \bibitem{wimpzillas}
D.~J.~H.~Chung, E.~W.~Kolb and A.~Riotto,
  {\em Superheavy dark matter},
  Phys.\ Rev.\ D {\bf 59} (1999) 023501
  [hep-ph/9802238];
 V.~Kuzmin and I.~Tkachev,
  {\em Ultrahigh-energy cosmic rays, superheavy long living particles, and matter creation after inflation},
  JETP Lett.\  {\bf 68} (1998) 271
   [Pisma Zh.\ Eksp.\ Teor.\ Fiz.\  {\bf 68} (1998) 255]
  [hep-ph/9802304].

\bibitem{davoudiasl}
H.~Davoudiasl and R.~N.~Mohapatra,
{\em On Relating the Genesis of Cosmic Baryons and Dark Matter},
New J. Phys. \textbf{14} (2012), 095011
[arXiv:1203.1247 [hep-ph]].

\bibitem{volkas}
K.~Petraki and R.~R.~Volkas,
{\em Review of asymmetric dark matter},
Int. J. Mod. Phys. A \textbf{28} (2013), 1330028
[arXiv:1305.4939 [hep-ph]].

\bibitem{nufit2020}
I.~Esteban, M.~C.~Gonzalez-Garcia, M.~Maltoni, T.~Schwetz and A.~Zhou,
{\em The fate of hints: updated global analysis of three-flavor neutrino oscillations},
[arXiv:2007.14792 [hep-ph]], NuFIT 5.0 (2020), www.nu-fit.org.

\bibitem{KATRIN}
M.~Aker {\it et al.} [KATRIN Collaboration],
  {\em Improved Upper Limit on the Neutrino Mass from a Direct Kinematic Method by KATRIN},
  Phys.\ Rev.\ Lett.\  {\bf 123} (2019) no.22,  221802
  [arXiv:1909.06048 [hep-ex]].

\bibitem{KATRIN2}
M.~Aker, A.~Beglarian, J.~Behrens, A.~Berlev, U.~Besserer, B.~Bieringer, F.~Block, B.~Bornschein, L.~Bornschein and M.~B\"ottcher, \textit{et al.}
{\em First direct neutrino-mass measurement with sub-eV sensitivity},
[arXiv:2105.08533 [hep-ex]].

\bibitem{kamlandzen}
A.~Gando {\it et al.} [KamLAND-Zen Collaboration],
 {\em Search for Majorana Neutrinos near the Inverted Mass Hierarchy Region with KamLAND-Zen},
  Phys.\ Rev.\ Lett.\  {\bf 117} (2016) no.8,  082503
   Addendum: [Phys.\ Rev.\ Lett.\  {\bf 117} (2016) no.10,  109903]
  [arXiv:1605.02889 [hep-ex]].

\bibitem{hannestad}
S.~Roy Choudhury and S.~Hannestad,
{\em Updated results on neutrino mass and mass hierarchy from cosmology with Planck 2018 likelihoods},
JCAP \textbf{07} (2020), 037
[arXiv:1907.12598 [astro-ph.CO]].

\bibitem{degouvea}
J.~M.~Berryman, A.~De Gouvêa, K.~J.~Kelly and Y.~Zhang,
{\em Lepton-Number-Charged Scalars and Neutrino Beamstrahlung},
Phys. Rev. D \textbf{97} (2018) no.7, 075030
[arXiv:1802.00009 [hep-ph]].

\bibitem{arkani}
N.~Arkani-Hamed, S.~Dimopoulos, G.~R.~Dvali and J.~March-Russell,
  {\em Neutrino masses from large extra dimensions},
  Phys.\ Rev.\ D {\bf 65} (2001) 024032
  [hep-ph/9811448].

\bibitem{neubert}
 Y.~Grossman and M.~Neubert,
  {\em Neutrino masses and mixings in nonfactorizable geometry},
  Phys.\ Lett.\ B {\bf 474} (2000) 361
  [hep-ph/9912408].

\bibitem{lindner}
K.~Dick, M.~Lindner, M.~Ratz and D.~Wright,
  {\em Leptogenesis with Dirac neutrinos},
  Phys.\ Rev.\ Lett.\  {\bf 84} (2000) 4039
  [hep-ph/9907562].

\bibitem{seesaw}
P.~Minkowski,
  {\em $\mu \to {\rm e}\,\gamma$ At A Rate Of One Out Of 1-Billion Muon Decays?},
  Phys.\ Lett.\  B {\bf 67} (1977) 421;
T. Yanagida, in Proceedings of the Workshop on Unified Theory and Baryon Number
of the Universe, eds. O. Sawada and A. Sugamoto (KEK, 1979) p.95;
  P.~Ramond, 
Invited talk given at Conference: C79-02-25
(Feb 1979) p.265-280, CALT-68-709,
  {\em The Family Group in Grand Unified Theories},
  hep-ph/9809459;
 M. Gell-Mann,
P. Ramond and R. Slansky, in Supergravity, eds. P. van Niewwenhuizen and D.
Freedman (North Holland, Amsterdam, 1979) Conf.Proc. C790927 p.315, PRINT-80-0576;
R.~Barbieri, D.~V.~Nanopoulos, G.~Morchio and F.~Strocchi,
Phys.\ Lett.\ B {\bf 90} (1980) 91;
R.~N.~Mohapatra and G.~Senjanovic,
Phys.\ Rev.\ Lett.\  {\bf 44} (1980) 912.

\bibitem{morisi}
D.~V.~Forero, S.~Morisi, M.~Tortola and J.~W.~F.~Valle,
{\em Lepton flavor violation and non-unitary lepton mixing in low-scale type-I seesaw},
JHEP \textbf{09} (2011), 142
[arXiv:1107.6009 [hep-ph]].

\bibitem{Antusch:2014woa}
S.~Antusch and O.~Fischer,
{\em Non-unitarity of the leptonic mixing matrix: Present bounds and future sensitivities},
JHEP \textbf{10} (2014), 094
[arXiv:1407.6607 [hep-ph]].

\bibitem{Chrzaszcz:2019inj}
M.~Chrzaszcz, M.~Drewes, T.~E.~Gonzalo, J.~Harz, S.~Krishnamurthy and C.~Weniger,
{\em A frequentist analysis of three right-handed neutrinos with GAMBIT},
Eur. Phys. J. C \textbf{80} (2020) no.6, 569
[arXiv:1908.02302 [hep-ph]].

\bibitem{Drewes:2013gca}
M.~Drewes, {\em The Phenomenology of Right Handed Neutrinos},
Int. J. Mod. Phys. E \textbf{22} (2013), 1330019
[arXiv:1303.6912 [hep-ph]].



\bibitem{flavoursymmetrylep}
E.~Bertuzzo, P.~Di Bari, F.~Feruglio and E.~Nardi,
{\em Flavor symmetries, leptogenesis and the absolute neutrino mass scale}, JHEP \textbf{11} (2009), 036
[arXiv:0908.0161 [hep-ph]].

\bibitem{SO10inspired}
A.~Y.~Smirnov,
  {\em Seesaw enhancement of lepton mixing},
  Phys.\ Rev.\ D {\bf 48} (1993) 3264
  [hep-ph/9304205];
W.~Buchmuller and M.~Plumacher,
  {\em Baryon asymmetry and neutrino mixing},
  Phys.\ Lett.\ B {\bf 389} (1996) 73 [hep-ph/9608308];
E.~Nezri and J.~Orloff,
  {\em Neutrino oscillations versus leptogenesis in $SO(10)$ models},
  JHEP {\bf 0304} (2003) 020
  [hep-ph/0004227];
F.~Buccella, D.~Falcone and F.~Tramontano,
  {\em Baryogenesis via leptogenesis in $SO(10)$ models},
  Phys.\ Lett.\ B {\bf 524} (2002) 241 [hep-ph/0108172];
G.~C.~Branco, R.~Gonzalez Felipe, F.~R.~Joaquim and M.~N.~Rebelo,
  {\em Leptogenesis, CP violation and neutrino data: What can we learn?},
  Nucl.\ Phys.\ B {\bf 640} (2002) 202 [hep-ph/0202030];

\bibitem{decrypting}
P.~Di Bari, L.~Marzola and M.~Re Fiorentin,
 {\em Decrypting $SO(10)$-inspired leptogenesis},
  Nucl.\ Phys.\ B {\bf 893} (2015) 122
  [arXiv:1411.5478 [hep-ph]].

\bibitem{SO10full}
P.~Di Bari and M.~Re Fiorentin,
{\em A full analytic solution of $SO(10)$-inspired leptogenesis},
JHEP \textbf{10} (2017), 029
[arXiv:1705.01935 [hep-ph]].

\bibitem{afs}
E.~K.~Akhmedov, M.~Frigerio and A.~Y.~Smirnov,
{\em Probing the seesaw mechanism with neutrino data and leptogenesis},
JHEP \textbf{09} (2003), 021
[arXiv:hep-ph/0305322 [hep-ph]].

\bibitem{2RH}
 S.~F.~King,
{\em Large mixing angle MSW and atmospheric neutrinos from single  RH
 neutrino dominance and $U(1)$ family symmetry},
  Nucl.\ Phys.\  B {\bf 576} (2000) 85  [arXiv:hep-ph/9912492];
P.~H.~Frampton, S.~L.~Glashow and T.~Yanagida,
 {\em Cosmological sign of neutrino CP violation},
  Phys.\ Lett.\  B {\bf 548} (2002) 119
  [arXiv:hep-ph/0208157].
P.~H.~Chankowski and K.~Turzynski,
  {\em Limits on $T_{\rm reh}$ for thermal leptogenesis with hierarchical neutrino masses},
  Phys.\ Lett.\  B {\bf 570} (2003) 198
  [arXiv:hep-ph/0306059];
A.~Ibarra and G.~G.~Ross,
  {\em Neutrino phenomenology: The case of two right handed neutrinos},
  Phys.\ Lett.\  B {\bf 591} (2004) 285.

\bibitem{weinberg}
S.~Weinberg,
  {\em Baryon and Lepton Nonconserving Processes},
  Phys.\ Rev.\ Lett.\  {\bf 43} (1979) 1566.


\bibitem{Giunti:2007ry}
C.~Giunti and C.~W.~Kim, {\em Fundamentals of Neutrino Physics and Astrophysics},
Oxford University Press (2007).

\bibitem{hirschwinter}
F.~Bonnet, M.~Hirsch, T.~Ota and W.~Winter,
 {\em Systematic study of the d=5 Weinberg operator at one-loop order},
  JHEP {\bf 1207} (2012) 153
  [arXiv:1204.5862 [hep-ph]].

\bibitem{typeII}
M.~Magg and C.~Wetterich,
  {\em Neutrino Mass Problem and Gauge Hierarchy},
  Phys.\ Lett.\  {\bf 94B} (1980) 61;
J.~Schechter and J.~W.~F.~Valle,
  {\em Neutrino Masses in $SU(2) \times U(1)$ Theories},
  Phys.\ Rev.\ D {\bf 22} (1980) 2227;
C.~Wetterich,
  {\em Neutrino Masses and the Scale of B-L Violation} ,
  Nucl.\ Phys.\ B {\bf 187} (1981) 343;
G.~Lazarides, Q.~Shafi and C.~Wetterich,
  {\em Proton Lifetime and Fermion Masses in an SO(10) Model},
  Nucl.\ Phys.\ B {\bf 181} (1981) 287;
 R.~N.~Mohapatra and G.~Senjanovic,
  {\em Neutrino Masses and Mixings in Gauge Models with Spontaneous Parity Violation},
  Phys.\ Rev.\ D {\bf 23} (1981) 165;
T.~P.~Cheng and L.~F.~Li,
  {\em Neutrino Masses, Mixings and Oscillations in SU(2) x U(1) Models of Electroweak Interactions},
  Phys.\ Rev.\ D {\bf 22} (1980) 2860.

\bibitem{typeIII}
R.~Foot, H.~Lew, X.~G.~He and G.~C.~Joshi,
  {\em Seesaw Neutrino Masses Induced by a Triplet of Leptons},
  Z.\ Phys.\ C {\bf 44} (1989) 441.

\bibitem{Cai:2017jrq}
Y.~Cai, J.~Herrero-Garc\'\i a, M.~A.~Schmidt, A.~Vicente and R.~R.~Volkas,
{\em From the trees to the forest: a review of radiative neutrino mass models},
Front. in Phys. \textbf{5} (2017), 63
[arXiv:1706.08524 [hep-ph]].


\bibitem{minimal}
S.~Blanchet and P.~Di Bari,
{\em The minimal scenario of leptogenesis},
New J. Phys. \textbf{14} (2012), 125012
[arXiv:1211.0512 [hep-ph]].


\bibitem{Xing:2020ald}
Z.~z.~Xing and Z.~h.~Zhao,
{\em The minimal seesaw and leptogenesis models},
Rept. Prog. Phys. \textbf{84} (2021) no.6, 066201
[arXiv:2008.12090 [hep-ph]].

\bibitem{harvey}
S.~Y.~Khlebnikov and M.~E.~Shaposhnikov,
{\em The Statistical Theory of Anomalous Fermion Number Nonconservation},
Nucl. Phys. B \textbf{308} (1988), 885-912;
J.~A.~Harvey, M.~S.~Turner, \pr{42}{1990}{3344}.

\bibitem{book}
P.~Di Bari, {\em Cosmology and the early Universe}, CRC Press, 2018.

\bibitem{flanz}
M.~Flanz, E.~A.~Paschos and U.~Sarkar,
{\em Baryogenesis from a lepton asymmetric universe},
Phys. Lett. B \textbf{345} (1995), 248-252
[erratum: Phys. Lett. B \textbf{384} (1996), 487-487; erratum: Phys. Lett. B \textbf{382} (1996), 447-447]
[arXiv:hep-ph/9411366 [hep-ph]].


\bibitem{Covi:1996wh}
L.~Covi, E.~Roulet and F.~Vissani,
{\em CP violating decays in leptogenesis scenarios},
Phys. Lett. B \textbf{384} (1996), 169-174
[arXiv:hep-ph/9605319 [hep-ph]].

\bibitem{buchplumi1}
W.~Buchmuller and M.~Plumacher,
{\em CP asymmetry in Majorana neutrino decays},
Phys. Lett. B \textbf{431} (1998), 354-362
[arXiv:hep-ph/9710460 [hep-ph]].

\bibitem{Pilaftsis:1997jf}
A.~Pilaftsis, {\em CP violation and baryogenesis due to heavy Majorana neutrinos},
Phys. Rev. D \textbf{56} (1997), 5431-5451
[arXiv:hep-ph/9707235 [hep-ph]].

\bibitem{plumacher}
M.~Plumacher,
{\em Baryogenesis and lepton number violation},
Z. Phys. C \textbf{74} (1997), 549-559
[arXiv:hep-ph/9604229 [hep-ph]].

\bibitem{orloff}
E.~Nezri and J.~Orloff,
{\em Neutrino oscillations versus leptogenesis in SO(10) models},
JHEP \textbf{04} (2003), 020
[arXiv:hep-ph/0004227 [hep-ph]].


\bibitem{predictions}
S.~Blanchet and P.~Di Bari,
{\em Flavor effects on leptogenesis predictions},
JCAP \textbf{03} (2007), 018
[arXiv:hep-ph/0607330 [hep-ph]].

\bibitem{di}
S.~Davidson and A.~Ibarra,
{\em A Lower bound on the right-handed neutrino mass from leptogenesis},
Phys. Lett. B \textbf{535} (2002), 25-32
[arXiv:hep-ph/0202239 [hep-ph]].


\bibitem{seesawmotion}
P.~Di Bari, M.~Re Fiorentin and R.~Samanta,
{\em Representing seesaw neutrino models and their motion in lepton flavour space},
JHEP \textbf{05} (2019), 011
[arXiv:1812.07720 [hep-ph]].


\bibitem{vissani}
F.~Vissani,
{\em Do experiments suggest a hierarchy problem?},
Phys. Rev. D \textbf{57} (1998), 7027-7030
[arXiv:hep-ph/9709409 [hep-ph]].

\bibitem{resonant}
A.~Pilaftsis and T.~E.~J.~Underwood,
{\em Resonant leptogenesis},
Nucl. Phys. B \textbf{692} (2004), 303-345
[arXiv:hep-ph/0309342 [hep-ph]].

\bibitem{Dev:2017wwc}
B.~Dev, M.~Garny, J.~Klaric, P.~Millington and D.~Teresi,
{\em Resonant enhancement in leptogenesis},
Int. J. Mod. Phys. A \textbf{33} (2018), 1842003
[arXiv:1711.02863 [hep-ph]].

\bibitem{Klaric:2021cpi}
J.~Klaric, M.~Shaposhnikov and I.~Timiryasov,
{\em Reconciling resonant leptogenesis and baryogenesis via neutrino oscillations},
[arXiv:2103.16545 [hep-ph]].

\bibitem{Blanchet:2009bu}
S.~Blanchet, Z.~Chacko, S.~S.~Granor and R.~N.~Mohapatra,
{\em Probing Resonant Leptogenesis at the LHC},
Phys. Rev. D \textbf{82} (2010), 076008
[arXiv:0904.2174 [hep-ph]].


\bibitem{Iso:2009nw}
S.~Iso, N.~Okada and Y.~Orikasa,
{\em The minimal B-L model naturally realized at TeV scale},
Phys. Rev. D \textbf{80} (2009), 115007
[arXiv:0909.0128 [hep-ph]].

\bibitem{Atre:2009rg}
A.~Atre, T.~Han, S.~Pascoli and B.~Zhang,
{\em The Search for Heavy Majorana Neutrinos},
JHEP \textbf{05} (2009), 030
[arXiv:0901.3589 [hep-ph]].

\bibitem{Deppisch:2015qwa}
F.~F.~Deppisch, P.~S.~Bhupal Dev and A.~Pilaftsis, {\em Neutrinos and Collider Physics},
New J. Phys. \textbf{17} (2015) no.7, 075019
[arXiv:1502.06541 [hep-ph]].

\bibitem{Cai:2017mow}
Y.~Cai, T.~Han, T.~Li and R.~Ruiz,
{\em Lepton Number Violation: Seesaw Models and Their Collider Tests},
Front. in Phys. \textbf{6} (2018), 40
[arXiv:1711.02180 [hep-ph]].

\bibitem{Albright:2003xb}
C.~H.~Albright and S.~M.~Barr,
{\em Leptogenesis in the type III seesaw mechanism},
Phys. Rev. D \textbf{69} (2004), 073010
[arXiv:hep-ph/0312224 [hep-ph]].


\bibitem{inverselinear}
A.~Abada, G.~Arcadi, V.~Domcke and M.~Lucente,
{\em Neutrino masses, leptogenesis and dark matter from small lepton number violation?},
JCAP \textbf{12} (2017), 024
[arXiv:1709.00415 [hep-ph]].


\bibitem{ARS}
E.~K.~Akhmedov, V.~A.~Rubakov and A.~Y.~Smirnov,
{\em Baryogenesis via neutrino oscillations},
Phys. Rev. Lett. \textbf{81} (1998), 1359-1362
[arXiv:hep-ph/9803255 [hep-ph]].

\bibitem{Abada:2018oly}
For recent results on ARS leptogenesis in connection to collider tests see:
A.~Abada, G.~Arcadi, V.~Domcke, M.~Drewes, J.~Klaric and M.~Lucente,
{\em Low-scale leptogenesis with three heavy neutrinos},
JHEP \textbf{01} (2019), 164
[arXiv:1810.12463 [hep-ph]].

\bibitem{Drewes:2021nqr}
For a recent. discussion of viable parameter space in ARS leptogenesis in combination with resonant leptogenesis see: 
M.~Drewes, Y.~Georis and J.~Klari\'c,
{\em Mapping the viable parameter space for testable leptogenesis},
[arXiv:2106.16226 [hep-ph]].

\bibitem{nardi1}
E.~Nardi, Y.~Nir, E.~Roulet and J.~Racker,
{\em The Importance of flavor in leptogenesis},
JHEP \textbf{01} (2006), 164
[arXiv:hep-ph/0601084 [hep-ph]].

\bibitem{riottolosada}
A.~Abada, S.~Davidson, F.~X.~Josse-Michaux, M.~Losada and A.~Riotto,
{\em Flavor issues in leptogenesis},
JCAP \textbf{04} (2006), 004
[arXiv:hep-ph/0601083 [hep-ph]].

\bibitem{reviewflavour}
P.~S.~B.~Dev, P.~Di Bari, B.~Garbrecht, S.~Lavignac, P.~Millington and D.~Teresi,
{\em Flavor effects in leptogenesis},
Int. J. Mod. Phys. A \textbf{33} (2018), 1842001
[arXiv:1711.02861 [hep-ph]].

\bibitem{Pilaftsis:2004xx}
A.~Pilaftsis,
{\em Resonant tau-leptogenesis with observable lepton number violation},
Phys. Rev. Lett. \textbf{95} (2005), 081602
[arXiv:hep-ph/0408103 [hep-ph]].

\bibitem{Pilaftsis:2005rv}
A.~Pilaftsis and T.~E.~J.~Underwood,
{\em Electroweak-scale resonant leptogenesis},
Phys. Rev. D \textbf{72} (2005), 113001
[arXiv:hep-ph/0506107 [hep-ph]].

\bibitem{vives}
O.~Vives,
{\em Flavor dependence of CP asymmetries and thermal leptogenesis with strong right-handed neutrino mass hierarchy},
Phys. Rev. D \textbf{73} (2006), 073006
[arXiv:hep-ph/0512160 [hep-ph]].

\bibitem{bounds}
S.~Blanchet and P.~Di Bari,
{\em New aspects of leptogenesis bounds},
Nucl. Phys. B \textbf{807} (2009), 155-187
[arXiv:0807.0743 [hep-ph]].

\bibitem{pascoliriotto}
S.~Pascoli, S.~T.~Petcov and A.~Riotto,
{\em Connecting low energy leptonic CP-violation to leptogenesis},
Phys. Rev. D \textbf{75} (2007), 083511
[arXiv:hep-ph/0609125 [hep-ph]].

\bibitem{branco}
G.~C.~Branco, R.~Gonzalez Felipe and F.~R.~Joaquim,
{\em A New bridge between leptonic CP violation and leptogenesis},
Phys. Lett. B \textbf{645} (2007), 432-436
[arXiv:hep-ph/0609297 [hep-ph]].

\bibitem{antuschblanchet}
S.~Antusch, S.~Blanchet, M.~Blennow and E.~Fernandez-Martinez,
{\em Non-unitary Leptonic Mixing and Leptogenesis},
JHEP \textbf{01} (2010), 017
[arXiv:0910.5957 [hep-ph]].

\bibitem{jturner}
K.~Moffat, S.~Pascoli, S.~T.~Petcov and J.~Turner,
{\em Leptogenesis from Low Energy $CP$ Violation},
JHEP \textbf{03} (2019), 034
[arXiv:1809.08251 [hep-ph]].

\bibitem{desimone}
A.~De Simone and A.~Riotto,
{\em On the impact of flavour oscillations in leptogenesis},
JCAP \textbf{02} (2007), 005
[arXiv:hep-ph/0611357 [hep-ph]].

\bibitem{bdraffelt}
S.~Blanchet, P.~Di Bari and G.~G.~Raffelt,
{\em Quantum Zeno effect and the impact of flavor in leptogenesis},
JCAP \textbf{03} (2007), 012
[arXiv:hep-ph/0611337 [hep-ph]].


\bibitem{geometry}
P.~Di Bari, {\em Seesaw geometry and leptogenesis},
Nucl. Phys. B \textbf{727} (2005), 318-354
[arXiv:hep-ph/0502082 [hep-ph]].



\bibitem{nardi2}
G.~Engelhard, Y.~Grossman, E.~Nardi and Y.~Nir,
{\em The Importance of $N_2$ leptogenesis},
Phys. Rev. Lett. \textbf{99} (2007), 081802
[arXiv:hep-ph/0612187 [hep-ph]].

\bibitem{2RHNlep}
S.~Antusch, P.~Di Bari, D.~A.~Jones and S.~F.~King,
{\em Leptogenesis in the Two Right-Handed Neutrino Model Revisited},
Phys. Rev. D \textbf{86} (2012), 023516
[arXiv:1107.6002 [hep-ph]].

\bibitem{riotto1}
P.~Di Bari and A.~Riotto,
{\em Successful type I Leptogenesis with SO(10)-inspired mass relations},
Phys. Lett. B \textbf{671} (2009), 462-469
[arXiv:0809.2285 [hep-ph]].

\bibitem{riotto2}
P.~Di Bari and A.~Riotto,
{\em Testing SO(10)-inspired leptogenesis with low energy neutrino experiments},
JCAP \textbf{04} (2011), 037
[arXiv:1012.2343 [hep-ph]].

\bibitem{fuller}
S.~Antusch, P.~Di Bari, D.~A.~Jones and S.~F.~King,
{\em A fuller flavour treatment of $N_2$-dominated leptogenesis},
Nucl. Phys. B \textbf{856} (2012), 180-209
[arXiv:1003.5132 [hep-ph]].

\bibitem{opportunity}
P.~Di Bari and R.~Samanta,
{\em The $SO(10)$-inspired leptogenesis timely opportunity},
JHEP \textbf{08} (2020), 124
[arXiv:2005.03057 [hep-ph]].



\bibitem{dw}
S.~Dodelson and L.~M.~Widrow,
{\em Sterile-neutrinos as dark matter},
Phys. Rev. Lett. \textbf{72} (1994), 17-20
[arXiv:hep-ph/9303287 [hep-ph]].

\bibitem{ad}
A.~Anisimov and P.~Di Bari,
{\em Cold Dark Matter from heavy Right-Handed neutrino mixing},
Phys. Rev. D \textbf{80} (2009), 073017
[arXiv:0812.5085 [hep-ph]].

\bibitem{asakablanchet}
T.~Asaka, S.~Blanchet and M.~Shaposhnikov,
{\em The $\nu$MSM, dark matter and neutrino masses},
Phys. Lett. B \textbf{631} (2005), 151-156
[arXiv:hep-ph/0503065 [hep-ph]].




\bibitem{Drewes:2016upu}
M.~Drewes, T.~Lasserre, A.~Merle, S.~Mertens, R.~Adhikari, M.~Agostini, N.~A.~Ky, T.~Araki, M.~Archidiacono and M.~Bahr, \textit{et al.}
{\em A White Paper on keV Sterile Neutrino Dark Matter},
JCAP \textbf{01} (2017), 025
[arXiv:1602.04816 [hep-ph]].

\bibitem{Boyarsky:2018tvu}
A.~Boyarsky, M.~Drewes, T.~Lasserre, S.~Mertens and O.~Ruchayskiy,
{\em Sterile neutrino Dark Matter},
Prog. Part. Nucl. Phys. \textbf{104} (2019), 1-45
[arXiv:1807.07938 [hep-ph]].

\bibitem{horiuchi}
S.~Horiuchi, P.~J.~Humphrey, J.~Onorbe, K.~N.~Abazajian, M.~Kaplinghat and S.~Garrison-Kimmel,
{\em Sterile neutrino dark matter bounds from galaxies of the Local Group},
Phys. Rev. D \textbf{89} (2014) no.2, 025017
[arXiv:1311.0282 [astro-ph.CO]].

\bibitem{shifuller}
X.~D.~Shi and G.~M.~Fuller,
{\em A New dark matter candidate: Nonthermal sterile neutrinos},
Phys. Rev. Lett. \textbf{82} (1999), 2832-2835
[arXiv:astro-ph/9810076 [astro-ph]].

\bibitem{dolgovhansen}
A.~D.~Dolgov and S.~H.~Hansen,
{\em Massive sterile neutrinos as warm dark matter},
Astropart. Phys. \textbf{16} (2002), 339-344
[arXiv:hep-ph/0009083 [hep-ph]].

\bibitem{bulbul}
E.~Bulbul, M.~Markevitch, A.~Foster, R.~K.~Smith, M.~Loewenstein and S.~W.~Randall,
{\em Detection of An Unidentified Emission Line in the Stacked X-ray spectrum of Galaxy Clusters},
Astrophys. J. \textbf{789} (2014), 13
[arXiv:1402.2301 [astro-ph.CO]].

\bibitem{boyarsky}
A.~Boyarsky, O.~Ruchayskiy, D.~Iakubovskyi and J.~Franse,
{\em Unidentified Line in X-Ray Spectra of the Andromeda Galaxy and Perseus Galaxy Cluster},
Phys. Rev. Lett. \textbf{113} (2014), 251301
[arXiv:1402.4119 [astro-ph.CO]].

\bibitem{abazajian}
K.~N.~Abazajian,
{\em Resonantly Produced 7 keV Sterile Neutrino Dark Matter Models and the Properties of Milky Way Satellites},
Phys. Rev. Lett. \textbf{112} (2014) no.16, 161303
[arXiv:1403.0954 [astro-ph.CO]].

\bibitem{dessert}
C.~Dessert, N.~L.~Rodd and B.~R.~Safdi,
{\em The dark matter interpretation of the 3.5-keV line is inconsistent with blank-sky observations},
Science \textbf{367} (2020), 1465
[arXiv:1812.06976 [astro-ph.CO]].

\bibitem{replies}
K.~N.~Abazajian,
{\em Technical Comment on ''The dark matter interpretation of the 3.5-keV line is inconsistent with blank-sky observations},
[arXiv:2004.06170 [astro-ph.HE]];
A.~Boyarsky, D.~Malyshev, O.~Ruchayskiy and D.~Savchenko,
{\em Technical comment on the paper of Dessert et al. ``The dark matter interpretation of the 3.5 keV line is inconsistent with blank-sky observations''},
[arXiv:2004.06601 [astro-ph.CO]];
C.~Dessert, N.~L.~Rodd and B.~R.~Safdi,
{\em Response to a comment on Dessert et al. `The dark matter interpretation of the 3.5 keV line is inconsistent with blank-sky observations'},
Phys. Dark Univ. \textbf{30} (2020), 100656
[arXiv:2006.03974 [astro-ph.CO]];
D.~Sicilian, N.~Cappelluti, E.~Bulbul, F.~Civano, M.~Moscetti and C.~S.~Reynolds,
{\em Probing the Milky Way's Dark Matter Halo for the 3.5 keV Line},
[arXiv:2008.02283 [astro-ph.HE]].

\bibitem{nuMSM}
T.~Asaka and M.~Shaposhnikov,
{\em The $\nu$MSM, dark matter and baryon asymmetry of the universe},
Phys. Lett. B \textbf{620} (2005), 17-26
[arXiv:hep-ph/0505013 [hep-ph]].

\bibitem{shap}
M.~Shaposhnikov,
{\em The nuMSM, leptonic asymmetries, and properties of singlet fermions},
JHEP \textbf{08} (2008), 008
[arXiv:0804.4542 [hep-ph]].



\bibitem{cdshap1}
L.~Canetti, M.~Drewes and M.~Shaposhnikov,
{\em Sterile Neutrinos as the Origin of Dark and Baryonic Matter},
Phys. Rev. Lett. \textbf{110} (2013) no.6, 061801
[arXiv:1204.3902 [hep-ph]].

\bibitem{cdshap2}
L.~Canetti, M.~Drewes, T.~Frossard and M.~Shaposhnikov,
{\em Dark Matter, Baryogenesis and Neutrino Oscillations from Right Handed Neutrinos},
Phys. Rev. D \textbf{87} (2013), 093006
[arXiv:1208.4607 [hep-ph]].

\bibitem{royshap}
A.~Roy and M.~Shaposhnikov,
{\em Resonant production of the sterile neutrino dark matter and fine-tunings in the [nu]MSM},
Phys. Rev. D \textbf{82} (2010), 056014
[arXiv:1006.4008 [hep-ph]].

\bibitem{ship}
S.~Alekhin, W.~Altmannshofer, T.~Asaka, B.~Batell, F.~Bezrukov, K.~Bondarenko, A.~Boyarsky, K.~Y.~Choi, C.~Corral and N.~Craig, \textit{et al.}
{\em A facility to Search for Hidden Particles at the CERN SPS: the SHiP physics case},
Rept. Prog. Phys. \textbf{79} (2016) no.12, 124201
[arXiv:1504.04855 [hep-ph]].

\bibitem{mathusla}
C.~Alpigiani \textit{et al.} [MATHUSLA],
{\em An Update to the Letter of Intent for MATHUSLA: Search for Long-Lived Particles at the HL-LHC},
[arXiv:2009.01693 [physics.ins-det]].

\bibitem{FASER:2019aik}
A.~Ariga \textit{et al.} [FASER],
{\em FASER: ForwArd Search ExpeRiment at the LHC},
[arXiv:1901.04468 [hep-ex]].


\bibitem{laine}
J.~Ghiglieri and M.~Laine,
{\em Sterile neutrino dark matter via GeV-scale leptogenesis?},
JHEP \textbf{07} (2019), 078
[arXiv:1905.08814 [hep-ph]].

\bibitem{Ghiglieri:2020ulj}
J.~Ghiglieri and M.~Laine,
{\em Sterile neutrino dark matter via coinciding resonances},
JCAP \textbf{07} (2020), 012
[arXiv:2004.10766 [hep-ph]].



\bibitem{XRISM} 
 [XRISM Science Team],
{\em Science with the X-ray Imaging and Spectroscopy Mission (XRISM)},
[arXiv:2003.04962 [astro-ph.HE]].

\bibitem{anisimov}
A.~Anisimov,
{\em Majorana Dark Matter},
[arXiv:hep-ph/0612024 [hep-ph]].

\bibitem{DiBari:2016guw}
P.~Di Bari, P.~O.~Ludl and S.~Palomares-Ruiz,
{\em Unifying leptogenesis, dark matter and high-energy neutrinos with right-handed neutrino mixing via Higgs portal}, 
JCAP \textbf{11} (2016), 044
[arXiv:1606.06238 [hep-ph]].

\bibitem{densitym}
P.~Di Bari, K.~Farrag, R.~Samanta and Y.~L.~Zhou,
{\em Density matrix calculation of the dark matter abundance in the Higgs induced right-handed neutrino mixing model},
JCAP \textbf{10} (2020), 029
[arXiv:1908.00521 [hep-ph]].

\bibitem{gwaves}
P.~Di Bari, D.~Marfatia and Y.~L.~Zhou,
{\em Gravitational waves from neutrino mass and dark matter genesis},
[arXiv:2001.07637 [hep-ph]].

\bibitem{lindnerrode}
C.~Jaramillo, M.~Lindner and W.~Rodejohann,
{\em Seesaw neutrino dark matter by freeze-out},
[arXiv:2004.12904 [hep-ph]].

\bibitem{Bhattacharya:2018ljs}
S.~Bhattacharya, I.~de Medeiros Varzielas, B.~Karmakar, S.~F.~King and A.~Sil,
{\em Dark side of the Seesaw},
JHEP \textbf{12} (2018), 007
[arXiv:1806.00490 [hep-ph]].

\bibitem{Chianese:2018dsz}
M.~Chianese and S.~F.~King,
{\em The Dark Side of the Littlest Seesaw: freeze-in, the two right-handed neutrino portal and leptogenesis-friendly fimpzillas},
JCAP \textbf{09} (2018), 027
[arXiv:1806.10606 [hep-ph]].

\bibitem{Ma:2006km}
E.~Ma,
{\em Verifiable radiative seesaw mechanism of neutrino mass and dark matter},
Phys. Rev. D \textbf{73} (2006), 077301
[arXiv:hep-ph/0601225 [hep-ph]].

\bibitem{Borah:2018rca}
D.~Borah, P.~S.~B.~Dev and A.~Kumar,
{\em TeV scale leptogenesis, inflaton dark matter and neutrino mass in a scotogenic model},
Phys. Rev. D \textbf{99} (2019) no.5, 055012
[arXiv:1810.03645 [hep-ph]].

\bibitem{first}
S.~Weinberg,
{\em The First Three Minutes. A Modern View of the Origin of the Universe}, New York, Basic Books, 1977.

\end{thebibliography}
\end{document}